\documentclass[twoside,12pt,usenames,dvipsnames]{article}
\usepackage{epsfig}
\usepackage{colortbl}
\usepackage{xcolor}
\usepackage{amsmath}
\usepackage{float}
\usepackage{mathtools}
\usepackage{relsize}
\usepackage{float}
\usepackage{braket}
\usepackage{hyperref}
\usepackage[T1]{fontenc}

\def\Journal#1#2#3#4{{#1} {#2} (#4) #3 }

\def\PRC{{\em Phys. Rev.} C}

\newcommand{\ma}{\color{magenta}}

\newcommand{\be}{\begin{equation}}
\newcommand{\ee}{\end{equation}}
\newcommand{\bea}{\begin{eqnarray}}
\newcommand{\eea}{\end{eqnarray}}

\newcommand{\vlwk}{$V_{{\rm low}\mbox{-}k}$}
\newcommand{\thetaeff}{$\Theta_{\rm eff}$}

\newcommand{\heff}{$H_{\rm eff}$}
\newcommand{\heffs}{$H_{\rm eff}$s}
\newcommand{\qbox}{$\hat{Q}$~box}
\newcommand{\tbox}{$\hat{\Theta}$~box}

\newcommand{\tptn}{$2\pi$-3NF}

\usepackage{cite}
\usepackage{color}
\usepackage{ulem}
\usepackage{amssymb}
\usepackage{bbold}
\newcommand{\Add}[1]{\textcolor{red}{#1}}

\newcommand{\vect}{\boldsymbol}
\DeclareSymbolFont{largesymbol}{OMX}{yhex}{m}{n}
\DeclareMathAccent{\Widehat}{\mathord}{largesymbol}{"62}

\begin{document}

\title{ \vspace{1cm} The role of three-nucleon potentials within the shell model: past and present}
\author{L.\ Coraggio,$^{1,2}$ G.\ De Gregorio,$^{1,2}$ T.\ Fukui,$^3$ \\
A.\ Gargano,$^{2}$  Y. Z.\ Ma,$^{4}$ Z. H. Cheng,$^{5}$ F. R. \ Xu,$^{5}$\\ 
\\
$^1$Dipartimento di Matematica e Fisica, Universit\`{a} degli Studi della Campania \\
``Luigi Vanvitelli'', viale Abramo Lincoln 5-I-81100 Caserta, Italy\\
$^2$Istituto Nazionale di Fisica Nucleare, Complesso Universitario di Monte \\
 S. Angelo, Via Cintia, I-80126 Napoli, Italy\\
 $^3$Faculty of Arts and Science, Kyushu University, Fukuoka 819-0395, Japan \\ 
$^4$Guangdong Provincial Key Laboratory of Nuclear Science, \\
Institute of Quantum Matter, South China Normal University, \\
Guangzhou 510006, China\\
$^5$School of Physics and State Key Laboratory of Nuclear Physics and Technology, \\
Peking University, Beijing 100871, China}

\maketitle
\begin{abstract}
We survey the impact of nuclear three-body forces on structure properties of nuclei within the shell model.
It has long been acknowledged, since the seminal works of Zuker and
coworkers, that three-body forces play a fundamental role in making
the monopole component of shell-model Hamiltonians, derived from realistic nucleon-nucleon potentials, able to reproduce the observed evolution of the shell structure.
In the vast majority of calculations, however, their effects have been
taken into account by shell-model practitioners by introducing {\it ad
hoc} modifications of the monopole matrix elements.
During last twenty years, a new theoretical approach, framed within
the chiral perturbation theory, has progressed in developing nuclear
potentials, where two- and many-body components are naturally and
consistently built in.
This new class of nuclear forces allows to carry out nuclear structure studies that are improving our ability to understand nuclear phenomena in a microscopic approach.
We provide in this work an update on the status of the nuclear shell
model based on realistic Hamiltonians that are derived from two- and
three-nucleon chiral potentials, focusing on the role of the
three-body component to provide the observed shell evolution and
closure properties, as well as the location of driplines.
To this end, we present the results of shell-model calculations and
their comparison with recent experimental measurements, which
enlighten the relevance of the inclusion of three-nucleon forces to
master our knowledge of the physics of atomic nuclei.
\end{abstract}

\tableofcontents

\section{Introduction}\label{intro}
The awareness of a defined role of many-body forces in the study of
nuclear systems traces back to the early stages of meson theory
\cite{Primakoff39}.
As a matter of fact, there is no guarantee that the picture of the
meson degrees of freedom to be frozen as soon as they have created the
interaction between two nucleons -- and then being responsible only for two-nucleon forces (2NFs) -- may work in any nuclear environment, in any
energy regime, and, more significantly, within any desired degree of
accuracy \cite{Friar84,Machleidt89}.

The theoretical progress in the construction of high-quality
nucleon-nucleon ($NN$) potentials able to reproduce
large sets of two-nucleon data \cite{Wiringa95,Machleidt01b}, as well
as the advancement of high-precision nuclear structure approaches
\cite{Barrett13,Carlson15}, has established that the sole use of
two-body nuclear forces does not provide a fully satisfactory
reproduction of the low-energy spectroscopy of light nuclear systems
\cite{Pudliner97}.

However, two main issues have slowed the development of nuclear
structure calculations employing three-nucleon forces (3NFs).
First, the little knowledge of a mechanism providing
many-body forces consistently with the nature of
the $NN$ interaction, as, for example, the pion-nucleon ($\pi N$)
scattering amplitude which is a fundamental quantity to construct 3NF
contributions in meson theory \cite{Machleidt89}.
Second, the difficulty to manage three-nucleon (3N) potentials
within a many-body system, whose solution requires formalisms that are
computationally extremely demanding.

As regards the first issue, a major breakthrough in the last two
decades has been the derivation of nuclear potentials in terms of the
chiral perturbation theory (ChPT) to build realistic $NN$
and 3N forces starting from a chiral Lagrangian.
This idea goes back to the seminal work of Weinberg
\cite{Weinberg79,Weinberg90,Weinberg91}, where the concept of an
effective field theory (EFT) has been introduced to study the
$S$-matrix for processes involving arbitrary numbers of low-momentum
pions and nucleons.
Within such an approach, the long-range component of the potential is
ruled by the symmetries of low-energy quantum chromodynamics (QCD) -- as the spontaneously
broken chiral symmetry -- and the short-range dynamics is absorbed into
a complete basis of contact terms that are proportional to low-energy
constants (LECs).
The LECs may be fitted to $NN$ and 3N data, but in a future they
could be determined by extrapolating them from lattice QCD (LQCD) results for light nuclear systems at the
physical $\pi$ mass \cite{Tews20,Drischler21}.

The main advantage of ChPT, as regards the need of consistency between
$NN$ and 3N potentials, is that it generates nuclear two- and
many-body forces on an equal footing
\cite{Weinberg92,vanKolck94,Machleidt11,Epelbaum20}.
In fact, most interaction vertices that appear in the three- and four-nucleon forces also occur in the two-nucleon ones.
Since the LECs associated to these vertices are shared with the chiral $NN$ potential, then consistency requires
that for the same vertices the same parameter values are used in the
many-body components of the nuclear Hamiltonian.

This new generation of two- and three-nucleon forces has been
successfully employed to study both spectroscopic properties and
scattering processes of light systems within the framework of the {\it ab initio}
no-core shell model (NCSM)  method
\cite{Barrett13,Navratil07a,Quaglioni07,Hupin13}.

However, the computational difficulty to manage a full treatment of
three-body correlations increases rapidly with the mass of the nuclei
under investigation making calculations unfeasible.
A successful approach to overcome such a hindrance is to resort to the
so-called normal-ordered decomposition of the three-body component of 
the nuclear Hamiltonian \cite{Hagen07a}. This is a convenient procedure in nuclear many-body methods which
 starts from an unperturbed reference state.
The basic idea is to use the Wick's theorem \cite{Wick50} and
re-arrange, with respect to the reference state, the three-body
component of the nuclear Hamiltonian into a sum of zero-, one-, two-,
and three-body terms \cite{Hebeler21}.
Then, a truncation is performed neglecting the residual three-body 
term,  which arises from the normal-ordering decomposition, and retaining only
the zero-, one-, and two-body contributions.
This approximation is obviously advantageous to simplify the
theoretical expressions characterizing different nuclear many-body
methods, and to drastically reduce the computational complexity. The validity of the normal-ordering approximation has been tested in light- and
medium-mass nuclei \cite{Roth12,Holt14,Gebrerufael16}, and it is currently a
building block of {\it ab initio} nuclear structure calculations where chiral $NN$  and 3N
potentials are employed
\cite{Hagen07a,Hagen07b,Carbone13a,Cipollone13,Hergert13,Soma14,Ekstrom15,Soma20}.

The question of the significance of including the effects of 3NFs in the derivation of the  effective shell-model  (SM) Hamiltonians \heff s~ becomes crucial when they are obtained by way of the many-body theory starting from realistic nuclear potentials
\cite{Suzuki80,Kuo95,Hjorth95,Stroberg19,Coraggio20c}.
In fact, for phenomenological \heff s~, where  the single-particle
(SP) energies and the two-body matrix elements (TBMEs) of the residual interaction are fitted or adjusted so as to
reproduce a certain set of spectroscopic observables
\cite{Cohen65,Brown88}, it is reasonable  to 
conclude that  3N forces are implicitly taken into account.

The first studies about the role of 3NFs in nuclear SM
calculations have been carried out by Zuker and coworkers
(see Ref. \cite{Caurier05}, where a complete list of reference can be found), who have extensively investigated the
characteristics of the TBMEs of the residual SM interaction
derived within the many-body perturbation theory (MBPT) from realistic 2NFs by Kuo and Brown \cite{Kuo66,Kuo68a}.
They have shown that these \heffs~need to be modified in their monopole component
to reproduce the experimental evolution of shell closures as a
function of the number of valence nucleons
\cite{Caurier94,Duflo99,Zuker00}. 
The inferred conclusion  is that this deficiency
traces back to the lack of a 3NF component in the nuclear realistic
Hamiltonian, which affects negatively the \heff~monopole component, as discussed in Ref. \cite{Zuker03} that was inspired by the NCSM results of Ref. \cite{Navratil02} indicating the need of 3NF in describing $p$-shell nuclei and in particular the ground state of $^{10}$B. 

This is not a negligible drawback for a SM calculation, since the ability to describe the
evolution of the nuclear spectroscopic properties along isotopic and
isotonic chains, and consequently the formation of magic numbers, is the
feature that has placed the SM in a central role within the
structure of atomic nuclei, and represents also its main success
\cite{Mayer49,Haxel49,Mayer55}.
Moreover, the \heff~monopole component affects also the evolution of
the calculated binding energies as a function of the number of valence nucleons
along isotopic and isotonic chains, and then the ability of SM
calculations to reproduce or predict correctly the edge of the nuclear
chart. This is the reason why it is fundamental that \heffs~should be able to
reproduce the observed shell evolution and closures.

The first SM study where the effects of 3NFs have been explicitly
included in the derivation of \heff~has been carried out by Otsuka and
coworkers in order to  reproduce  the oxygen-isotope
dripline \cite{Otsuka10}. In this work, the authors aimed to study what are the 
underlying conditions to reproduce the limit of oxygen isotopes as bound systems,
which is experimentally very close to the stability line
\cite{Janssens09,Kanungo09}.
To this end, they derived an effective Hamiltonian for the
$sd$-shell model space within  the MBPT framework\cite{Hjorth95},  starting from a realistic $NN$
potential as well as from an $NN$+3N potential.
The two-body component of the nuclear Hamiltonian was chosen to be a
potential constructed by way of the ChPT at
next-to-next-to-next-to-leading order (N$^3$LO) \cite{Entem02}, whose
high-momentum components were renormalized by way of the
\vlwk~procedure \cite{Bogner02}, while the three-body contribution was given by
the Fujita-Miyazawa $\Delta$-exchange force
\cite{Fujita57} or  the chiral 3NF term at N$^2$LO.
In order to manage the 3N component in SM calculations, its
contribution was evaluated by way of the previously mentioned
normal-ordering approximation, and such a contribution was shown to be
crucial to obtain $^{24}$O as the last bound oxygen isotope.

Starting from the work of Ref. \cite{Otsuka10} extensive studies have then carried out
on the role of  3NFs in SM calculations to reproduce
the spectroscopy and the binding energies of nuclei belonging to the
$sd$-shell region \cite{Holt13,Hebeler15,Simonis16,Simonis17,Tsunoda20}, as well
as to investigate and predict the nuclear structure of heavy calcium
isotopes \cite{Holt12,Holt14,Hebeler15,Simonis17}.
For all these works, the \heffs~have been derived
starting from two- and three-body potentials built up within the
chiral perturbative expansion and softened by way of the $V_{\rm low-k}$
technique \cite{Bogner02} or the similarity renormalization-group
(SRG) approach \cite{Bogner07,Bogner10}.

Chiral $NN$ and 3N forces have been also the starting point of non-perturbative
approaches to the derivation of \heff, such as the SM coupled
cluster (SMCC) \cite{Jansen14,Sun18,Sun21} and the valence-space
in-medium SRG (VS-IMSRG) \cite{Stroberg17,Morris18}, and a
comprehensive review about tackling the problem of deriving \heff~
within {\it ab initio} methods can be found in
Ref. \cite{Stroberg19} where an extension of the normal-ordering approximation using a multi-reference state is also outlined. In particular, calculations within the VS-IMSRG approach have validated
the need of 3NFs to reproduce the experimental behavior of ground-state (g.s.)
energies of oxygen and calcium isotopes \cite{Stroberg19}, and have
been employed to provide theoretical insight in many experimental
works
\cite{Brodeur17,Garnsworthy17,Reiter17,Steppenbeck17,Henderson18,Izzo18,Leistenschneider18,Michimasa18,Mougeot18,Randhawa19,Liu19}.
 
In 2018, the authors of the present work have started a research plan
aimed to employ chiral two- and three-body potentials both in standard
SM calculations as well as in Gamow SM (GSM) \cite{Michel02,Michel-2009,Li21}, the latter being focused on the description of weakly-bound nuclei by coupling  bound
 to resonant states (Gamow states).
The main goal of such an investigation is to single out the role of
3NF among the main components of the residual SM interaction, starting
from nuclear potentials based on ChPT and linked to the QCD, namely
the fundamental  theory of strong interactions \cite{vanKolck99}. 

In the first work of our project, we have studied the spectroscopy of $p$-shell nuclei
\cite{Fukui18} by employing an
\heff~derived from an $NN$ potential obtained within the ChPT at N$^3$LO
\cite{Entem02}, and including also one- and two-body components of a
normal-ordered 3N chiral potential at N$^2$LO.
The matrix elements of the 3NF have been constructed using consistently
the same LECs belonging to the components of the 2NF, the only two
extra LECs $c_D,c_E$ -- which are attached to the 3N $1$-$\pi$
exchange and contact terms, respectively -- have been chosen to be the
same as fixed in Ref. \cite{Navratil07b}.
With this choice of the $NN$ and 3N forces we have intended to benchmark  our SM results with those of \textit{ab initio} NCSM for the same class of nuclei \cite{Navratil07b,Maris13} in order to verify the quality of the approximations which characterize our calculations.

The results reported in that paper evidenced a substantial agreement
between the SM and NCSM calculations, and a positive assessment for
the MBPT to provide reliable \heffs~starting
from $NN$ and 3N forces. Furthermore, it is worth emphasizing that in Ref. \cite{Fukui18} we have shown that the 3NF is essential to make explicit the spin-orbit splitting of the $0p$ orbitals.

The success of this work has been the starting point for further SM studies, and
the focus has been shifted to nuclei belonging to $fp$ shell, which
provide the best paradigm to investigate the role of 3NFs to
reproduce the observed shell evolution of isotopic chains belonging to
this region.
In Ref. \cite{Ma19}, it has been shown that 3NFs are crucial to
obtain the shell closure at $N=28$, and  to provide SP-energy  splittings which are consistent with the
interpretation of $^{56}$Ni as a doubly magic nucleus.
This feature has been the cornerstone to obtain the evolution of
the excitation energies of the yrast $J=2^+$ states of calcium,
titanium, chromium, iron, and nickel isotopes in agreement with the
experimental one \cite{Ma19}.

The paper on $fp$ nuclei has triggered two following studies on the location of the neutron dripline for calcium \cite{Coraggio20e} and titanium \cite{Coraggio21} isotopic chains, both aimed to assess the relevance of accounting for induced many-body forces,  which originate from the 2NF in nuclei with more than two valence nucleons (see  Section \ref{sec-II.1}).
For a thorough SM investigation of Ca
and Ti isotopes, which are experimentally found to be stable beyond $N=40$, these two works adopt   a model space larger
than the standard $fp$ shell including the neutron $0g_{9/2}$ orbital.
Such an enlargement of the number of configurations of the nuclear
wave functions, which pushes the performance of the available SM codes
to their computational limits, has been proven to be important to
improve the agreement between theory and experiment.

As mentioned previously, the role of 3NFs has been investigated in the
framework of the GSM \cite{Michel02,Michel-2009,Li21} for weakly-bound nuclei. 
In Ref. \cite{Ma20a}, the three-body component of the nuclear
Hamiltonian and continuum states have been shown to be fundamental to
reproduce the dripline position and the unbound properties of oxygen
isotopes beyond the dripline.
The GSM, with the inclusion of 3NFs, reproduces the
experimental resonance widths of the $^{24}$O excited states, and predicts
the particle-emission widths for other resonant states  in $^{24,25,26}$O.

The GSM with chiral $NN$ and 3N forces has been employed also to study the
structure of the halo nucleus $^{17}$Ne, and it has been observed that the
repulsive behavior of the 3NF is decisive to raise the energy of $^{16}$F
over the threshold of the proton emission, leading to the Borromean
nature of $^{17}$Ne \cite{Ma20b}.

In a recent work, the focus has been spotted on the mirror-symmetry
breaking of $sd$-shell mass region, and it has been found that, for
$Z(N)=8$ isotopes (isotones), 2NFs only cannot provide the correct
binding energies, nucleon separation energies and excitation spectra
\cite{Zhang22}. 

Let us now outline the structure of the present work.
In the following section, we sketch out the essentials of our
understanding of nuclear many-body forces and the framework of
deriving nuclear potentials in terms of  chiral EFT.
In Section \ref{shell-model} we present the nuclear SM,
which is the many-body method where our calculations are grounded on,
and the theoretical aspects related to the derivation of
 effective SM operators starting from realistic nuclear forces.
In Section \ref{applications}, we review a large variety
of the latest nuclear structure calculations that have been performed
in terms of SM   employing realistic two- and
three-body potentials. Results for nuclei ranging from light and weakly-bound systems up
to intermediate-mass nuclei belonging to the $fp$ shell  are presented, with the aim 
to investigate  the role of 3NFs in explaining
phenomena such as the location of neutron dripline of
isotopic chains, the Borromean structure of halo nuclei, as well as
the shell evolution  as a function  of the number of valence nucleons.
The last section is devoted to summarizing our considerations about the
current status of SM with three-body potentials, and to looking
over the future evolution of this kind of approach to the study of
nuclear structure problems.
In the appendix \ref{appendix1}, we present  a few of the theoretical details of
the methods which are needed to employ 3NFs in SM
calculations.

\section{Three-body forces}\label{3BF}
\subsection{\it Backgrounds   \label{sec-II.1}}
As mentioned in the Introduction, the first example of
three-body force was the electromagnetic force introduced by
Primakoff and Holstein in their seminal paper in
Ref. \cite{Primakoff39}, where they introduced a three-body term in a
non-relativistic Schr\"odinger equation to account for the creation of
relativistic particle-antiparticle pairs.
In the very same paper, the authors employed the meson theory of the
nuclear force among nucleons, analogously to the electromagnetic one
among charged particles, to introduce the contribution of a
three-nucleon potential.

Actually, a theoretical framework that assumes  the
meson exchange as the source of the two-nucleon potential but, at the same time, freezes the meson degrees of freedom out of many-body systems is not a consistent picture \cite{Machleidt89}.
This means that meson exchange should be also considered the source of three-, four-, ..., many-nucleon forces, whose relevance with respect to the two-nucleon one is somehow model-dependent.

There are a few indications that the inclusion of a 3NF, alongside 
a 2NF, may improve the quality of theoretical calculations in reproducing the observables of three-nucleon systems.
The role of 3NFs may be considered circumstantial, in other words, there is
evidence that some specific theoretical results cannot be improved,
when compared with data, by considering only a two-body component for the
nuclear potential \cite{Friar84}.

More precisely, the introduction of a 3NF may lower the discrepancy
between the experimental and theoretical binding energies (BEs), charge
radii, and charge form factors of $^3$H and $^3$He in terms of
non-relativistic nuclear models.
Calculations of three-nucleon systems in terms of a realistic
2NF only suffer a bad correlation between BE and charge
radius. In fact, 2NFs that reproduce the experimental dimension of the nucleus
underestimate the BE and, viceversa, if the theory reproduces the
correct BE then the radius is too small.
Another observable whose reproduction  evidences the need of
resorting to three-body forces is the charge form factor of the
three-nucleon system $^3$He \cite{Camsonne17}.
As a matter of fact, the $^3$He charge density, which can be obtained
from the experimental charge form factor as a function of the momentum
transfer, exhibits a  ``hole'' in the inner part of the nuclear volume
at variance with a flat profile  that
is obtained with calculations performed with 2NF only (see Fig. 3 in
Ref. \cite{Friar84}).

The inclusion of 3NFs, as well as of two-body electromagnetic
meson-exchange currents \cite{Lina86,Frois87}, to calculate
$^3$H, $^3$He charge form factors has been proved to be crucial in
order to reproduce data \cite{Carlson98}.
Both of them - 3NF and meson-exchange currents - originate from the
same source of many-body nuclear forces: nucleons are not point-like
particles, and their quark structure needs to be accounted effectively
in terms of meson exchange and extra degrees of freedom.

These concepts of extra degrees freedom and meson exchange among
nucleons introduce us to the basic question: what is the definition of
a 3NF in terms of microscopic degrees of freedom?

A sound definition of nuclear many-body forces is the one which can
be found in Refs. \cite{Friar84,Machleidt89}, namely we should express
them as irreducible functions of coordinates or momenta of N nucleons
by way of irreducible Feynman diagrams which cannot be generated by
iterating $NN$ interactions.
The contributions that are represented by diagrams constructed
merely as products of 2NF vertices are then named induced many-body
forces, and they are taken into account by the diagonalization
of the nuclear Hamiltonian of the many-body method employed to
calculate energies and wave functions.

\begin{figure}[h]
\begin{center}
\includegraphics[scale=0.70,angle=0]{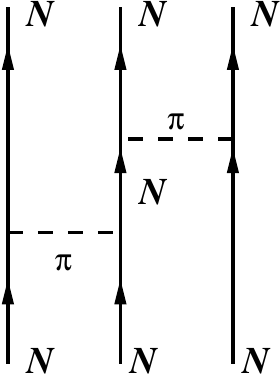}
\caption{Induced 3NF constructed iterating two one-pion
  exchange 2NF.}
\label{correlation}
\end{center}
\end{figure}

To exemplify, in Fig. \ref{correlation} we have reported a
diagram which, in a nuclear model based on a $NN$ potential
including one-pion exchange component, corresponds to an induced 3NF that is distinct from a genuine 3NF.

The distinction between many-body forces and correlations is based on
the choice of the internal degrees of freedom of a nucleus.
For example, if the adopted nuclear model freezes out the $\Delta$-isobar
degree of freedom, then many-body diagrams that include the $\Delta$
as an intermediate state between two-body vertices are to be
considered as a component of a real many-body force.

\begin{figure}[h]
\begin{center}
\includegraphics[scale=0.70,angle=0]{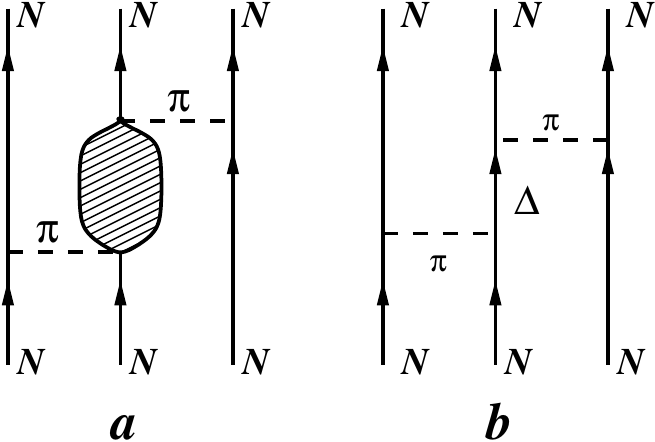}
\caption{Diagram {\it a}: $2\pi$-exchange three-nucleon force. Diagram
  {\it b}: Fujita-Miyazawa force. See text for details.}
\label{2pi3N}
\end{center}
\end{figure}

The most commonly considered 3NF in meson-exchange nuclear models is
the so-called $2\pi$-exchange three-nucleon force (\tptn), which is
 reported as diagram {\it a} in Fig. \ref{2pi3N}.

The first one is to approach the problem by using the current algebra
and partially-conserved axial-current (PCAC) constraints to extrapolate
the off-mass-shell parametrization of the $\pi$-N scattering amplitude
from on-mass-shell properties \cite{Coon81}.
This way  has generated the well-known
Tucson-Melbourne 3N potential \cite{Ellis85,Coon93}, which
has been employed for the calculations of infinite nuclear matter \cite{Ellis85}.

The other way to develop 3NFs is to resort to field theory by carrying
out a diagrammatic expansion.
A relevant component of such a model is the celebrated Fujita-Miyazawa
force \cite{Fujita57}, where a component of the off-shell amplitude is
the scattering into an intermediate $\Delta$-isobar state, that is
reported as diagram {\it b} in Fig. \ref{2pi3N}.
This pioneering potential has been the first attempt to frame nuclear
3NFs in terms of meson exchange, and has been considered for
calculations of the nuclear equation of state of infinite nuclear
matter (EOS) \cite{Dickhoff82,Muther85} as well as for nuclear
structure studies for finite nuclei \cite{Otsuka10}. It is worth mentioning that a good description of the nuclear matter saturation properties can be  also provided by a relativistic meson exchange potential, as discussed in Ref. \cite{Sammarruca12}.

The above mentioned  approaches suffer both the lack of a proper hierarchy in the
perturbative expansion of the nuclear Hamiltonian and tight
connection between the derivation of the $NN$ and 3N forces.
This drawback has been overcome in the last twenty years with the
advent of a new theoretical approach to the derivation of nuclear
forces, based on the ChPT, where the starting point is a chiral
Lagrangian which traces back to the work of Weinberg
\cite{Weinberg79,Weinberg90,Weinberg91}, calling in the concept of EFT
for $S$ matrix in processes with an arbitrary numbers of low-momentum
pions and nucleons.

In the following section, the approach to many-nucleon forces in the
framework of ChPT will be sketched out in its essentials.

\subsection{\it   Chiral effective field theory for three-nucleon
  forces  \label{sec-II.2}}
As mentioned in Section \ref{intro}, the
derivation of high-precision nuclear potentials based on ChPT
represents a major breakthrough in the last two decades
\cite{Entem02,Epelbaum05,Epelbaum09,Machleidt11,Epelbaum20}.
Nowadays, this class of theoretical potentials is widely employed to
link the fundamental theory of strong interactions, QCD, to nuclear many-body systems.

The derivation  of nuclear forces starting from a chiral Lagrangian
is framed within the EFT, by employing an arbitrary number of
low-momentum pions and nucleons.
The long-range forces are then ruled by the symmetries of
low-energy QCD and, particularly, by the spontaneously broken chiral
symmetry.
The short-range dynamics is absorbed into a complete basis of
contact terms that are proportional to LECs,
which are determined in order to reproduce few-body-system data.
 
In our perspective -- that focuses on the role of 3NFs in the study of the
structure of finite nuclei -- the main advantage of ChPT is that it
generates nuclear two- and many-body forces on an equal
footing \cite{Weinberg92,vanKolck94,Machleidt11}.
Moreover, most interaction vertices that appear in the 3NF and in the
four-nucleon force (4NF) also occur in the 2NF.
This means that the parameters carried by these vertices, as well as
the LECs of the 2NF contact terms,  are determined by the construction
of the chiral 2NF, and consistency then requires that the same
parameter values should be inserted at the same vertices in the
many-body-force contributions.

\begin{figure}[h]
\begin{center}
\includegraphics[scale=1.0,angle=0]{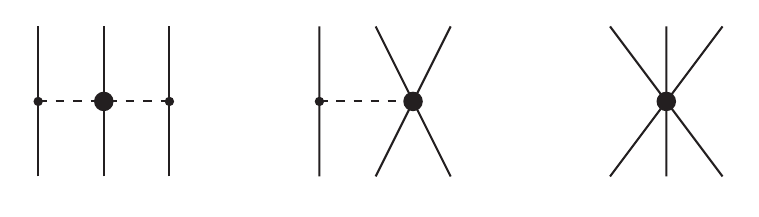}
\caption{The three-nucleon potential at N$^2$LO.
From left to right: 2PE, 1PE, and contact
diagrams.}
\label{n2lo3f}
\end{center}
\end{figure}

When the chiral perturbation expansion is performed, the first
non-vanishing 3NF occurs at N$^2$LO
\cite{Machleidt11,Epelbaum20}.
At this order, there are three 3NF topologies:  two-pion exchange
(2PE), one-pion exchange  plus a $NN$-contact interaction (1PE), and 
pure 3N-contact interaction. The three topologies are reported in Fig. \ref{n2lo3f}.

The 2PE 3N-potential is given by
\begin{equation}
v_{3N}^{(2\pi)} = 
\left( \frac{g_A}{2f_\pi} \right)^2
\frac12 
\sum_{i \neq j \neq k}
\frac{
( \vect{\sigma}_i \cdot \vect{q}_i ) 
( \vect{\sigma}_j \cdot \vect{q}_j ) }{
( q^2_i + m^2_\pi )
( q^2_j + m^2_\pi ) } \;
F^{ab}_{ijk} \;
\tau^a_i \tau^b_j.
\label{eq_3nf_nnloa}
\end{equation}
\noindent
Here, $\vect{\sigma}_i(\vect{\tau}_i)$ is the Pauli spin (isospin) matrix of nucleon $i$, and the transferred momentum is $\vect{q}_i \equiv \vect{p}_i' - \vect{p}_i$, with $\mathbf{p}_i$ and $\mathbf{p'}_i$ being the initial and final momenta, respectively. The other quantities entering the expression are the axial coupling constant $g_A$, the pion mass $m_\pi$, and the pion-decay constant $f_\pi= 92.4$ MeV, by using  natural units, namely $c=\hbar=1$.
In the above expression we have used the definition 

\begin{equation}
F^{ab}_{ijk} = \delta^{ab}
\left[ - \frac{4c_1 m^2_\pi}{f^2_\pi}
+ \frac{2c_3}{f^2_\pi} \; \vect{q}_i \cdot \vect{q}_j \right]
+ 
\frac{c_4}{f^2_\pi}  
\sum_{c} 
\epsilon^{abc} \;
\tau^c_k \; \vect{\sigma}_k \cdot ( \vect{q}_i \times \vect{q}_j).
\label{eq_3nf_nnlob}
\end{equation}

\noindent
It is worth noticing that the 2PE contribution to the structure of
the 3NF does not contain any new parameters with respect to those
appearing in the expression of the N$^2$LO 2NF, because the LECs
$c_1$, $c_3$, and $c_4$ have to be determined when fitting the 2NF to
the data of the two-nucleon system.

The 1PE contribution is
\begin{equation}
v_{3N}^{(1\pi)} = 
-\frac{c_D}{f^2_\pi\Lambda_\chi} \; \frac{g_A}{8f^2_\pi} 
\sum_{i \neq j \neq k}
\frac{(\vect{\sigma}_i \cdot \vect{q}_j )(\vect{\sigma}_j \cdot \vect{q}_j)}{
 q^2_j + m^2_\pi }
\vect{\tau}_i \cdot \vect{\tau}_j ,
\label{eq_3nf_nnloc}
\end{equation}
with $\Lambda_\chi = 700$ MeV.

The 3N contact potential reads
\begin{equation}
v_{3N}^{(\rm ct)} = 
\frac{c_E}{f^4_\pi\Lambda_\chi}
\; \frac12
\sum_{j \neq k} 
 \mbox{\boldmath $\tau$}_j \cdot \mbox{\boldmath $\tau$}_k.
\label{eq_3nf_nnlod}
\end{equation}

The last two 3NF terms involve the two new LECs $c_D$ and $c_E$,
which do not appear in the two-body problem.
There are many ways to constrain these two parameters.
For example, the triton binding energy and the $nd$ doublet scattering
length $^2a_{nd}$ or the $^4$He binding energy can be used.
However, because it is well-known that there is a correlation between these
observables,  an optimal over-all fit of the properties of light
nuclei is needed, as it has been done in Ref. \cite{Navratil07b}.
Another approach to fix $c_D$ and $c_E$ is to consider the consistency
of interactions and currents in chiral
EFT \cite{Gardestig06,Gazit08}, since  $c_D$ that
appears in $v_{3N}^{1\pi}$ is also involved in the two-nucleon contact
term in the $NN$ axial current operator derived up to
N$^2$LO.
Therefore, $c_D$ may be constrained using the accurate experimental
value of one observable from weak processes involving two- or
few-nucleon systems.
This procedure has been followed in Ref. \cite{Coraggio14a}, where  the triton $\beta$-decay half-life, in
particular its Gamow-Teller (GT) component, was used.
The same  choice  was already adopted in a variety of previous studies to
constrain the two-body axial current operator
\cite{Marcucci12}.

Because of the great success of ChPT nuclear forces in nuclear
structure calculations, the N$^2$LO three-body potential is the one
that has been mostly employed in SM calculations with realistic 3NFs.
The N$^2$LO 3N potential is defined in momentum space and, in order
to employ this potential in the derivation of the effective SM interaction, its matrix
elements must be calculated in the harmonic-oscillator (HO) basis
following a procedure that has been first reported in
Ref. \cite{Navratil07b}, for a potential characterized by a local
regulator in momentum space.

As concerns 3NFs beyond N$^2$LO, their contributions remain unclear.
At N$^3$LO, the two-pion exchange 3NFs produce rather weak effects for the nucleon-deuteron system~\cite{Ishikawa07}, while the short-range 3NFs may be significant~\cite{https://doi.org/10.48550/arxiv.2302.03468}. 
The short-range 3NFs at N$^4$LO, where thirteen LECs newly appear, may give sizeable contributions~\cite{Girlanda19,PhysRevC.105.054004}.

In Appendix \ref{appendix1}, one can find the details on the
calculation of the N$^2$LO 3NF matrix elements in the HO basis, but
employing a nonlocal regulator function.

\section{Shell model} \label{shell-model}
\subsection{\it Generalities} \label{sec-III.1}

The nuclear SM represents one the most powerful tools for  understanding the structure of atomic nuclei, in which the complexity of the nuclear many-body problem is reduced by considering only a limited number of microscopic degrees of freedom while the missing ones are taken into account by employing effective operators.

This model is based on the assumption that each nucleon moves in a mean field created by the remaining $A-1$ nucleons. The mean field gives rise to a shell structure, composed of single-particle states ({\it orbitals}) grouped in shells, which are well separated in energy from each other. To a first approximation, the nucleus can be considered as an inert core, made up of shells filled  with neutrons and protons paired to total angular momentum $J = 0$, with the remaining nucleons ({\it valence nucleons}) located in orbitals on top of the inert core. Within the SM, which is also called the ``interacting shell model'' to underline the difference with the simple independent-particle description, the valence nucleons interact in a truncated  space ({\it valence space}) spanned in general by a single proton and/or neutron shell above the inert core. Then, all the orbitals above the valence space are regarded as empty and constitute the external space. It is worth pointing out that in modern SM calculations larger valence spaces are considered, including proton and neutron cross-shell excitations, to describe some features of exotic neutron-rich nuclei. 

The SM is nowadays a well-established approach to investigate nuclei in different mass regions, as testified by the large number of successful calculations carried out during the 70 years since its introduction. The real beginning of the SM dates back to the end of 1940s and is connected to the publication of the papers by Mayer \cite{Mayer49} and Haxel {\it et al.} \cite{Haxel49}, even if the existence of shell structure was already evidenced during the previous two decades. It was, in fact, realized that nuclei with specific numbers of protons and/or neutrons ({\it magic numbers}) are more stable than others, which can be interpreted as a manifestation of an independent particle behavior. However, it was crucial for explaining the experimental regularities associated with the magic numbers the addition of a strong attractive spin-orbit term to the central mean field, as proposed in Refs. \cite{Mayer49,Haxel49}.

It became immediately clear that the description of nuclei only in terms of a simple mean-field potential was a very crude approximation, and the inclusion of the interaction between valence nucleons was indispensable to break up the degeneracies when considering systems with two or more particles outside doubly-closed nuclei.  Therefore, soon after the publication of the Mayer and Haxel's papers, a variety of two-body interactions was developed --  essentially  for  single-orbital configurations -- using central forces with different radial dependencies and including spin and isospin terms.   A review of these first SM calculations can be found in Ref. \cite{Talmi03}.

The construction of interactions to be used in SM  has always been a central issue within this approach, and still it is.  The use of phenomenological interactions,   which contain a certain
number of parameters adjusted to reproduce the experimental data, has been largely adopted in SM calculations, but at the same time significant efforts have been focused on the derivation of microscopic  interactions starting from realistic nuclear forces, namely from the bare  nuclear potentials determined from scattering experiments.  

Before discussing this point  in more detail, we would like to introduce the effective SM Hamiltonian, that in the coupled representation can be written as

\begin{equation}
H_{\rm eff }= \sum_{a} \epsilon_{a} {N}_{a}  - \frac{1}{4} \sum_{a b c d J} \langle a b; J \mid V_{\rm eff} \mid c d; J \rangle (-1)^{J} [a^ {\dag}_{a} a^ {\dag}_{b}]^{J } \cdot [\tilde{a}_{c} \tilde{a}_{d}]^{J},
\label{SMH}
\end{equation}

\noindent
where the symbol $(\cdot)$ indicates the scalar product as usual, while the latin indices run over the orbitals of the neutron and proton valence spaces and stand for $(nlj)$, with $n$ being the radial quantum number, $l$ and $j$, respectively, the orbital and total angular momentum. We have used  the  $a^{\dag}_{a m_{a}}$ and   $\tilde{a}_{a m_{a}}= (-1)^{j_{a} +m_{a}}  a_{a -m_{a}}$ operators which, respectively, creates  and annihilates  one particle in a state  of the underlying  mean field, with $m_{a}$  associated to the $z$ component of $j_{a}$. The one-body component of $H_{\rm eff}$ is written in terms of the SP energies  and the number operator $N_{a}= (-1)^{j_a} a^{\dag}_{a} \cdot \tilde{a}_{a}$. The TBMEs  are antisymmetrized but unnormalized.

The eigenvalue problem of the Hamiltonian (\ref{SMH}) can be solved by employing as basis states appropriate combinations of the Slater determinants

\begin{equation}
 [\, \underbrace {a^{\dag}_{am_a} a^{\dag}_{bm_b} a^{\dag}_{cm_c} \dots}_{\mathfrak{N}}\,]
|C\rangle,
\label{SLA}
\end{equation}

\noindent
where the set of SP orbitals ($a,b,c \ldots$) corresponds to a given configuration and   $\mathfrak{N}$ is the number of valence nucleons. The unperturbed doubly-closed core, $|C \rangle$,  can be explicitly written as  

\begin{equation}
| C \rangle = \prod_{a m_{a}\in~ filled~ shells} a^{\dagger}_{a m_a}| 0 \rangle~.
\end{equation}

In present days, there are several open source codes  for performing  SM calculations, such as 
 NuShellX \cite{NuShellX},  BIGSTICK \cite{BIGSTICK}, ANTOINE or NATHAN  \cite{Caurier05}, KSHELL \cite{KSHELL}, MFDN \cite{Shao18} and others (see \cite{Shimizu22} for more details).

 Some of them are developed for massive parallel computation and, therefore, can run on high-performance computing clusters. These type of codes are able to handle up to $\sim 100$  billion dimensions, making it possible to approach nuclei  with many valence nucleons in large valence spaces. 

One component of $H_{\rm eff}$ relevant for the following discussion is given by the monopole interaction, which, using explicitly the indices  $\tau,\tau'$  for  proton or  neutron, takes the form

\begin{equation}
H_{\rm mon }= \sum_{a\tau} \epsilon_{a \tau}  { N}_{a \tau}  + \frac{1}{2} \sum_{a b \tau \tau' }   \frac{{\bar V}_{a b}^ {\tau \tau'}  {N}_{a \tau} ({N}_{b \tau'} - \delta_{a b} \delta_{\tau \tau'})} {\sum_{J} \hat J^{2}}
\label{MonI}
\end{equation}

\noindent
where the angular momentum $J$ runs on  all allowed values and  the notation $\hat {J}= \sqrt{2 J +1}$  is used.

\noindent
The matrix elements ${\bar V}_{a b}^ {\tau \tau'}$ are defined as
\begin{equation}
{\bar V}_{a b}^ {\tau \tau'} =	\frac{ \sum_J {\hat J} \langle a \tau  b \tau'; J \mid V_{\rm eff} \mid  a \tau b \tau'; J  \rangle}{\sum_J {\hat J}}~, 
\label{Mon}
\end{equation}

\noindent
and represent  the angular-momentum averaged TBMEs, or centroids of the interaction.

The monopole interaction corresponds to  the  spherical mean field as extracted  from the SM Hamiltonian  in open-shell nuclei \cite{Caurier05} and therefore governs  the evolution   of the SP energies along isotopic and isotonic chains. Starting from the monopole interaction,     effective single particle energies (ESPEs)  are defined

\begin{equation}
{\rm ESPE} (a \tau)= \epsilon_{a \tau} +  \sum_{b  \tau^{\prime}}   {\bar V}_ {a b }^{\tau \tau^{\prime}} n_{b}^{\tau^{\prime}},
\label{ESPE}
\end{equation}

\noindent
which, as compared to the  SP energies of the SM Hamiltonian (Eq. (\ref{SMH})), incorporate the  mean effects from other nucleons outer the inert core (see, for instance, Ref. \cite{Otsuka20}).  Here the g.s. occupation number of the proton/neutron  
$b$ orbital is denoted by $n_{b}^{\tau^{\prime}}$.

Once   energies  $E_\alpha$  and wave  functions $|\psi_\alpha\rangle$  of the system under consideration are determined by solving the eigenvalue problem for the Hamiltonian of Eq. (\ref{SMH}), it is possible to  compute the  matrix elements of operators which are related to physical quantities, like electromagnetic transitions and  decay strengths, or appear in form factors needed to evaluate reaction cross sections. The matrix elements for one-body operators can be written as

\begin{equation}
\langle \psi_{f} || \Theta^{\lambda}_{\rm eff} ||  \psi_{i} \rangle =  \sum_{a b} \frac{ \langle \psi_{f} || [a^{\dag}_{a} \tilde{a}_{b}]^{\lambda} ||  \psi_{i} \rangle}{\hat{ \lambda}} 
  \langle a || \Theta^{\lambda}_{\rm eff }| | b \rangle = \sum_{a b} {\rm OBTD}(fi a b \lambda) \langle a || \Theta^{\lambda}_{\rm eff }| | b \rangle ,
  \label{OBO}
\end{equation}

\noindent 
where the one-body transition densities (OBTDs) represent in a compact form the nuclear structure information on the initial and final states involved  in the process. Note that we have added the label ``eff'' to  $\Theta^{\lambda}$ to point out that a different operator  with respect to the bare one should be used with SM wave functions, since, as for the Hamiltonian,  renormalizations due to the adopted truncated space are needed.

As mentioned above, the choice of the Hamiltonian is one of the most crucial issues  in the SM  approach, which  attracted the attention of the nuclear community  from the very beginning.  At first,  interest was essentially addressed to the development  of phenomenological  schematic interactions, which are  parametrized  functions of nucleon coordinates with very simple or more complicated structures, depending on the  included  exchange operators.  Between them, we mention the $\delta$ and  pairing forces, which, even if very simple,  account for the short-range nature of  the residual interaction and  its tendency  to correlate nucleons in zero-coupled pairs ($J^{\pi }=0^{+}$).

Contemporaneously, there were efforts to derive  the SM Hamiltonian from a realistic interaction between nucleons. 
It was soon evident that the peculiar properties of the bare potential,  containing a strong repulsive core at short distances, prevent the description of
the nucleus in terms of mean field and consequently within the SM framework. However,  the  introduction of the $G$ matrix by  Keith Brueckner and coworkers  \cite{Brueckner54} was the first milestone for the development of a microscopic interpretation of the SM.
In the   $G$ matrix,  strong short-range correlations are renormalized  by summing  all two-particle ladder-type interactions. It can be, therefore,  used to perform Hartree-Fock self-consistent calculations or taken,  in principle, as residual interaction in  SM calculations.  A further relevant step  along this line is represented by the paper of Kuo and Brown \cite{Kuo66}, in which  a new effective interaction was derived for the $sd$ shell starting from the $G$ matrix by a perturbative expansion  including terms up to second order in $G$, namely the  so-called $\it bubble$ diagrams  corresponding to one particle-one hole ($1p-1h$)  core-polarization excitations.

These pioneering works made it evident that the  interaction used in SM  calculations  cannot be the bare one between free nucleons. In fact, the SM Hamiltonian is defined in a reduced space, and should  therefore account  -- in an effective way --  for the omitted degrees of freedom, namely  for the excitations of core particles into the valence and external spaces as well as for the excitations of valence particles in the external space. 
Then, as testified by the large number  of papers published on the subject, great attention was dedicated  to the derivation of  the effective  SM Hamiltonian within a perturbative  approach  and  to the assessment of its role in the study of nuclear structure.
Nowadays, substantial progress has been made regarding the  starting bare  potential   as well as  the many-body technique for constructing the effective interaction, that will  be   presented in detail in  forthcoming sections.

Within the class of phenomenological interactions, in  which the introduced parameters  are adjusted to reproduce a selected set of experimental data, an alternative   way  to the schematic interactions,  introduced by  Talmi in Ref. \cite{Talmi63},  consists in considering the Hamiltonian matrix elements  themselves  as free parameters. This approach is quite successful, and a comprehensive discussion and
presentation of such a  procedure can be found in Refs. \cite{Talmi03,Caurier05,ABrown01,Brown19,Otsuka20}. Here, we report just a few examples  of this kind of interactions, as the $p$-shell interaction  by Cohen and Kurath \cite{Cohen65},  the so-called universal interaction developed for the $sd$ shell by Brown  {\it et al.} \cite{Brown88,Brown06}, or   the  GXPF1 \cite{Honma04} and JUN45  \cite{Honma09} interaction by M. Honma {\it et al.}, respectively,  for the $fp$   and   $f_{5/2}pg_{9/2 }$ valence spaces.
For large valence spaces, the number of matrix elements  increases drastically and they are determined by choosing a starting Hamiltonian, as for instance the $G$ matrix of a realistic $NN$ potential,  and using  the linear combination  method of Ref. \cite{Brown88},  where only  selected linear combinations of matrix elements  are varied to fit experimental data.
All these phenomenological  interactions have been largely used in nuclear  structure studies providing  a successful description  of  a certain variety of phenomena.

It is worthwhile to point out that in the vast majority of SM  calculations, only two-body interactions have been used.  As mentioned in the Introduction, the explicit inclusion of  3NFs has historically been neglected in SM treatment due to the ambiguity in producing a 3N term consistent with the $NN$ one, as well as to the difficulty in handling such a term  in  many-body systems.
On the other hand, the good agreement between theory and experiment obtained  with SM calculations employing phenomenological interactions suggests  that  the
effects of  3NFs can be   empirically taken into account.  We already mentioned that  Zucker and coworkers  \cite{Zuker03} argued that the main effects of 3NFs concentrate in the monopole component of the effective interaction.

As a matter of fact, modifications of the monopole component were first proposed, without a clear connection with 3NFs, to  cure  the deficiencies of effective Hamiltonians derived from realistic $NN$ potentials related to their  bad saturation and shell formation properties \cite{Poves81}. The  phenomenological adjustment of monopole terms in  realistic  effective  Hamiltonians has been  largely applied by the Strasbourg-Madrid group to  various mass regions  providing  interactions able to give an accurate description of the nuclear spectroscopy. Examples of monopole corrected interactions are   KB3  \cite{Poves81} in the $fp$ shell,
 SDPF-U \cite{Nowacki09} in the $sdfp$ shell, and LNPS \cite{Lenzi10} including the $fp$ shell for protons and the $f_{5/2}pg_{9/2}d_{5/2}$  orbitals for neutrons.

In the last decade SM calculations  that explicitly take into account  3NF have been carried out  and they will be discussed  in detail in the following sections.

\subsection{\it The derivation of realistic effective interactions and
  operators} \label{effint} 
The  SM is grounded on the ansatz that each nucleon belonging to a nucleus moves
independently in a spherically symmetric auxiliary potential, which
accounts for the average interaction with the other protons and
neutrons.
This potential is usually described by a Woods-Saxon (WS) or an HO potential
including a spin-orbit term.
Actually, it is clear that the
residual interaction between valence nucleons, which is not explicitly
included in the one-body auxiliary potential, has to be considered to
describe quantitatively the low-energy structure of nuclei with two or
more valence nucleons confined to move in the valence space.
In fact, the action of the residual interaction generates a  mixing
of different configurations thus removing the degeneracy of
states belonging to the same configuration.

In general, when considering only the valence nucleons interacting in the reduced number of orbitals of the valence space, one is left  with the problem to construct a SM
Hamiltonian and decay operators  defined in a truncated
space, but whose matrix elements should account for the neglected degrees of freedom.
Namely, we need an effective Hamiltonian \heff~and effective decay
operators \thetaeff.


We start by sketching out the fundamentals of the derivation of $H_{\rm eff}$ in a formal way, by  considering, without loss of generality, an $A$-body Hamiltonian with no transitional invariance. This is appropriate in SM calculations performed including only one proton and/or neutron major shell above the closed core.  A purely intrinsic Hamiltonian will be introduced in Section \ref{sec-III.3}. The full Hilbert-space eigenvalue problem is then written as

\begin{equation}
H | \Psi_{\alpha} \rangle  = E_{\alpha} | \Psi_{\alpha} \rangle \label{eq1},
\end{equation}

\noindent
where
\begin{equation}
 H=H_0+H_1  \label{defh},
\end{equation}
and

\begin{equation}
H_0= \sum_{i=1}^A \left(\frac{p_i^2}{2m}+U_i \right)  \label{defh0},
\end{equation}
\begin{equation}
H_1=\sum_{i<j=1}^{A} V^{NN}_{ij}-\sum_{i=1}^AU_i \label{defh1}~. 
\end{equation}
As mentioned before, to introduce the SM framework we must
consider an auxiliary one-body potential $U$ to break up the nuclear
Hamiltonian as the sum of a one-body term $H_0$, which describes the
independent motion of the nucleons, and  the residual interaction $H_1$.

Without any loss of generality and for
the sake of simplicity, we assume that the interaction between the
nucleons is described only by a two-body force, and neglect 3NF
contributions.
The generalization of the formalism to include a three-body potential
will be  considered later. 

The solution of Eq. (\ref{eq1}) requires the diagonalization of the infinite matrix
$H$, a task that is obviously unfeasible. Then, one has to reduce this
huge matrix to a smaller one \heff~ -- defined in a model space made of the only  configurations allowed by the valence nucleons within  the  adopted valence 
 space -- by requiring that its eigenvalues belong to the set of 
the eigenvalues of $H$. This  model space is defined in terms of a  subset of eigenvectors of $H_{0}$, $|\Phi_{i}\rangle$, namely as  appropriate combinations of the Slater determinants of Eq. (\ref{SLA}),  written, in general, in the angular momentum-coupled scheme.  

It is worth introducing the projection operators $P$ and $Q=1-P$,
which project from the complete Hilbert space onto the  model space
and its complementary space, respectively. The operator $P$ can be expressed by way of the states $\Phi_{i}$  as 

\begin{equation}
P= \sum_{i=1}^d | \Phi_i \rangle \langle \Phi_i |, \label{Peq1}
\end{equation}

\noindent 
where $d$ is the dimension of the model space.
The projection operators $P$ and $Q$ then satisfy the properties
\begin{equation}
P^2=P, ~~ Q^2=Q, ~~ PQ=QP=0 . \label{Peq2}
\end{equation}

The aim of the effective SM interaction theory is to reduce the
eigenvalue problem of Eq. (\ref{eq1}) to a model-space eigenvalue
problem
\begin{equation}
H_{\rm eff}P | \Psi_{\alpha} \rangle  = E_{\alpha} P | \Psi_{\alpha}
\rangle \label{eq3},
\end{equation}

\noindent 
where $\alpha = 1, \ldots,d$ and \heff~is defined only in the model space. 

As mentioned in Section \ref{sec-III.1}, there are two ways to tackle the problem of deriving \heff, namely by
\begin{enumerate}
  \item employing a phenomenological approach,
  \item starting from bare realistic nuclear forces and then resorting
    to the many-body theory.
\end{enumerate}

\noindent
 References about  the phenomenological  approach and some examples of this kind of interactions were given in Section \ref{sec-III.1}, while here we discuss in some detail the second way.

Nowadays, novel non-perturbative methods have been developed to derive a effective SM Hamiltonian
starting from the bare nuclear interaction,
like valence-space in-medium SRG (VS-IMSRG) \cite{Morris15},
SM coupled cluster (SMCC) \cite{Sun18}, or NCSM with a core
\cite{Lisetskiy08,Lisetskiy09,Dikmen15,Smirnova19}, all of them based
on similarity transformations.
These non-perturbative approaches are rooted in many-body
theory and provide somehow different paths to \heff.
However, they can all be derived in a general theoretical framework, that
consists in expressing  \heff~as the result of a similarity transformation acting
on the original Hamiltonian 

\begin{equation}
    H_{\mathrm{eff}} = Pe^{{\ensuremath{\mathcal{G}}}}He^{-{\ensuremath{\mathcal{G}}}}P,
\end{equation}

\noindent
where the transformation is parametrized as the exponential of a
generator ${\ensuremath{\mathcal{G}}}$, which needs to satisfy the
decoupling condition

\begin{equation}\label{decouple}
QH_{\mathrm{eff}}P =0.
\end{equation}

An extended and up-to-date presentation of non-perturbative approaches
to the derivation of \heff~from realistic nuclear interactions can be
found in Ref. \cite{Stroberg19}, showing how the different methods can
be derived in such a general framework and describing the
approximation schemes that have to be employed in each of them.

As mentioned in the Introduction, in the present review we focus on  the MBPT  approach to \heff, since it is at the moment the one that has been most largely adopted in SM calculations where the role of 3NFs has been investigated.

\subsubsection{\it The perturbative expansion of effective shell-model
  Hamiltonian}\label{perturbativeh}
Here, we introduce the formalism of the perturbative derivation of the
effective SM  Hamiltonian \heff, using the similarity
transformation introduced by Lee and Suzuki \cite{Lee80,Suzuki80}.
The starting point is the Schr\"odinger equation for the $A$-nucleon
system in the whole Hilbert space as defined in Eq.(\ref{eq1}).

Then, by following Eqs. (\ref{defh})-(\ref{defh1}), we introduce an auxiliary one-body potential $U$ to break up the
nuclear Hamiltonian as the sum of an unperturbed one-body term $H_0$,
that describes the independent motion of the nucleons, and the
residual interaction Hamiltonian $H_1$.

The robust energy gap between the shells allows considering as inert
the $A-\mathfrak{N}$ core nucleons, which fill the energy orbitals below the
Fermi surface.
The SP states that are accessible to the $\mathfrak{N}$ valence nucleons are those
included in the major shell placed in energy above the closed core, and constitute the valence space.
The configurations allowed by the valence nucleons within this valence
space define a reduced Hilbert space, the so-called model space,  by way of a finite subset of $d$
eigenvectors of $H_0$.
The  operators  that  project the wave functions from the complete
Hilbert space onto the model space and its complementary space are,
respectively, $P$ and $Q$,  satisfying the properties of Eq. (\ref{Peq2}).

As already mentioned before, the goal is to reduce the eigenvalue
problem of Eq. (\ref{eq1}) to the model-space eigenvalue problem of Eq. (\ref{eq3}).
Therefore, we need to obtain a new Hamiltonian $\mathcal{H}$
whose eigenvalues are the same of the Hamiltonian $H$ for the
$A$-nucleon system, but satisfying the decoupling equation between the
model space $P$ and its complement $Q$

\begin{equation}
Q \mathcal{H} P=0, \label{deceq1}
\end{equation}

\noindent
in order to guarantee that the effective Hamiltonian is $H_{\rm eff}=
P \mathcal{H} P$.

Clearly, the Hamiltonian $\mathcal{H}$ has to be obtained by way of a
similarity transformation defined in the whole Hilbert space

\begin{equation}
\mathcal{H}=X^{-1} H X .
\end{equation}

\noindent
The class of transformation operators $X$ such that $\mathcal{H}$ satisfies  the decoupling
equation (\ref{deceq1}) is infinite, and Lee and Suzuki
\cite{Lee80,Suzuki80} have suggested an operator $X$ defined as
$X=e^{\omega}$.
Without loss of generality, $\omega$ is chosen to satisfy the
following properties:

\begin{equation}
\omega= Q \omega P , 
\label{omegapro1}
\end{equation}
\begin{equation}
P \omega P= Q \omega Q = P \omega Q =0, 
\label{omegapro2}
\end{equation}

\noindent
with Eq. (\ref{omegapro1}) implying that 

\begin{equation}
\omega ^2 = \omega^3 = ~...~=0 . 
\label{omegapro3}
\end{equation}

\noindent
The above equation leads us to write the operator $X$ as $X=1+ \omega$,
and, consequently, \heff~takes the form

\begin{equation}
H_{\rm eff} = P \mathcal{H} P = PHP +PH Q \omega , \label{defheff}
\end{equation}

\noindent
while the decoupling equation (\ref{deceq1}) can be expressed as 

\begin{equation}
Q H P + Q H Q \omega - \omega P H P - \omega P H Q \omega = 0
. \label{deceq2} 
\end{equation}

\noindent 
This is a non-linear matrix equation and can  easily provide a solution for $\omega$ as long
as the Hamiltonian $H$ is explicitly expressed in the whole Hilbert space.
Actually, this is not feasible and some sort of approximations would be necessary.

A successful way to solve Eq. (\ref{deceq2}) and derive $H_{\rm eff}$ for SM
calculations is to introduce  a vertex function, the \qbox, which is
suitable for a perturbative expansion.
In the following, we make explicit the $\hat{Q}$-box approach assuming, for the sake of
simplicity,  that the unperturbed Hamiltonian $H_0$ is
degenerate for those states  belonging to the model space,

\begin{equation}
P H_0 P =\epsilon_0 P. \label{dege}
\end{equation}
This limitation may be overcome by introducing multi-energy $\hat{Q}$ boxes, which are able to account for non-degenerate spaces \cite{Kuo95}.
Actually, this approach is quite complicated for practical
applications, but recently two methods have been introduced,  which may be implemented straightforwardly  in the derivation of \heffs~with eigenstates of $H_0$ non-degenerate in the model space.
The details of these procedures are reported in
Refs. \cite{Takayanagi11a,Suzuki11}.

Starting from Eq. (\ref{defheff}), the effective
interaction $H^{\rm eff}_1=H_{\rm eff} - P H_0 P$ can be written in terms of $\omega$ as

\begin{equation}
H^{\rm eff}_1 = P \mathcal{H} P - P H_0 P = P H_1 P + P H_1 Q \omega
, \label{eqqq}
\end{equation}

\noindent
where we have used the diagonality of $H_0$ in the $P$ and $Q$ states which implies

\begin{equation}
QHP= QH_1P + QH_0P = QH_1P,
\label{qhp}
\end{equation}

\begin{equation}
PHQ= PH_1Q + PH_0Q = PH_1Q.
\end{equation}

\noindent
Similarly,  the decoupling equation (\ref{deceq2}) takes the form
\begin{equation}
Q H_1 P + Q H Q \omega - \omega (P H_0 P + P H_1 P + P H_1 Q \omega) = 
Q H_1 P + QHQ \omega - \omega ( \epsilon_0 P + H_1^{\rm eff}) = 0
, \label{deceq3} 
\end{equation}

\noindent
which leads to a new identity for the operator $\omega$

\begin{equation}
\omega = Q \frac{1}{\epsilon_0 -QHQ} Q H_1 P - Q \frac{1}{\epsilon_0
    -QHQ} \omega H^{\rm eff}_1. \label{omegaq}
\end{equation}

\noindent
Finally, by inserting Eq. (\ref{omegaq}) into the identity (\ref{eqqq}) we obtain a recursive equation  for $H^{\rm eff}_1$ 

\begin{equation}
H^{\rm eff}_1 (\omega) = P H_1 P + P H_1 Q \frac{1}{\epsilon_0 - Q H Q} Q
  H_1 P - P H_1 Q \frac{1}{\epsilon_0 - Q H Q} \omega H^{\rm eff}_1
  (\omega) . \label{eqsemifinal}
\end{equation}

\noindent
Then, by defining the $\hat{Q}$-box vertex function as

\begin{equation}
\hat{Q} (\epsilon) = P H_1 P + P H_1 Q \frac{1}{\epsilon - Q H Q} Q
H_1 P , \label{qbox}
\end{equation}

\noindent
the recursive equation (\ref{eqsemifinal}) can be written as

\begin{equation}
H^{\rm eff}_1 (\omega) = \hat{Q}(\epsilon_0) - P H_1 Q \frac{1}{\epsilon_0
  - Q H Q} \omega H^{\rm eff}_1 (\omega) . \label{eqfinal}
\end{equation}

Lee and Suzuki suggested two possible iterative schemes to solve
Eq. (\ref{eqfinal}), which are based on the calculation of the $\hat{Q}$ box
and its derivatives, known as the Krenciglowa-Kuo (KK) and the Lee-Suzuki
(LS) techniques \cite{Suzuki80}.

Let us start from the KK iterative method, which traces back to the coupling of
Eqs. (\ref{eqfinal}) and (\ref{omegaq}), providing the  relation

\begin{equation}
 H^{\rm eff}_1 (\omega_n) = 
\sum_{m=0}^{\infty} \left[-P H_1 Q \left( \frac{-1}{\epsilon_0 -QHQ}
  \right)^{m+1} QH_1P \right] \left[ H^{\rm eff}_1 (\omega_{n-1})
\right]^m .
\label{eqa}
\end{equation}

\noindent
The quantity inside the square brackets of Eq. (\ref{eqa}), which is
commonly dubbed as $\hat{Q}_m(\epsilon_0)$, is proportional to the
$m$-th derivative of the \qbox~calculated in $\epsilon=\epsilon_0$

\begin{equation}
\hat{Q}_m(\epsilon_0) = -P H_1 Q \left( \frac{-1}{\epsilon_0 -QHQ}
  \right)^{m+1} QH_1P = \frac{1}{m!} \left[ \frac{d^m \hat{Q} 
(\epsilon)} {d \epsilon^m} \right]_{\epsilon=\epsilon_0} .
\label{eqb}
\end{equation}

\noindent
We may then rewrite Eq. (\ref{eqa}), according to the above identity,
as

\begin{equation}
 H^{\rm eff}_1 (\omega_n) =
\sum_{m=0}^{\infty} \frac{1}{m!} \left[ \frac{d^m \hat{Q} (\epsilon)} {d
    \epsilon^m} \right]_{\epsilon=\epsilon_0} \left[ H^{\rm eff}_1
  (\omega_{n-1}) \right]^m  = 
\sum_{m=0}^{\infty} \hat{Q}_m(\epsilon_0) \left[ H^{\rm eff}_1
  (\omega_{n-1}) \right]^m .
\label{eqc}
\end{equation}

\noindent
The starting point of the KK iterative method is the assumption
$H^{\rm eff}_1 (\omega_0)=\hat{Q} (\epsilon_0)$, leading to rewrite
Eq. (\ref{eqc}) as 

\begin{equation}
H^{\rm eff}_{1} = \sum_{i=0}^{\infty} F_i, \label{kkeq}
\end{equation}

\noindent
where

\begin{eqnarray}
F_0 & = & \hat{Q}(\epsilon_0)  \nonumber \\
F_1 & = & \hat{Q}_1(\epsilon_0)\hat{Q}(\epsilon_0)  \nonumber \\
F_2 & = & \hat{Q}_2(\epsilon_0)\hat{Q}(\epsilon_0)\hat{Q}(\epsilon_0) + 
\hat{Q}_1(\epsilon_0)\hat{Q}_1(\epsilon_0)\hat{Q}(\epsilon_0)  \nonumber \\
~~ & ... & ~~ 
\label{kkeqexp}
\end{eqnarray}

The above expression is a different form of the well-known
folded-diagram expansion of the effective Hamiltonian as introduced by
Kuo and Krenciglowa, since in Ref. \cite{Krenciglowa74} it has been
demonstrated the  operatorial identity

\begin{equation}
\hat{Q}_1\hat{Q}= - \hat{Q} \int \hat{Q},
\end{equation}

\noindent
where the integral sign corresponds to the so-called folding operation
as introduced by Brandow in Ref. \cite{Brandow67}.

An alternative approach to the solution of Eq. (\ref{eqfinal}) is to
resort to the LS technique.
This can be carried out by rearranging Eq.  (\ref{eqfinal}) in order to obtain
an explicit expression of the effective Hamiltonian $H^{\rm eff}_1$ as
a function of the operators $\omega$ and $\hat{Q}$ \cite{Suzuki80}

\begin{equation} 
H^{\rm eff}_1 (\omega) = \left( 1 + P H_1 Q \frac{1}{\epsilon_0 - Q H Q}
  \omega \right)^{-1} \hat{Q} (\epsilon_0) .
\end{equation}

\noindent
The iterative form of this equation is

\begin{equation} 
H^{\rm eff}_1 (\omega_n) = \left( 1 + P H_1 Q
\frac{1}{\epsilon_0 - Q H Q} \omega_{n-1} \right)^{-1} \hat{Q}
(\epsilon_0),
\end{equation}

\noindent
while  an iterative expression of Eq. (\ref{omegaq}) is given by

\begin{equation} 
\omega_n = Q \frac{1}{\epsilon_0 -QHQ} Q H_1 P - Q \frac{1}{\epsilon_0
    -QHQ} \omega_{n-1}H^{\rm eff}_1(\omega_n) .
\end{equation}

\noindent
The starting point of the procedure is to choose $\omega_0=0$, so that
we may write

\begin{eqnarray}
H^{\rm eff}_1 (\omega_1) & = & \hat{Q} (\epsilon_0) \nonumber \\
\omega_1 & = &  Q \frac{1} {\epsilon_0 -Q H Q} Q H_1 P . \nonumber
\end{eqnarray}

\noindent
Using some algebra,  the following identity can be demonstrated

\begin{equation}
 \hat{Q}_1 (\epsilon_0) = - P H_1 Q \frac{1}{\epsilon_0 - QHQ} Q
\frac{1}{\epsilon_0 - QHQ} Q H_1 P  =  - P H_1 Q \frac{1}{\epsilon_0 - QHQ}
\omega_1 ,
\end{equation}

\noindent
and  for the iteration step $n=2$ we have

\begin{eqnarray}
H^{\rm eff}_1 (\omega_2) & = & \left( 1 + P H_1 \frac{1} {\epsilon_0 -
    QHQ} \omega_1 \right)^{-1} \hat{Q}(\epsilon_0)  =  
  \frac{1}{1 - \hat{Q}_1(\epsilon_0)} \hat{Q} (\epsilon_0), \nonumber \\
\omega_2 & = & Q \frac{1}{\epsilon_0 - QHQ} Q H_1 P -  Q
\frac{1}{\epsilon_0 -QHQ} \omega_{1}H^{\rm eff}_1(\omega_2)  .
\end{eqnarray}

\noindent
Finally, the LS iterative expression of \heff~is

\begin{equation}
H^{\rm eff}_1 (\omega_n) = \left[
1 - \hat{Q}_1 (\epsilon_0) \sum_{m=2}^{n-1} \hat{Q}_m (\epsilon_0)
\prod_{k=n-m+1}^{n-1} H^{\rm eff}_1 (\omega_k)
\right]^{-1} \hat{Q} (\epsilon_0)  . \label{eqLS}
\end{equation}

It is important noting that the KK and LS iterative techniques,
even if they have been both conceived to solve the decoupling equation
(\ref{deceq3}), do not provide necessarily the same \heff.
Suzuki and Lee have shown that by way of the KK iterative approach one obtains 
eigenstates that have 
a  large overlap with the model space.
On the other side, when   \heff~is derived by employing the LS
technique, its eigenvalues  are the lowest in energy among those
belonging to the set of the full Hamiltonian $H$ \cite{Suzuki80}.

The heart of the matter is now the calculation of the $\hat{Q}$-box~
vertex function defined in Eq. (\ref{qbox}). Within a perturbative framework, the term $1/(\epsilon - Q H Q)$ appearing in Eq. (\ref{qbox}) should
be expanded as a power series
\begin{equation}
\frac{1}{\epsilon - Q H Q} = \sum_{n=0}^{\infty} \frac{1}{\epsilon -Q
  H_0 Q} \left( \frac{Q H_1 Q}{\epsilon -Q H_0 Q} \right)^{n} .
\end{equation}

\noindent
It is common to employ a diagrammatic approach of the perturbative
expansion by representing the \qbox~as a collection of irreducible
Goldstone diagrams --  diagrams with  at least one line between two successive
vertices not belonging to the model space -- that  have  at least one $H_1$-vertex and are linked to minimum one external valence line
\cite{Kuo71}.

Usually, the derivation of \heff~is performed for systems with one and
two valence nucleons. Single valence-nucleon nuclei supplies the one-body component of $H_{\rm eff}$, $H_{\rm eff}^{1b}$, namely  the theoretical
effective SP energies, while the TBMEs  are obtained from the effective Hamiltonian for systems with two-valence nucleons, we indicate by $H_{\rm eff}^{2b}$.
In particular, the TBMEs are obtained by way of  a subtraction procedure, which consists in removing
from $H_{\rm eff}^{2b}$ the diagonal component of  $H_{\rm eff}^{1b}$ \cite{Shurpin83}.

In Ref. \cite{Kuo81}, the topic of the calculation of $\hat {Q}$-box diagrams in
the angular momentum-coupled representation is extensively treated.
It should be noted that in literature the effective  SM
Hamiltonians are derived accounting for $\hat{Q}$-box diagrams
up to at most the third order in perturbation theory, and their complete list
is reported in Ref. \cite{Coraggio12a}.
This limitation is dictated by the computational cost of performing
calculations that include complete higher-order sets of diagrams.

The diagrammatic reported in Refs. \cite{Hjorth95,Coraggio12a} is
constrained to the derivation of \heffs~for one- and two-valence
nucleon systems, but  many-body diagrams must be included to obtain
\heff s for systems with three or more valence nucleons.

At present, few codes can perform the diagonalization of SM Hamiltonians including a three body component, like BIGSTICK \cite{BIGSTICK} and MFDN \cite{Shao18}. They are however mainly oriented to NCSM and limited to light nuclei.

\begin{figure}[h]
\begin{center}
\includegraphics[width=6cm]{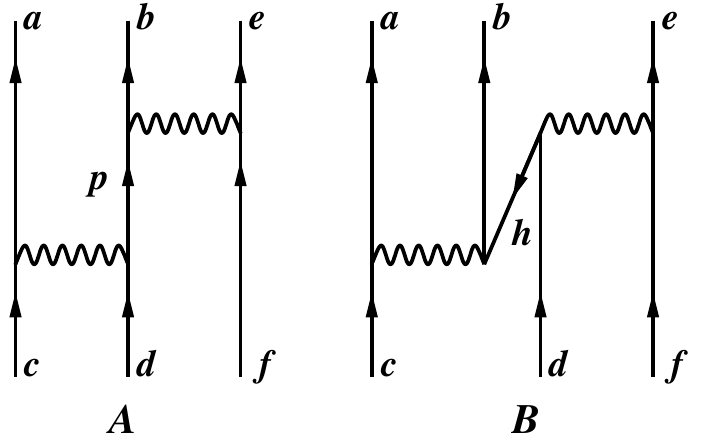}
\caption{Second-order three-body diagrams. The sum over the
  intermediate lines runs over particle and hole states outside the
 valence space, shown by A and B, respectively. For the sake of
  simplicity, for each topology we report only one of the diagrams
  which correspond to the permutations of the external lines.}
\label{diagram3corr}
\end{center}
\end{figure}

Then, in order to include the contribution of $\hat{Q}$-box diagrams
with at least three incoming and outcoming valence particles in 
\heff, one can resort to the same approximation used to manage the input  of 3NFs, namely  the so-called normal-ordering decomposition of the three-body component of a
many-body Hamiltonian \cite{HjorthJensen17}.
To this end, we start with a \qbox~including second-order three-body diagrams,  which, for nuclei with more than two-valence nucleons, account for the interaction via the two-body
force of the valence nucleons with core excitations  as well as with
virtual intermediate nucleons scattered above the valence space (see
Fig. \ref{diagram3corr}).

According to the definition of 3NFs  introduced in Section
\ref{sec-II.1}, the contributions reported in Fig. \ref{diagram3corr}
are proper three-body forces, since the intermediate states are
orbitals belonging to shells outside the valence space and,
consequently, they cannot be constructed by iterating 2NF diagrams
with valence-space external lines. The analytical expressions of these diagrams are in Ref. \cite{Polls83}.

\begin{figure}[h]
\begin{center}
\includegraphics[width=5.5cm]{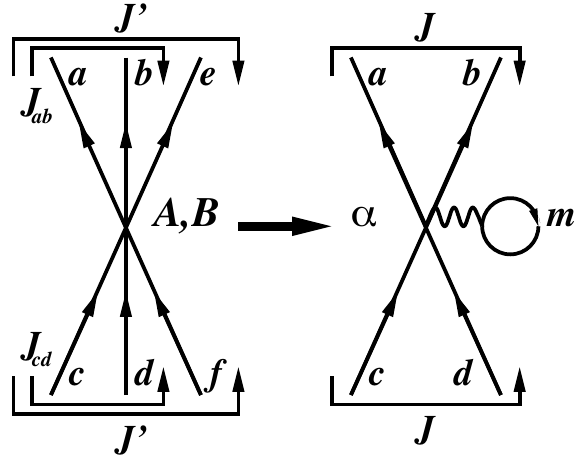}
\caption{Density-dependent two-body contribution that is
  obtained from a three-body one. The graph $\alpha$ is obtained by
  summing over one incoming and outgoing particle of the three-body
  graphs $A,B$ reported in Fig. \ref{diagram3corr}.}
\label{3bf}
\end{center}
\end{figure}

As discussed in Ref. \cite{Stroberg19}, the main advantage of
expressing many-body operators in normal-ordered form is to include as
much information as possible from the higher-particle-rank operators
into the lower-rank operators.
Then, after the normal-ordered decomposition, the approximation consists in
neglecting the residual three-body component and, consequently, \heff~may be
employed in standard SM codes.
Such a procedure allows to obtain, starting from the calculation of
each $(A,B)$ topology in Fig. \ref{diagram3corr}, nine one-loop
diagrams represented by the graph $(\alpha)$ in Fig. \ref{3bf}.

\noindent
Their explicit form, in terms of the three-body graphs $(A,B)$, is
\begin{equation}
\label{correq}
V_{ab,cd,J}^\alpha
= 
\sum_{m J'} \rho_{m} \frac{\hat{J}'^{2}}{\hat{J}^{2}}
{}_{\substack{\\A}\!}\Braket{\left(a,b\right),m;JJ'
 \left|V^{A,B}\right|\left(c,d\right),m;JJ'}_{A},
\end{equation}
where the summation over $m$ runs in the valence space, and
$\rho_{m}$ is the unperturbed occupation density of the orbital $m$ according to the number of valence nucleons.
The definition of the antisymmetrized but unnormalized three-body states,
$\Ket{\left(a,b\right),c;JJ'}_{A}$, can be found in Appendix~\ref{appendix1}.

Finally, the perturbative expression of the $\hat{Q}$ box contains one- and
two-body diagrams up to third order in $V_{NN}$, and a
density-dependent two-body contribution that includes the effect of
three-body diagrams at second-order in $V_{NN}$ \cite{Ellis77,Polls83}.

Obviously, this means that a specific effective SM
Hamiltonian has to be derived for a given system, depending on the number of valence
protons and neutrons, and  the obtained set of \heffs~differs  only in the TBMEs.
The role played by density-dependent \heffs~in the calculation of
g.s. energies in nuclei with many-valence nucleons has been
investigated in Refs. \cite{Ma19,Coraggio20e,Coraggio21}, and
discussed in Section \ref{sec-IV.2.2}.

Now we draw our attention on  the calculation of \heff~accounting also
for the contributions of the 3NF component of a realistic nuclear
potential, such as, for example, the N$^2$LO 3N potential reported in
Fig. \ref{n2lo3f}.
In SM calculations where \heff~has been derived perturbatively, this
contribution is  introduced at first-order in many-body
perturbation theory only for the one- and two-valence nucleon
systems.

\begin{figure}[h]
\begin{center}
\includegraphics[scale=1.0,angle=0]{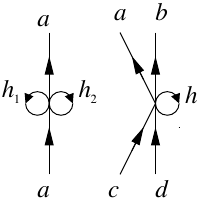}
\caption{First-order one- and two-body diagrams with a
  three-body-force vertex. See text for details.}
\label{1b2b3bf}
\end{center}
\end{figure}

\noindent 
In Fig. \ref{1b2b3bf}, both  first order one- and two-body diagrams of \qbox~from a 3N potential are shown, and their explicit expressions are
\begin{equation}
\label{1b3bfeq}
\epsilon_a^{\mathrm{(3NF)}}
=  \mathlarger{\sum}_{\substack{h_1,h_2{}\\J_{12}J}} ~
\frac{\hat{J}^2}{2 \hat{j_a}^2} 
{}_{\substack{\\A}\!}\Braket{\left(h_1,h_2\right),a;J_{12}J
 \left|V_{3N}\right|\left(h_1,h_2\right),a;J_{12}J}_{A},
\end{equation}

\begin{equation}
\label{2b3bfeq}
V_{ab,cd,J}^{\mathrm{(3NF)}}
= 
\mathlarger{\sum}_{h,J'} ~
\frac{\hat{J'}^2}{\hat{J}^2} 
{}_{\substack{\\A}\!}\Braket{\left(a,b\right),h;JJ'
 \left|V_{3N}\right|\left(c,d\right),h;JJ'}_{A},
\end{equation}
where the indices $h$  refer to core states, while the three-body matrix element on the right hand side of both equations, expressed within the proton-neutron formalism, is antisymmetrized but not normalized, and
its explicit form is given, for example, by Eq.~\eqref{pn3BMEs}.

The three-body component of a many-body Hamiltonian is therefore written in terms of one- and two-body pieces, which correspond, respectively, to interactions among one-valence and two-core nucleons, or two-valence and one-core nucleon, with coefficients given by the expressions in Eqs. (\ref{1b3bfeq}) and (\ref{2b3bfeq}). It is worth noting that these two pieces arise from the normal-ordering two-body decomposition of the 3NF with respect to the core as reference state. This means that the 3NF among valence nucleons is neglected, which may lead to underestimation of the 3NF repulsion and to overbinding that acquire more relevance with an increasing number of valence particles, as pointed out in Refs.
\cite{Stroberg17,Stroberg19}. In Ref. \cite{Stroberg17}, an ensemble or mixed-state reference was introduced to account for these interactions in an approximate way within the VS-IMSRG framework, which it is shown to cure this deficiency.
In this connection, it will be certainly relevant to investigate the relative importance of the missing 3NF contributions in our approach.  However, we would mention here, as discussed in detail in section \ref{sec-IV.1}, that of our  SM results for $p$ shell nuclei obtained by using an $NN$ +3N potential  are in close agreement with those of NCSM.  In particular, we are able to reproduce the experimental sequence of observed states in $^{10}$B, as it is done by NCSM calculations. The agreement between the results of the SM and NCSM models including 3NFs is on the overall of the same quality of that obtained when using the $NN$ force only. Larger discrepancies are found only for energies of high excited states, which may be related to our approximation in the 3NF treatment.


In concluding this section, it may be useful to give a summary of the diagrams we include in the derivation of our SM effective Hamiltonians. We arrest the $\hat Q$-box expansion to the one- and two-body Goldstone diagrams at third order in the $NN$ potential, which are explicitly shown in the Appendices of Ref. \cite{Coraggio12a}, while we consider only diagrams at first order in the 3N force whose expressions are given in Eqs. \ref{1b3bfeq} and \ref{2b3bfeq}. In addition, to account for the progressive filling of the valence space in systems with more than two-valence particles we include the density-dependent two-body contributions of Eq. \ref{correq} arising from the second-order three-body diagrams of Fig. \ref{diagram3corr}.
\subsubsection{\it The perturbative expansion of effective shell-model
 decay operators}\label{effopsec}
Besides the calculation of energy spectra, SM wave functions
may provide also the matrix elements of operators $\Theta$ which are related to 
physical observables, such as electromagnetic transition rates, multipole moments,
etc.

As it has been previously pointed out, the wave functions $|
\psi_{\alpha} \rangle$ obtained diagonalizing \heff~are not the true
ones $| \Psi_{\alpha} \rangle$, but their projections onto the model space
($|\psi_{\alpha} \rangle = P | \Psi_{\alpha} \rangle $).
Then, it is necessary to renormalize $\Theta$ in order to account for
the neglected degrees of freedom belonging to configurations outside the model space.
Formally, an effective operator $\Theta_{\rm eff}$ has to be derived,
such that
\begin{equation}
\langle \tilde{\Psi}_{\alpha} | \Theta | \Psi_{\beta} \rangle = \langle
\tilde{\psi}_{\alpha} | \Theta_{\rm eff} | \psi_{\beta} \rangle \label{effop}.  
\end{equation}

The perturbative expansion of effective operators has been approached
in the early attempts to employ realistic potentials for SM
calculations by L. Zamick for the problematics of electromagnetic
transitions \cite{Mavromatis66,Mavromatis67,Federman69} and by
I. S. Towner for the study of the quenching of spin-dependent
decay-operator matrix elements \cite{Towner83,Towner87}.

A formally improved structure to the derivation of non-Hermitian
effective operators has been elaborated by Suzuki and Okamoto in
Ref. \cite{Suzuki95}, where they have  introduced an expansion
formula for the effective operators in terms of a vertex function
\tbox~that, analogously to the \qbox~in the effective Hamiltonian
theory, is the building block for constructing effective operators.

We outline now some details about this procedure.
According to Eq. (\ref{defheff}) and keeping in mind that $\omega
\equiv Q \omega P$, \heff~may be written as
\begin{equation}
H_{\rm eff} = P H (P + \omega),
\end{equation}
\noindent
so that the true eigenstates $|\Psi_\alpha \rangle$ and their
orthonormal counterparts $\langle \tilde{\Psi}_\alpha |$ are given by

\begin{equation}
|\Psi_\alpha \rangle = (P + \omega) |\psi_\alpha \rangle ~~~~~,~~~~~
\langle \tilde{\Psi}_\alpha | = \langle \tilde{\psi}_\alpha | (P +
\omega^\dagger \omega ) (P + \omega^\dagger) . 
\end{equation}

\noindent
Actually, a general effective operator expression in the bra-ket
representation is written as
\begin{equation}
\Theta_{\rm eff} = \sum_{\alpha \beta} |\psi_\alpha \rangle \langle
\tilde{\Psi}_\alpha | \Theta | \Psi_\beta \rangle \langle
\tilde{\psi}_\beta | ,
\end{equation}
where $\Theta$ is a given time-independent Hermitian operator.
Then, \thetaeff~in an operator form is
\begin{equation}
\Theta_{\rm eff} = (P + \omega^\dagger \omega )^{-1} (P +
\omega^\dagger) \Theta (P + \omega).
\end{equation}

\noindent
It is worth noting that Eq. (\ref{effop}) holds whatever it is the
normalization of $|\Psi_\alpha \rangle$ and $|\psi_\alpha \rangle$,
but if the true eigenvectors are normalized, then $\langle
\tilde{\Psi}_\alpha | = \langle \Psi_\alpha | $  and $|\psi_\alpha
\rangle $ should be normalized as
\begin{equation}
\langle \tilde{\psi}_\alpha |(P + \omega^\dagger\omega)| \psi_\alpha \rangle = 1.
\end{equation}

To calculate \thetaeff, it is convenient to introduce the vertex
function \tbox, which is defined as
\begin{equation}
  \hat{\Theta} = (P + \omega^\dagger) \Theta (P + \omega),\label{thetaboxdef}
\end{equation}
\noindent
in order to factorize \thetaeff~as follows
\begin{equation}
\Theta_{\rm eff} = (P + \omega^\dagger \omega )^{-1} \hat{\Theta}.
\end{equation}

Therefore, to derive \thetaeff~one needs to calculate both $\hat{\Theta}$
and $\omega^\dagger \omega$.
Let us  tackle the first issue: according to Eq. (\ref{thetaboxdef})
and to the following expression  of $\omega$ in terms of \heff
\begin{equation}
\omega = \sum_{n=0}^\infty (-1)^n \left (\frac{1}{\epsilon_0 - QHQ}
\right )^{n+1}QH_1P(H_1^{\rm eff})^n,
\end{equation}
the following relation can be written for $\hat{\Theta}$ 
\begin{equation}
  \hat{\Theta} = \hat{\Theta}_{PP} + (\hat{\Theta}_{PQ} + h.c.) +
  \hat{\Theta}_{QQ},
\end{equation}
where
\begin{equation}
  \hat{\Theta}_{PP} = P \Theta P,
\end{equation}
\begin{equation}
  \hat{\Theta}_{PQ} = P \Theta \omega P = \sum_{n=0}^\infty
  \hat{\Theta}_n (H_1^{\rm eff})^n,
\end{equation}
\begin{equation}
  \hat{\Theta}_{QQ} = P \omega^\dagger \Theta \omega P =
  \sum_{n,m=0}^\infty (H_1^{\rm eff})^n
  \hat{\Theta}_{nm} (H_1^{\rm eff})^m,
\end{equation}
\noindent
and $\hat{\Theta}_m$, $\hat{\Theta}_{mn}$ are defined as
\begin{eqnarray}
\hat{\Theta}_m & = & \frac {1}{m!} \frac {d^m \hat{\Theta}
 (\epsilon)}{d \epsilon^m} \biggl|_{\epsilon=\epsilon_0} , \\
\hat{\Theta}_{mn} & = & \frac {1}{m! n!} \frac{d^m}{d \epsilon_1^m}
\frac{d^n}{d \epsilon_2^n}  \hat{\Theta} (\epsilon_1 ;\epsilon_2)
\biggl|_{\epsilon_1= \epsilon_0, \epsilon_2  = \epsilon_0},
\end{eqnarray}

\noindent
with
\begin{eqnarray}
\hat{\Theta} (\epsilon) = & P \Theta P + P \Theta Q
\frac{1}{\epsilon - Q H Q} Q H_1 P , ~~~~~~~~~~~~~~~~~~~\label{thetabox} \\
\hat{\Theta} (\epsilon_1 ; \epsilon_2) = & P H_1 Q
\frac{1}{\epsilon_1 - Q H Q} Q \Theta Q \frac{1}{\epsilon_2 - Q H Q} Q
H_1 P.
\end{eqnarray}

\noindent 
By way of definition (\ref{eqb}), the product $\omega^\dagger \omega$ takes the form
\begin{equation}
\omega^\dagger \omega = -\sum_{n=1}^\infty\sum_{m=1}^\infty((H_1^{\rm
  eff})^\dagger)^{n-1}\hat{Q}(\epsilon_0)_{n+m-1}(H_1^{\rm
  eff})^{m-1}.
\end{equation}
Using now the expression of $H_1^{\rm eff}$ in terms of the \qbox~and
its derivatives -- Eqs. (\ref{kkeq}) and (\ref{kkeqexp}) -- the above quantity may be
rewritten as 
\begin{equation}
\omega^\dagger \omega = - {\hat{Q}_1 + (\hat{Q}_2\hat{Q} + h.c.) +
  (\hat{Q}_3\hat{Q}\hat{Q} + h.c.) + (\hat{Q}_2\hat{Q}_1\hat{Q} +
  h.c.) + \cdots } \label{omega+omega} 
\end{equation}

\noindent
Melting together Eqs. (\ref{thetabox}) and (\ref{omega+omega}), the
final perturbative expansion form of the effective operator 
$\Theta_{\rm eff}$ is
\begin{equation}
\Theta_{\rm eff}  =  (P + \hat{Q}_1 + \hat{Q}_1 \hat{Q}_1 + \hat{Q}_2
\hat{Q} + \hat{Q} \hat{Q}_2 + \cdots) \times (\chi_0+ \chi_1 + \chi_2
+\cdots), \label{effopexp1} 
\end{equation}
where
\begin{eqnarray}
\chi_0 &=& (\hat{\Theta}_0 + h.c.)+ \hat{\Theta}_{00},  \label{chi0} \\
\chi_1 &=& (\hat{\Theta}_1\hat{Q} + h.c.) + (\hat{\Theta}_{01}\hat{Q}
+ h.c.), \\
\chi_2 &=& (\hat{\Theta}_1\hat{Q}_1 \hat{Q}+ h.c.) +
(\hat{\Theta}_{2}\hat{Q}\hat{Q} + h.c.) +
(\hat{\Theta}_{02}\hat{Q}\hat{Q} + h.c.)+  \hat{Q} 
\hat{\Theta}_{11} \hat{Q}. \label{chin} \\
&~~~& \cdots \nonumber
\end{eqnarray}

It is worth to evidence the link existing between \heff, derived
in terms of the \qbox, and any effective operator. 
This is achieved by inserting the identity $\hat{Q}  \hat{Q}^{-1} =
\mathbf{1}$ in Eq. (\ref{effopexp1}), and consequently obtaining the
following expression:

\begin{eqnarray}
\Theta_{\rm eff} & = & (P + \hat{Q}_1 + \hat{Q}_1 \hat{Q}_1 + \hat{Q}_2
\hat{Q} + \hat{Q} \hat{Q}_2 + \cdots) \hat{Q}  \hat{Q}^{-1} \times
                       (\chi_0+ \chi_1 + \chi_2 +\cdots)  \nonumber
  \\
~ & = & H_{\rm eff} \hat{Q}^{-1}  (\chi_0+ \chi_1 + \chi_2 +\cdots).
\label{effopexp2}
\end{eqnarray}

The $\chi_n$ series must be arrested to a finite order, and the
starting point is the derivation of a perturbative expansion of
$\hat{\Theta}_0 \equiv \hat{\Theta}(\epsilon_0)$ and
$\hat{\Theta}_{00} \equiv \hat{\Theta}(\epsilon_0;\epsilon_0)$,
including diagrams up to a finite order in the perturbation theory,
consistently with the $\hat{Q}$-box expansion. 
The issue of the convergence of the $\chi_n$ series and of the
perturbative expansion of $\hat{\Theta}_0$ and $\hat{\Theta}_{00}$ has
been investigated in Refs. \cite{Coraggio18,Coraggio19a,Coraggio20a},
and in Ref. \cite{Coraggio20c}, which report details about the
calculation of the diagrams appearing in the $\hat{\Theta}_0$
expansion for a one-body operator $\Theta$.

\subsection{\it Gamow shell model with three-body forces \label{sec-III.3}}

 As the number of protons or neutrons in the nucleus increases to the existence limit of the dripline, exotic phenomena such as halo and resonances can emerge. With extreme proton-neutron imbalance, the nuclei around the dripline are weakly bound or unbound. They belong to open quantum systems  in which the coupling to the continuum can be significant and should be properly treated. Due to the large spatial distributions of wave functions in resonance and continuum states, using a standard spatially-localized HO basis would not be a good option. In the past two decades, several methods have been developed to overcome this challenge. The conventional SM has been extended to include the continuum effect, e.g., SM embedded in the continuum \cite{Bennaceur1999289, Okolowicz03, Rotureau2005, Volya2005} and the continuum-coupled SM \cite{Tsukiyama15}.

\begin{figure}[!ht]
    \begin{center}
    \includegraphics[width=0.4\textwidth]{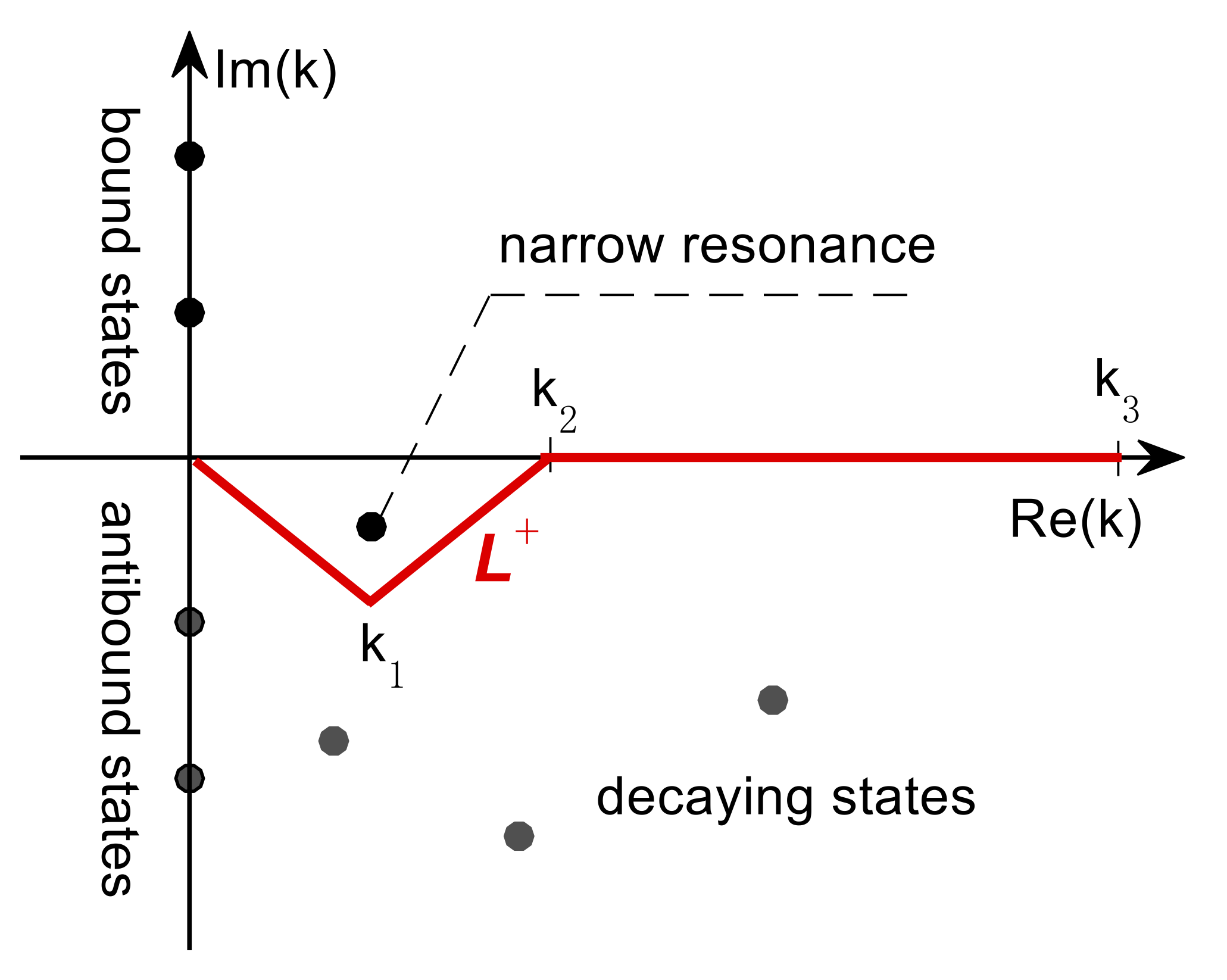}
    \caption{\label{fig:Berggren} Schematic Berggren complex-$k$ plane. The bound, resonant and scattering states construct the Berggren completeness relation. The contour $L^+$ has to be chosen in such a way that all the discrete narrow resonant states are contained in the domain between $L^+$ and the real-$k$ axis \cite{Sun17}. }
    \end{center}
\end{figure}

  An elegant treatment of the continuum coupling is to use the Berggren representation \cite{Berggren1968} in which the one-body Schr\"odinger equation is generalized to a complex-momentum (complex-$k$) plane. This creates naturally bound, resonance and continuum single-particle states on an equal footing, see Fig. \ref{fig:Berggren}. The many-body Hamiltonian can be expressed in the Berggren basis, and tackled by many-body methods in the complex-$k$ space. The SM has been successfully extended to the complex-$k$ Berggren basis, leading to the so-called  GSM, in which the continuum effect is included at the basis level. In the first GSM applications to atomic nuclei, phenomenological interactions were used with the potential parameters determined by fitting nuclear structure data \cite{IdBetan02, Michel02, Michel03, Michel19, Papadimitriou11, Michel20, Li21b}. Soon, the GSM was developed with realistic nuclear interactions \cite{Hagen2006, Tsukiyama2009, Papadimitriou2013, Sun17, Hu2020, Li2021a}. Meanwhile, the Berggren technique was also used in CC \cite{Hagen2007, Hagen2010, Hagen2012} and  IMSRG approaches \cite{Hu19}. 

 Along with the continuum coupling, 3NFs also play an important role in the descriptions of exotic nuclei. In Refs.~\cite{Ma20a,Ma20b,Zhang22}, the realistic GSM was extended with 3NFs considered. The inclusions of both the continuum coupling and 3NFs can give a better quantitative description of exotic nuclei~\cite{Ma20a}. With 3NFs included,  two steps of development towards a full self-consistent {\it ab initio} GSM have been made: i) the WS potential has been used to generate the complex-$k$ Berggren basis, and the many-body GSM with realistic 2NFs and 3NFs has been performed in the WS basis~\cite{Ma20a,Ma20b}, whose WS parameters have been  determined by fitting data; ii) for a more self-consistent calculation, the Berggren basis has been created starting from realistic interaction itself using the complex-$k$ Gamow Hartree-Fock (GHF) method, and the complex GSM has been performed in the GHF basis. The 3NF has been  included in both the GHF and GSM calculations, as done in Ref.~\cite{Zhang22}. 

Similar to standard SM  calculations, an auxiliary one-body potential $U$ is usually introduced into the Hamiltonian to obtain a one-body term $H_0$  describing the independent motions of the nucleons  and a residual interaction $H_1$. However, here we rewrite the Hamiltonian \eqref{defh} to include the 3NF and remove the center of mass kinetic energy thus  obtaining an intrinsic transitionally invariant Hamiltonian,   
\begin{equation}\label{eq:hm1}
    \begin{aligned}
      H&=\sum_{i<j}\frac{(\boldsymbol{p}_i-\boldsymbol{p}_j)^2}{2mA}+\hat{V}_{\text{NN}}+\hat{V}_{\text{3N}}\\
       &=\left[ \sum_{i=1}^A\left(\frac{p_i^2}{2m}+U_i\right) \right]
         +\left[\sum^A_{i<j}\left({V}_{\text{NN}}^{(ij)}-\frac{\boldsymbol{p_i}\cdot \boldsymbol{p_j}}{mA}\right)-\sum^A_{i=1}\left(U_i+\frac{p_i^2}{2mA}\right)+\sum^A_{i<j<k}{V}_{\text{3N}}^{(ijk)}\right]\\
       &=H_0+H_1,
    \end{aligned}
\end{equation}
To generate the Berggren basis, $U$ is usually taken as the WS potential produced by the core for the GSM calculation with a core~\cite{Ma20a,Ma20b}. The radial wave functions of the Berggren SP states are obtained by solving the SP Schr\"odinger equation in the complex-$k$ space with the WS potential $U$. The Berggren SP states form a complete set of basis states with discrete bound, resonant and continuum scattering states, as shown in Fig. \ref{fig:Berggren}. Due to the complexity and computational task of many-body calculations with full 3NF, as discussed in the Introduction and in Section \ref{perturbativeh}, we adopt the normal-ordering approximation with neglecting the residual three-body term~\cite{Roth12, Fukui18, Ma19, Ma20a, Ma20b}. 
The normal-ordered zero-body and one-body terms are absorbed into the core Hamiltonian which generates the Berggren basis, while the final GSM Hamiltonian has a two-body form but with normal-ordered two-body term of the 3NF included (see Eq.(\ref{2b3bfeq})). 
With the overlap method of wave functions~\cite{Sun17, Li2020}, the Hamiltonian can be transferred to the complex-$k$ Berggren basis.

 With the Hamiltonian matrix elements given in the Berggren basis, the $\hat{Q}$-box folded-diagram method is used to construct the realistic complex effective interaction for a chosen model space. In general, such kind of model space should include relevant bound, resonant and continuum states. Therefore, the basis states are certainly not degenerate, and we exploit an extension of the  Kuo-Krenciglowa (EKK) method ~\cite{Takayanagi11}  to the complex-$k$ space to build the effective GSM interaction. At last, the complex-symmetry GSM Hamiltonian is diagonalized in the model space using the Jacobi-Davidson method~\cite{Michel20a}.

 Although the Hamiltonian is intrinsic, the GSM wavefunction is not factorized into the center-of-mass (CoM) and intrinsic parts. This means that the effect from the CoM motion has not been removed exactly. In the HO basis, the CoM effect can be treated using the Lawson method~\cite{Gloeckner74}. In principle, one can generalize the Lawson method, for example, in the real-energy CC~\cite{Hagen2009} and IMSRG~\cite{Hergert16} calculations which adopted the Hartree-Fock basis. Unfortunately, the generalization is not valid in the complex-energy Berggren basis due to the fact that the $R^2$ matrix elements ($R$ is the CoM position) cannot be regularized in resonance and continuum states, which are not square integrable in fact. However, we have well discussed in the previous works~\cite{Sun17, Hu19, Hu2020} that the CoM effect is not significant in low-lying states.

\section{Applications and comparison with experiment}\label{applications}
In this section, we shall review recent results of SM calculations based on effective Hamiltonians derived from realistic two- and three-body potentials. 
The aim of the presentation is essentially to highlight the role of 3NFs in explaining phenomena such as the location of the neutron dripline in isotopic chains and the shell evolution as a function of the number of valence nucleons, as well as to investigate the combined effects of 3NFs and the coupling with continuum for the description of unbound 
nuclei and unbound resonance states in the vicinity of the dripline or of the Borromean structure of halo nuclei. We shell focus on systems ranging from the very light-mass 
$p$-shell nuclei to those with intermediate mass belonging to the $fp$ shell.

In the subsection devoted to $p$-shell nuclei, we have found it useful to evidence the validity of our perturbative approach in deriving the effective SM  Hamiltonian. To this end, in the first part we illustrate the convergence properties of the $\hat Q$-box vertex function as concerns the truncation of the intermediate-state space, the order-by-order convergence, and the dependence on the HO parameter. The analyses is based on an $NN$ potential since as concerns 3NFs only first-order contributions of the normal-ordered one- and two-body  parts  are taken into account. Then, once shown how the introduced approximations can be taken under control, we compare our SM results arising from the chiral $NN$-only and $NN$+3N forces with the corresponding ones obtained by the {\it ab initio} NCSM.

\subsection{\it Benchmark calculations in the $0p$-shell region \label{sec-IV.1}}
Benchmark calculations are very important to test methods as well as
computational approaches. More specifically, they may be helpful to understand to what
extent a many-body method is working and to estimate the impact of the
necessary truncations and approximations that have been introduced.
In Ref. \cite{Coraggio12a}, an extensive study has been carried out 
 to compare the results of SM calculations for $p$-shell nuclei
obtained by using  an effective Hamiltonian derived 
from an $NN$ chiral potential  at N$^3$LO
\cite{Entem02,Machleidt11} with those obtained with the  {\it ab
  initio} NCSM \cite{Navratil04,Navratil07a,Navratil07b,Maris13}.
In Ref. \cite{Fukui18}, this investigation has been extended by including
in the derivation of \heff~for $p$-shell nuclei  the one- and
two-body components of a normal-ordered 3N chiral potential at
N$^2$LO.

The work in Ref. \cite{Coraggio12a} has also tried to assess the
behavior of the perturbative expansion of \heff~with respect to
the dimension of the intermediate state space  and the
order-by-order convergence, when starting from chiral potentials.
As regards this  aspect, we recall that the $Q$ space appearing in
Eq. \eqref{qbox}  is the complement of the model space in the whole
Hilbert space, therefore it is composed by an infinite number of
configurations.
It is clearly unfeasible to employ an infinite $Q$ space, and
consequently the perturbative expansion of the \qbox~ implies a
truncation of the space of the intermediate states, which belong to the
$Q$ space by definition.
The common procedure is to employ an energy truncation, which consists in 
including only those intermediate states whose unperturbed
excitation energy is smaller than a fixed value $E_{\rm max}$
expressed in terms of the number of oscillator quanta $N_{\rm max}$,
namely as an integer multiple of the HO parameter
$\hbar \omega$

\[
  E_{\rm max} = N_{\rm max} \hbar \omega.
\]

\begin{figure}[h]
\begin{center}
\includegraphics[width=10cm]{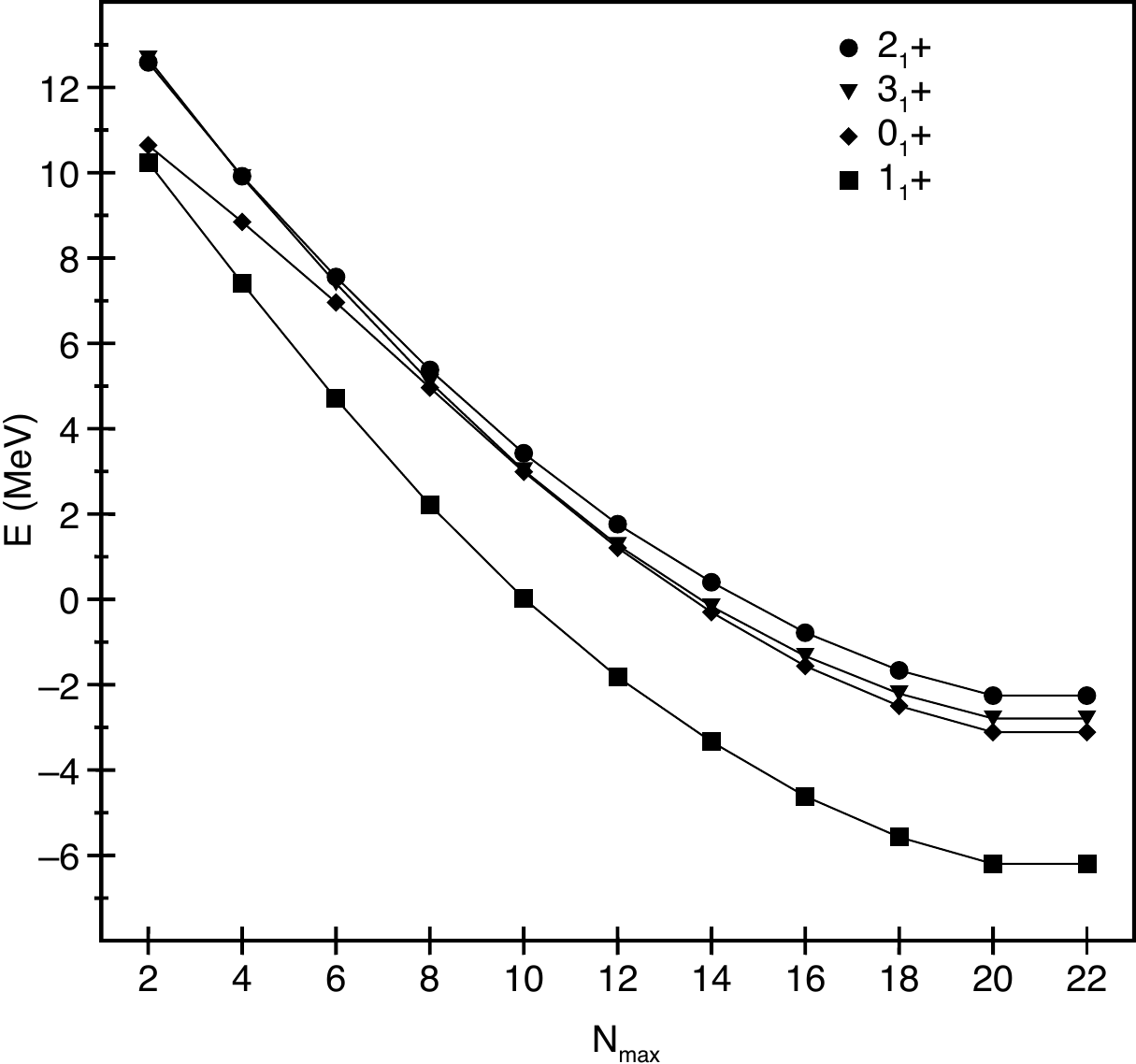}
\end{center}
\caption{Theoretical energies of $^{6}$Li yrast states relative to
  $^4$He, obtained with the N$^3$LO $NN$ potential, as a function of
  $N_{\rm max}$ \cite{Coraggio12a}.}
\label{intN3LO}
\end{figure}

In Ref. \cite{Coraggio12a} the theoretical energies of the yrast
states in $^6$Li, corresponding to the absolute energy values relative to $^4$He, have been calculated as a function
of $E_{\rm max}$ (see Fig. \ref{intN3LO}), using an effective
Hamiltonian derived from the N$^3$LO $NN$ potential and including in
the $\hat{Q}$-box diagrams up to the second order in $H_1$.

As can be seen in Fig. \ref{intN3LO}, the convergence is reached when 
 intermediate states at least up to $E_{\rm max}=20~ \hbar
\omega$, with $\hbar \omega=19$ MeV, are included.
This value of the HO parameter is close to the one
provided by the expression \cite{Blomqvist68} $\hbar \omega =
45A^{-1/3}-25A^{-2/3}$ for $A=4$.
This is no surprise if we consider that the $NN$ potential is
characterized (in momentum-space representation) by a certain cutoff momentum $\Lambda$, which  is also the maximum relative
momentum of the two-nucleon system.
Consequently, the maximum value of the energy corresponding to  the relative motion of two nucleons is

\begin{equation}
E_{\rm max} = \frac{ \hbar^2 \Lambda^2}{M},
\end{equation}

\noindent
where $M$ is the nucleon mass.
This relation may be rewritten in terms of $N_{\rm max}$ and $\hbar
\omega$,

\begin{equation}
N_{\rm max} \hbar \omega = \frac{ 
\hbar^2 \Lambda^2}{M}. \label{MSTA}
\end{equation}

Equation \eqref{MSTA} constrains the value of $N_{\rm max}$ for a chosen HO
parameter and depends on the cutoff $\Lambda$ of the $NN$ potential.
The chiral N$^3$LO potential under consideration is characterized by a
cutoff $\Lambda = 2.5, 2.6$ fm$^{-1}$ \cite{Entem02,Machleidt11}, and, therefore,
 if $\hbar \omega=$19 MeV, one should include in the
${\hat Q}$-box expansion the contributions of the $Q$-space  configurations at
least up to $N_{\rm max}=16$.
It is worth pointing out that the N$^3$LO potential is multiplied by
a  smooth regulator function with a gaussian shape \cite{Entem03}.
This characteristic slows down the convergence behavior of the $NN$
potential and justifies the need to include a larger number of
$Q$-space configurations.

Actually, in Ref. \cite{Coraggio12a} the convergence with respect to
the number of intermediate states has been studied also considering an
$NN$ potential with a sharp cutoff as regulator function, a chiral
potential dubbed N$^3$LOW \cite{Coraggio07b}.
This potential, whose cutoff is $\Lambda = 2.1$ fm$^{-1}$, is
characterized by a faster convergence, and the convergence is reached
at $N_{\rm max}=10$, which is consistent with the relationship
\eqref{MSTA} for $\hbar \omega=19$ MeV.

Another important aspect to be studied is the order-by-order
convergence properties of the \heff~ expansion, namely the dependence of
SM results on the order at which the perturbative expansion of the
$\hat{Q}$ box is arrested.
It is worth mentioning that this relevant topic has been first
investigated by Barrett and Kirson in the pioneering period of the
perturbative expansion of \heff~\cite{Barrett70}, and  it has been reprised in
Ref. \cite{Coraggio12a}  within the
$\hat{Q}$-box approach,  using $\hat{Q}$ boxes at second  ($H^{\rm eff}_{\rm
  2nd}$) and third  ($H^{\rm eff}_{\rm 3rd}$) order in perturbation theory.
Moreover, in order to estimate the value to which the perturbative
series may converge,  \heff~has been derived by calculating the
Pad\`e approximant $[2|1]$ \cite{Baker70,Ayoub79} of the
$\hat{Q}$ box ($H^{\rm eff}_{\mbox{\tiny{Pad\`e}}}$).

\begin{figure}[h]
\begin{center}
\includegraphics[width=10cm]{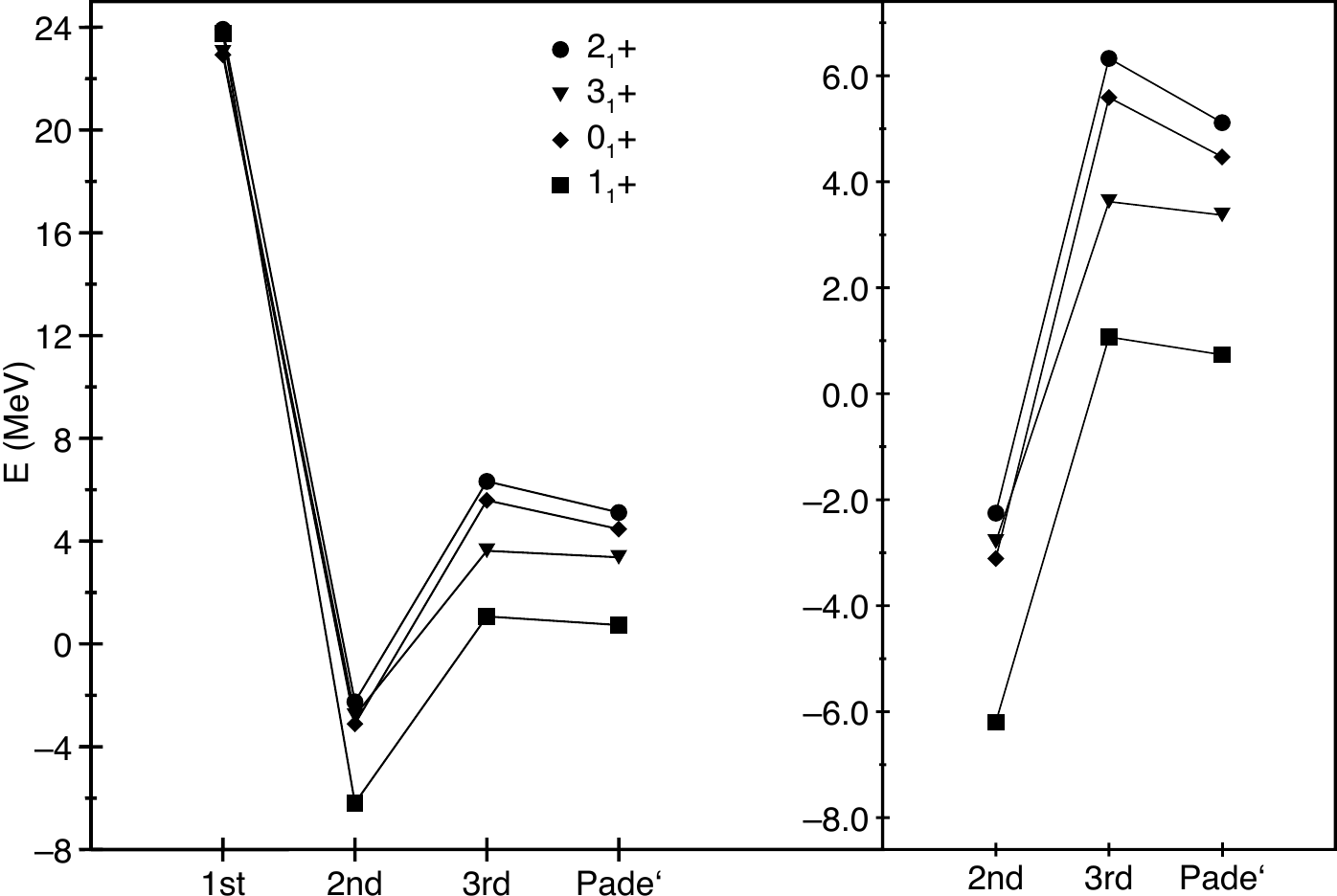}
\caption{Theoretical energies of $^{6}$Li yrast states relative to
  $^4$He, obtained with $H^{\rm eff}_{\rm 1st}$, $H^{\rm eff}_{\rm
    2nd}$, $H^{\rm eff}_{\rm 3rd}$, and $H^{\rm
    eff}_{\mbox{\tiny{Pad\`e}}}$ derived from the N$^3$LO $NN$ potential
  (see text for details). In the right side of the figure, where an
  expanded scale is adopted, the $H^{\rm eff}_{\rm 1st}$ results are
  omitted \cite{Coraggio12a}.}
\label{N3LO3rd}
\end{center}
\end{figure}

In Fig. \ref{N3LO3rd} the energies of $^{6}$Li yrast states with
respect to $^{4}$He, obtained with the N$^3$LO $NN$ potential, are reported
as calculated with $H^{\rm eff}_{\rm 1st}$, $H^{\rm  eff}_{\rm 2nd}$,
$H^{\rm eff}_{\rm 3rd}$, and $H^{\rm eff}_{\mbox{ \tiny{Pad\`e}}}$, $H^{\rm eff}_{\rm 1st}$
representing the one-body first-order $\hat{Q}$-box diagrams plus the $NN$ bare potential.

There are a couple of aspects that should be evidenced from the
inspection of Fig. \ref{N3LO3rd}:

\begin{enumerate}
\item the large gap between the results at first order in perturbation
  theory and those at higher orders shows that the employment of a
  bare $NN$ potential  in SM calculations, without any
  renormalization due to long-range correlations, leads to a poor
  description of the physics of atomic nuclei;
\item the results for $^6$Li obtained with $H^{\rm eff}_{\rm 3rd}$ are very
  close to those with $H^{\rm eff}_{\mbox{\tiny{Pad\`e}}}$, and this
  supports the hypothesis that SM calculations may have a weak
  dependence on higher-order $\hat{Q}$-box perturbative terms.
\end{enumerate} 

\noindent
On the above grounds, in all subsequent SM calculations the effective
Hamiltonians have been derived by calculating the Pad\`e approximant
$[2|1]$ of the $\hat{Q}$ box.

Finally, it is worth to examine another aspect that has been
evidenced in Ref. \cite{Coraggio12a}.
We recall that, because of Eqs. (\ref{defh}-\ref{defh1}), there could be
a dependence of \heff~on the choice of the auxiliary potential
$U$ introduced to construct the SP basis employed to expand the matrix elements of the input interaction.
More precisely, since we choose the HO basis, the results of the SM
calculations may depend on the choice of the HO parameter $\hbar
\omega$.
This is due to  the  approximations inherent our calculations, namely, as discussed above, to the truncation of the space of intermediate states and
 of the perturbative expansion of the $\hat{Q}$-box at a
certain order.

\begin{figure}[h]
\begin{center}
\includegraphics[width=8 cm]{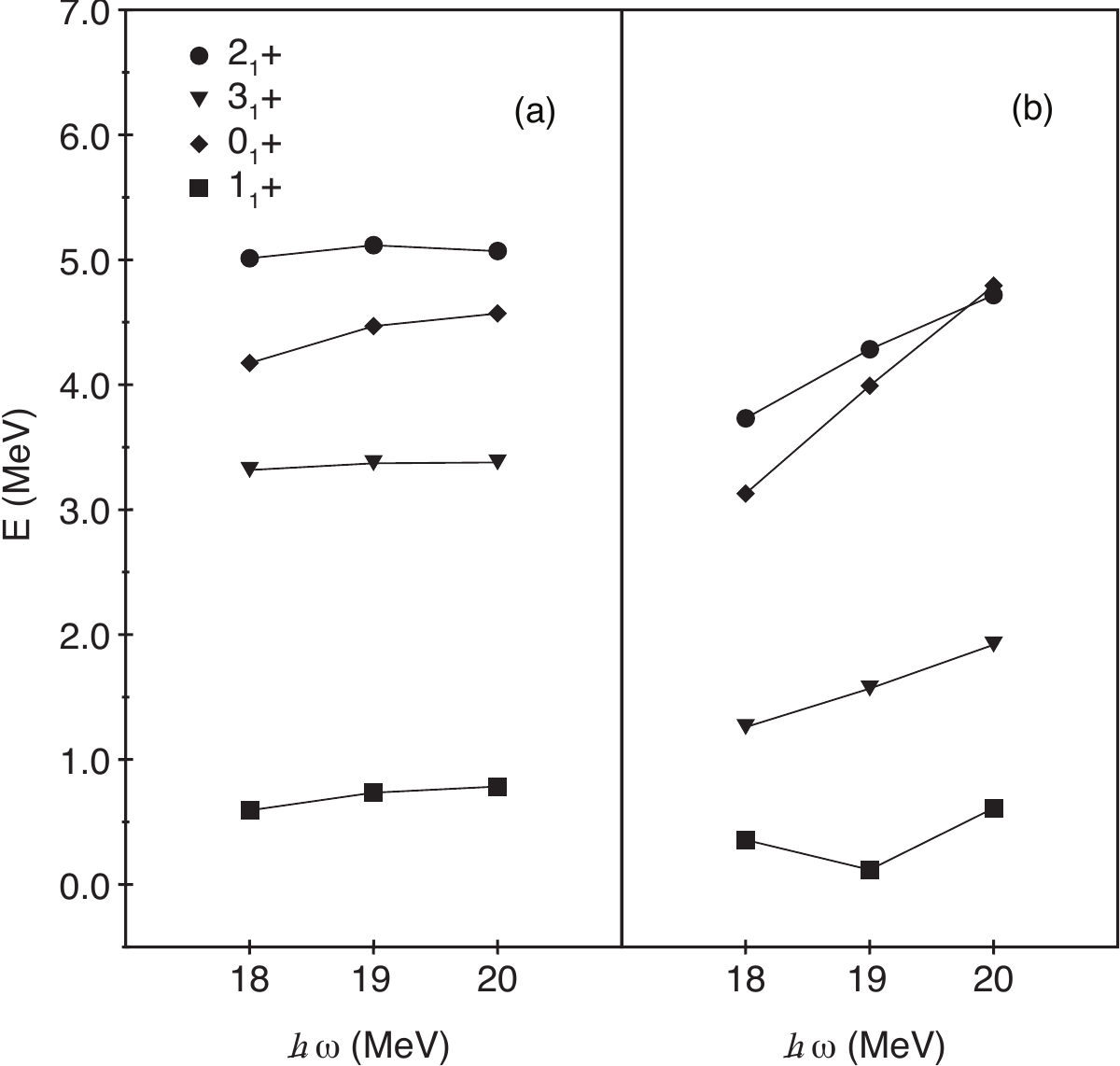}
\caption{Theoretical energies of $^{6}$Li yrast states relative to
  $^4$He, obtained with the N$^3$LO $NN$ potential, as a function of $\hbar
  \omega$ \cite{Coraggio12a}. See text for details.}
\label{hf1}
\end{center}
\end{figure}

In Fig. \ref{hf1}, the theoretical energies of the yrast states in
$^6$Li are reported as a function of $\hbar \omega$ for three effective Hamiltonians derived from the
N$^{3}$LO potential  by using as  HO parameter 
$\hbar \omega$= 18, 19, and 20 MeV.
 The panel (a) of  Fig. \ref{hf1} refers to effective Hamiltonians derived including all
third-order diagrams in the $\hat{Q}$ box, and then calculating its
Pad\`e approximant $[2|1]$.
On the other hand, the spectra in panel (b) are obtained by retaining in the $\hat{Q}$ box only the first-order
($V\mbox{-}U$)-insertion diagram (see Fig. 1 in
Ref. \cite{Coraggio12a}) and neglecting higher-order terms of the same
class of diagrams, and again calculating its Pad\`e approximant
$[2|1]$.
These results  show very clearly that ($V\mbox{-}U$)-insertion diagrams play a crucial role to reduce the
dependence on the HO parameter.

\begin{figure}[h]
\begin{center}
\includegraphics[width=8cm]{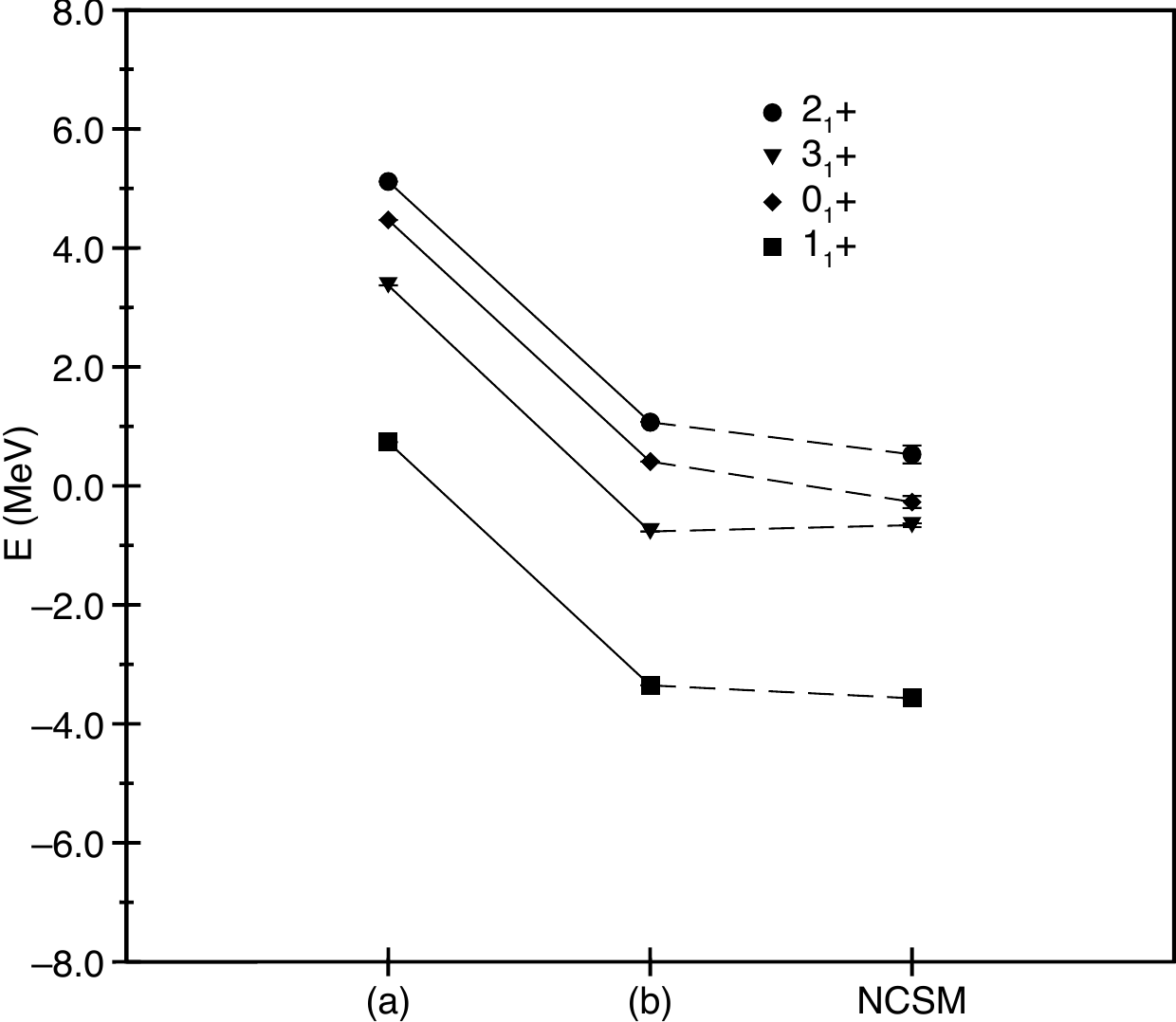}
\caption{Theoretical energies of $^{6}$Li yrast states relative to
  $^4$He, obtained with N$^3$LO $NN$ potential. (a) SM calculation
  with an effective intrinsic plus center of mass Hamiltonian . (b)
  SM calculation with an effective intrinsic Hamiltonian. (c) NCSM calculation \cite{Coraggio12a}.}
\label{6Lincsm}
\end{center}
\end{figure}

Once a complete survey of all possible sources of approximations
induced by the perturbative expansion has been completed, a comparison of the
SM results with those provided by the {\it ab initio} NCSM can be
performed \cite{Navratil04,Navratil07a}.

Actually, it should be pointed out that the results reported
 in Fig. \ref{intN3LO}-\ref{hf1} have been obtained
starting from the  $A$-body Hamiltonian  of Eqs. \eqref{defh}-\eqref{defh1}, which is not translationally
invariant, while NCSM calculations employ a purely intrinsic Hamiltonian. 
Therefore, to compare our SM results with the NCSM ones we have to remove the center of mass kinetic energy from Eqs.\eqref{defh}-\eqref{defh1}, namely we have to  use the Hamiltonian defined in Eq. \eqref{eq:hm1}.

The calculated energies of the yrast states in $^6$Li relative to
$^4$He are reported in Fig. \ref{6Lincsm}; the results labelled with
(a) refer to a SM calculation with an effective Hamiltonian
derived from Eqs. \eqref{defh}-\eqref{defh1}, the spectrum (b) corresponds to an effective
Hamiltonian derived from the translationally invariant Hamiltonian of
Eq. \eqref{eq:hm1} retaining only the $NN$ component. 
The NCSM spectrum (c) is obtained considering the calculated binding
energy of $^{6}$Li in Ref. \cite{Navratil07b} with respect to the $^{4}$He ground
state energy \cite{Navratil07a}, and the $^{6}$Li excitation
energies reported in Ref. \cite{Navratil04}.
The results in Fig. \ref{6Lincsm} evidence how relevant is to employ a
purely intrinsic Hamiltonian to compare correctly the ground-state
energies of SM and NCSM.
This choice of the Hamiltonian does not affect, however, the energy spacings,
and it should be noted that the difference between
the not transitionally invariant and intrinsic Hamiltonians rapidly decreases with growing $A$.

\begin{figure}[h]
\begin{center}
\includegraphics[width=18cm]{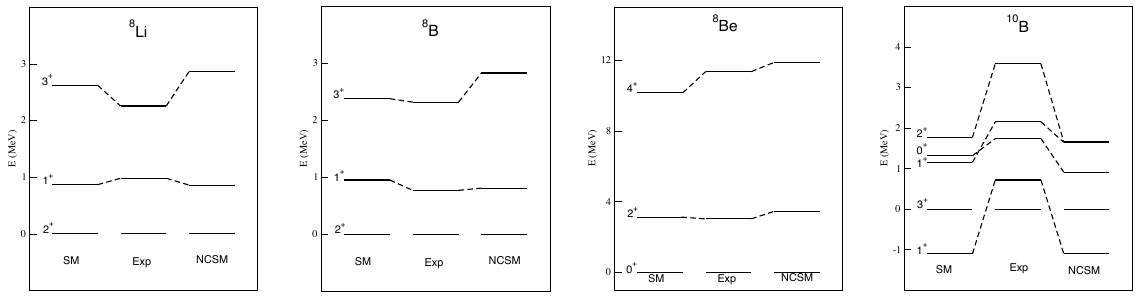}
\caption{Theoretical and experimental spectra for $^{8}$Li, $^{8}$B,
  $^{8}$Be, and $^{10}$B. The theoretical energies have been obtained
  using the N$^3$LO $NN$ potential within SM and NCSM calculations. Figure adapted from Ref. \cite{Fukui18}.}  
\label{spectra01}
\end{center}
\end{figure}

Moreover, we can conclude that the agreement between the SM and NCSM
results for $^6$Li, which is a two-valence nucleon
system with respect to $^4$He core, is quite good.
This conclusion may be extended also to calculations of nuclei with a
number of valence nucleons  larger than two.

\begin{figure}[h]
\begin{center}
\includegraphics[width=14cm]{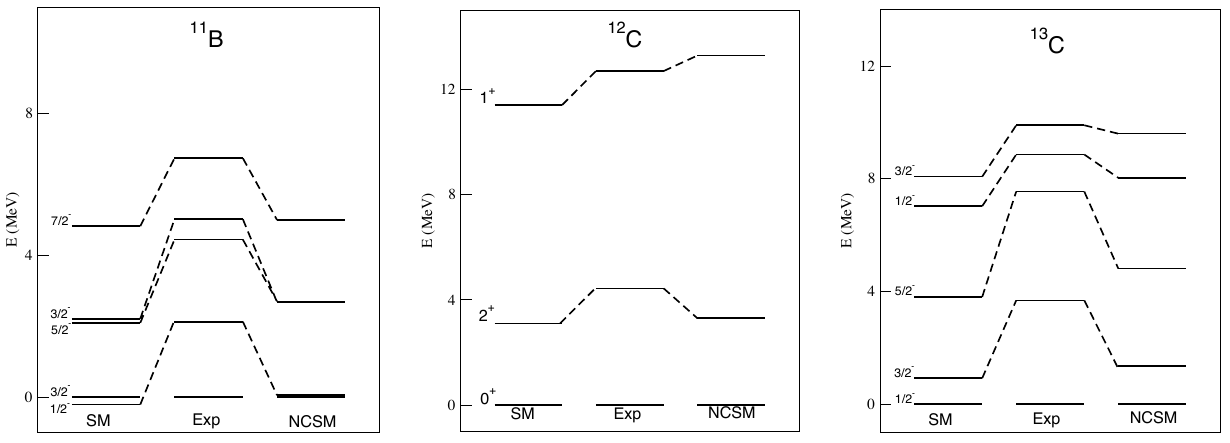}
\caption{ Same as in Fig. \ref{spectra01}, but for $^{11}$B, $^{12}$C,
  and $^{13}$C. Figure adapted from \cite{Fukui18}.}  
\label{spectra02}
\end{center}
\end{figure}

In Figs. \ref{spectra01} and \ref{spectra02} the low-energy
excitation spectra of $^8$Li, $^8$B, $^8$Be, $^{10}$B, $^{11}$B,
$^{12}$C, and $^{13}$C, calculated with SM \cite{Fukui18} and NCSM
\cite{Navratil07b,Maris13} are reported  and compared with experiment \cite{ensdf}.
From the inspection of these two figures we see that, as
regards the excitation spectra of many-valence nucleon systems, the
comparison between SM calculations with \heff~derived within the
perturbative approach and {\it ab initio} calculations is quite satisfactory.

\begin{figure}[h]
\begin{center}
\includegraphics[width=15cm]{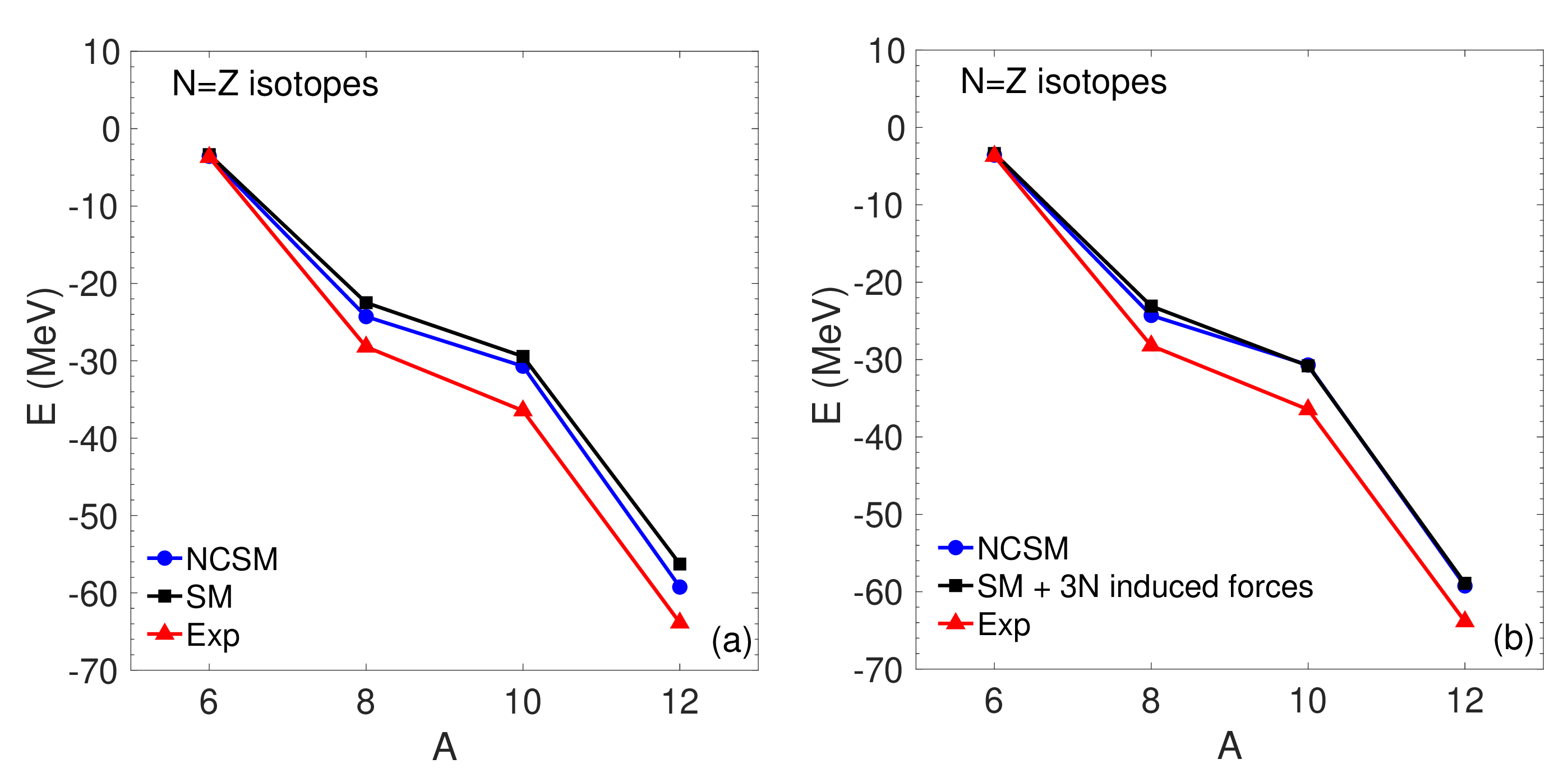}
\caption{
Experimental ground-state energies for $N = Z$ nuclei with mass $6 \leq A
  \leq 12$  are compared with theoretical values obtained using the N$^3$LO $NN$ potential within the NCSM and  SM. SM results refer to calculations  (a) without and (b) with   contributions from 3N induced forces \cite{Coraggio12a, Fukui18}.}
\label{gsencsm}
\end{center}
\end{figure}

In Fig. \ref{gsencsm} (a), the ground-state energies, relative to $^4$He,
for the $N=Z$ nuclei with mass $6 \leq A \leq 12$ calculated within the SM
(dot-dashed line) \cite{Coraggio12a} are
compared with those of  NCSM calculations (dotted line) and the experimental
ones (continuous line) \cite{Audi12}.

\noindent

We see that discrepancies between SM and NCSM results increase with
the number of valence nucleons, and this may be ascribed to the fact
that  many-body ($>2$) components of $H_{\rm eff}$ have not been taken into account.

As mentioned in Section \ref{perturbativeh}, for nuclei with a number of
valence nucleons larger than two, the \qbox~should contain diagrams
with at least three incoming and outcoming valence particles, as for
example the second-order three-body diagrams in
Fig. \ref{diagram3corr}.
In order to include the effects of these contributions, in Ref. \cite{Fukui18} the monopole component of
the diagrams in Fig. \ref{diagram3corr} has been calculated and added
to the theoretical g.s. energies.
The results of this procedure are reported in Fig. \ref{gsencsm} (b), where the new 
calculated SM g.s. energies (black
squares) are compared with both the experimental ones (red triangles)
and those obtained with NCSM (blue bullets) \cite{Fukui18}.
As it can be seen, the comparison between SM and NCSM has been
efficiently improved with respect to that of Fig. \ref{gsencsm} (a), the
largest discrepancy being about $4\%$ for $^8$Be.


So far, we have shown that the derivation of \heff~via a
perturbative expansion of the $\hat{Q}$-box vertex function provides SM
results that are in a satisfactory agreement with those of the {\it
  ab initio} method NCSM, when accounting for realistic $NN$ potential
only.
In Ref. \cite{Fukui18} a step forward has been made by including in
the derivation of \heff~ contributions from a chiral 3NF \cite{Navratil07a}.

It is worth recalling that in the chiral perturbative expansion the
3N potentials appear from N$^2$LO on, and  at this order the 3N
potential consists of three components (see Fig. \ref{n2lo3f}), which
are the 2PE, the
1PE, and the contact terms. The adopted intrinsic Hamiltonian is defined in Eq. \eqref{eq:hm1}.

As reported in Section \ref{sec-II.2}, a great advantage of ChPT is
that it generates nuclear two- and many-body forces on an equal footing
\cite{Weinberg92,vanKolck94,Machleidt11}, namely most interaction
vertices that appear in the 3NF also occur in the $NN$ potential. 
The parameters carried by these vertices are fixed in the construction
of the chiral 2NF, and for the N$^2$LO 3N potential they are the
LECs $c_1$, $c_3$, and $c_4$, appearing in
$v_{3N}^{(2\pi)}$.
However, the 3N 1PE term and the contact interaction are
characterized by two extra LECs (known as $c_D$ and $c_E$), which cannot be constrained by two-body observables,
and should be fitted  by reproducing  observables in systems with mass $A>2$.

The goal of the work of Ref. \cite{Fukui18} has been to benchmark SM
calculations, now including  also the contribution from a N$^2$LO 3N
potential, against those obtained with NCSM
\cite{Navratil07b,Maris13}, and consequently the adopted $c_D, c_E$
values are -1, -0.34, respectively, as those employed in Ref. \cite{Navratil07b} (see Fig. 1 in
 \cite{Navratil07b}).

As already mentioned before, \heff~is calculated introducing the
contribution of the N$^2$LO 3N potential at first-order in many-body
perturbation theory only for one- and two-valence nucleon systems.
The contribution at first order to the single-particle and two-body
components of the \qbox~from a three-body potential are shown in
Fig. \ref{1b2b3bf} and their expression is reported in
Eqs. \eqref{1b3bfeq} and \eqref{2b3bfeq}.
In Section \ref{perturbativeh} we have also pointed out that these
expressions  give  the
coefficients which multiply the one-body and two-body terms,
respectively, arising from the normal-ordering decomposition of the
three-body component of a many-body Hamiltonian
\cite{HjorthJensen17}.

\begin{figure}[h]
\begin{center}
\includegraphics[width=17cm]{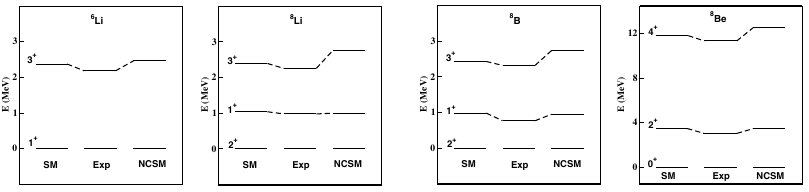}
\caption{Theoretical and experimental spectra for $^{6}$Li, $^{8}$Li,
  $^{8}$B, and $^{8}$Be. The theoretical energies have been obtained
  using the N$^3$LO $NN$ plus N$^2$LO 3N potentials within SM and
  NCSM calculations. Figure adapted from Ref. \cite{Fukui18}.}
\label{SM_NCSM_3b_1}
\end{center}
\end{figure}

\begin{figure}[h]
\begin{center}
\includegraphics[width=17cm]{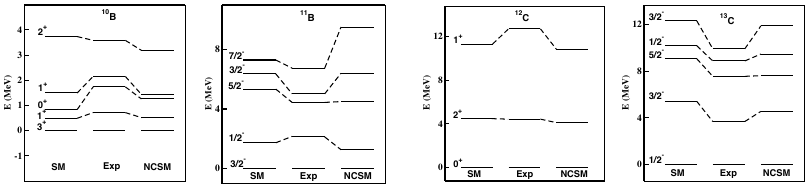}
\caption{Same as in Fig. \ref{SM_NCSM_3b_1}, but for $^{10}$B, $^{11}$B,
  $^{12}$C, and $^{13}$C. Figure adapted from Ref. \cite{Fukui18}.}
\label{SM_NCSM_3b_2}
\end{center}
\end{figure}

In Figs. \ref{SM_NCSM_3b_1} and \ref{SM_NCSM_3b_2}, we show the
low-energy spectra of $^6$Li, $^8$Li, $^8$B, $^8$Be, $^{10}$B,
$^{11}$B, $^{12}$C, and $^{13}$C, calculated within the SM framework, now
including also the contributions from the N$^2$LO 3N potential.
They are compared with the experimental ones \cite{ensdf} and the NCSM
results \cite{Navratil07b,Maris13}.

From the inspection of Figs. \ref{SM_NCSM_3b_1} and \ref{SM_NCSM_3b_2}, the results obtained with
\heff~derived by expanding perturbatively the \qbox~and the  NCSM ones are
in a close agreement, as in the case with only the $NN$ potential.
Moreover, the theory with the  3NF compares far better with experiment, as
can be seen in all the reported spectra.
In this regard, it is paramount, among other observations, to note
that the experimental sequence of observed states in $^{10}$B is
restored, and the degeneracies of
$J^{\pi}={1}/{2}^{-}_{1}, {3}/{2}^{-}_{1}$ and
$J^{\pi}={3}/{2}^-_2, {5}/{2}^-_1$ states in $^{11}$B are
removed.
This supports the crucial role played by the 3N potential to
improve the spectroscopic description of $p$-shell nuclei.

\subsection{\it Approaching the weakly bound systems \label{sec-IV.2}}
\subsubsection{\it The limit of oxygen isotopes \label{sec-IV.2.1}}
The oxygen isotopic chain exhibits, in addition to the conventional doubly magic $^{16}$O, the occurrence of two new shell closures in $^{22}$O \cite{Thirolf00} and $^{24}$O \cite{Kanungo09, Hoffman08,Hoffman09} with $N=14$, and 16, respectively. Of particular interest is that the dripline is located at $^{24}$O \cite{Thoennessen04} very close to stability line, the so-called ``Oxygen anaomaly''. The heaviest experimentally identified isotopes, $^{25}$O and 
$^{26}$O, are indeed unbound with respect to neutron emission \cite{Caesar13,Lunderberg12}. As a matter of fact, the doubly-closed nature of $^{24}$O was 
suggested, before to be experimentally confirmed,  by various SM calculations employing phenomenological interactions, which were also able to explain the occurrence of the neutron dripline at $N=16$ \cite{Utsuno99,Otsuka01b,Brown05a}.
In these papers, it was shown that the $N= 16$ shell gap arises from an upward shift of the $0d_{3/2}$ orbital, whose energy increases rapidly becoming close to zero while neutrons fill the $sd$ shell. As for the $0s_{1/2}$ orbital, these calculations indicate that it remains bound and the $0d_{5/2}-1s_{1/2}$   spacing  opens up from  $N=8$ and 14 making  $N=14$ a magic number. 
On the other hand, realistic SM calculations based on 2NF-only predict that the $0d_{3/2}$ orbital comes down in energy with increasing number of valence neutrons and remains well bound in $^{24}$O and beyond, putting the neutron dripline in an incorrect position.
\begin{figure}
    \begin{center}
    \includegraphics[width=0.8\textwidth]{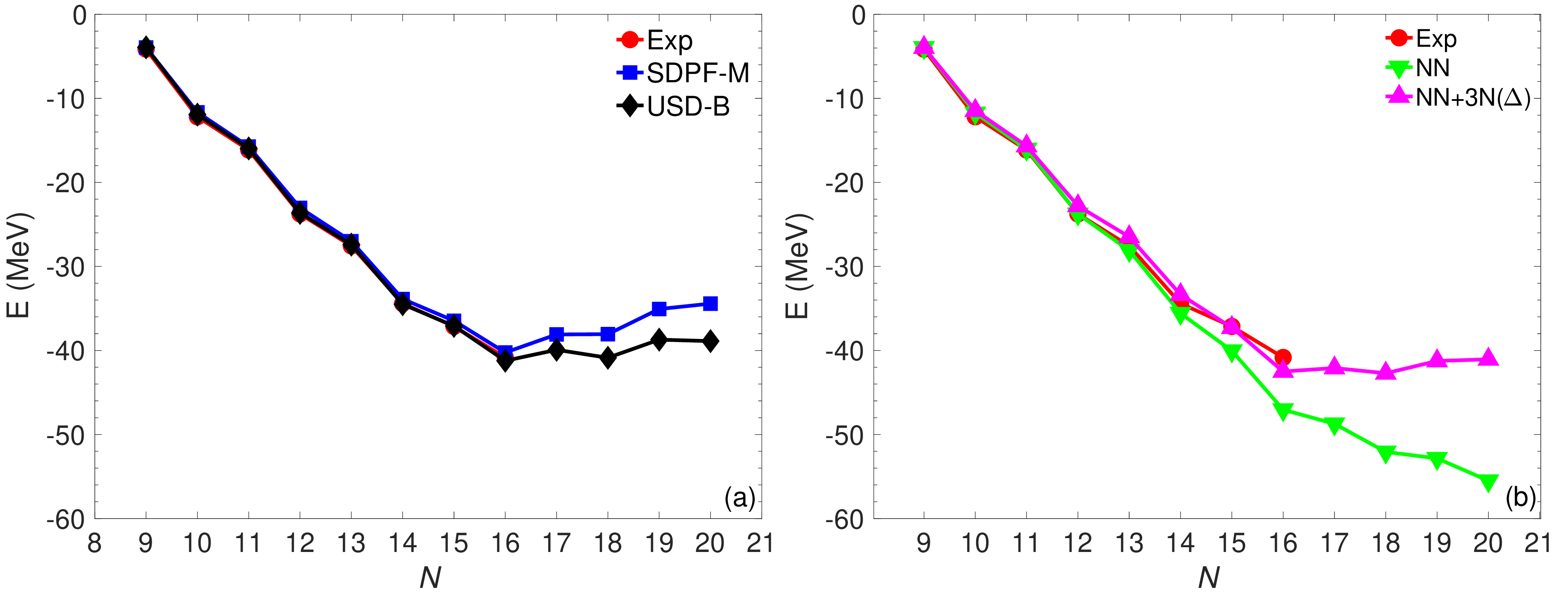}
    \caption{\label{Otsuka} Ground-state energies of oxygen isotopes measured from $^{16}$O, including experimental values of the bound $^{16-24}$O. Energies obtained from (a) phenomenological forces SDPF-M \cite{Utsuno99} and USD-B \cite{Brown06}, (b) a $G$ matrix and including Fujita-Miyazawa 3N forces due to $\Delta$ excitations. Figure adapted from Ref. \cite{Otsuka10}.}
    \end{center}
\end{figure}

The first study that has explicitly included the 3NF in the derivation of the effective SM Hamiltonian for oxygen isotopes  has been carried out by Otsuka and coworkers \cite{Otsuka10}. In the Introduction, we referred to this work by mentioning calculations using  a chiral $NN$ N$^{3}$LO potential  plus 3N contributions. However, Ref.  \cite{Otsuka10} also reports on the comparison with calculations employing  phenomenological and realistic $NN$ interactions, which more clearly highlights the effects of  3N forces on the limit of oxygen isotopes.

Specifically, g.s. energies of the oxygen isotopes  are also obtained by using  i) phenomenological SDPF-M \cite{Utsuno04} and USD-B \cite{USDB} forces; ii) $NN$ effective interactions derived by way of the MBPT at second order from a  $G$ matrix and including Fujita-Miyazawa 3N forces due to $\Delta$ excitations. 

The comparison between results of these calculations is illustrated in Fig. \ref{Otsuka}.
We see that the energies from phenomenological interactions well agree with experiment and the minimum value correctly locates at  $N=16$. By contrast, the energies based on the $NN$ force-only do not stop to decrease putting the dripline at $N=20$, regardless of the renormalization procedure employed for the $NN$ potential (the same results are obtained by using the $V_{\rm low-k}$ instead of the G matrix).
Then, the effects introduced by including the Fujita-Miyazawa 3N contributions, which become more relevant with increasing neutron number, correct the behavior of the binding energies bringing a significant raise from $N=16$ to 18.
 
Results of Ref. \cite{Otsuka10}, have been confirmed by more recent calculations based on large many-body spaces and with improved MBPT and nonperturbative valence-space Hamiltonians. A summary of these calculations performed  by using different {\it ab initio} approaches, including references, is  in Ref. \cite{Hebeler15}, where it is shown that  they all predict the correct dripline position at $^{24}$O when  large many-body  spaces are adopted with binding energy differing only  within a few percentage. 

As concerns oxygen isotopes beyond the dripline, the description of their binding energies and excitation spectra  requires to consider the coupling with the continuum in addition to  the 3NF contribution. This issue is presented within the GSM framework in Section \ref{sec:O_dripline}, where also unbound resonance states in $^{24}$O are discussed. The role played by the coupling with the continuum and the 3NF contribution is also evidenced for the proton-rich Borromean $^{17}$Ne  in Section \ref{sec:O_17Ne}.

\subsubsection{\it 3NF and continuum in neutron-rich oxygen isotopes} \label{sec:O_dripline}
Neutron-rich oxygen isotopes have been attracting many interests from both experiment and theory, not only because of the famous ``Oxygen Anomaly''~\cite{Otsuka10}, discussed above, but also their many peculiar phenomena. The $^{26}$O ground state has been found barely unbound with two-neutron separation energy of only $-18$ keV \cite{Kondo16}, and an excited state in the unbound resonant isotope $^{25}$O has been observed in a recent experiment \cite{Jones2017}. Even though many theoretical works \cite{Otsuka10, Hergert13b, Bogner14, Jansen14} have shown the importance of 3NFs in nuclear structure calculations to reproduce the oxygen dripline position, there still exists a strong demand to consider the coupling to the continuum, which is crucial to the understanding of the loose structures of exotic nuclei. Using the GSM with chiral $NN$ and 3N forces, we have investigated oxygen isotopes up to beyond the neutron dripline, in which both effects from the 3NF and continuum are considered.

For neutron-rich oxygen isotopes, we choose the doubly magic system $^{16}$O as  core, with its ground-state Slater determinant as reference state for the normal-ordering decomposition of 3NF. The Berggren basis is generated by the WS potential. We adopt the universal WS parameters \cite{Dudek1981}, but reduce the depth parameter by 2.3 MeV to obtain a reasonable $0d_{3/2}$ resonance width compared with the experimental width extracted in $^{17}$O. This WS potential gives bound $0d_{5/2}$ and $1s_{1/2}$ orbitals and a resonant $0d_{3/2}$ orbital. In this Berggren basis produced by the WS potential, we construct the effective GSM interaction for the model space \{$0d_{5/2}$, $1s_{1/2}$, $0d_{3/2}$ pole plus continuum\}, using the MBPT of $\hat{Q}$-box folded diagrams, named EKK~\cite{Takayanagi11}. The detail about the complex-$k$ MBPT can be found in our previous papers~~\cite{Sun17, Li2020}. The continuum effect enters the model through both the complex effective interaction and the active model space which includes continuum partial waves. As discussed in the previous Section \ref{sec-III.3}, the complex symmetric GSM Hamiltonian is diagonalized in the complex model space via the Jacobi-Davidson method.

\begin{figure}[!ht]
    \begin{center}
    \includegraphics[width=0.5\textwidth]{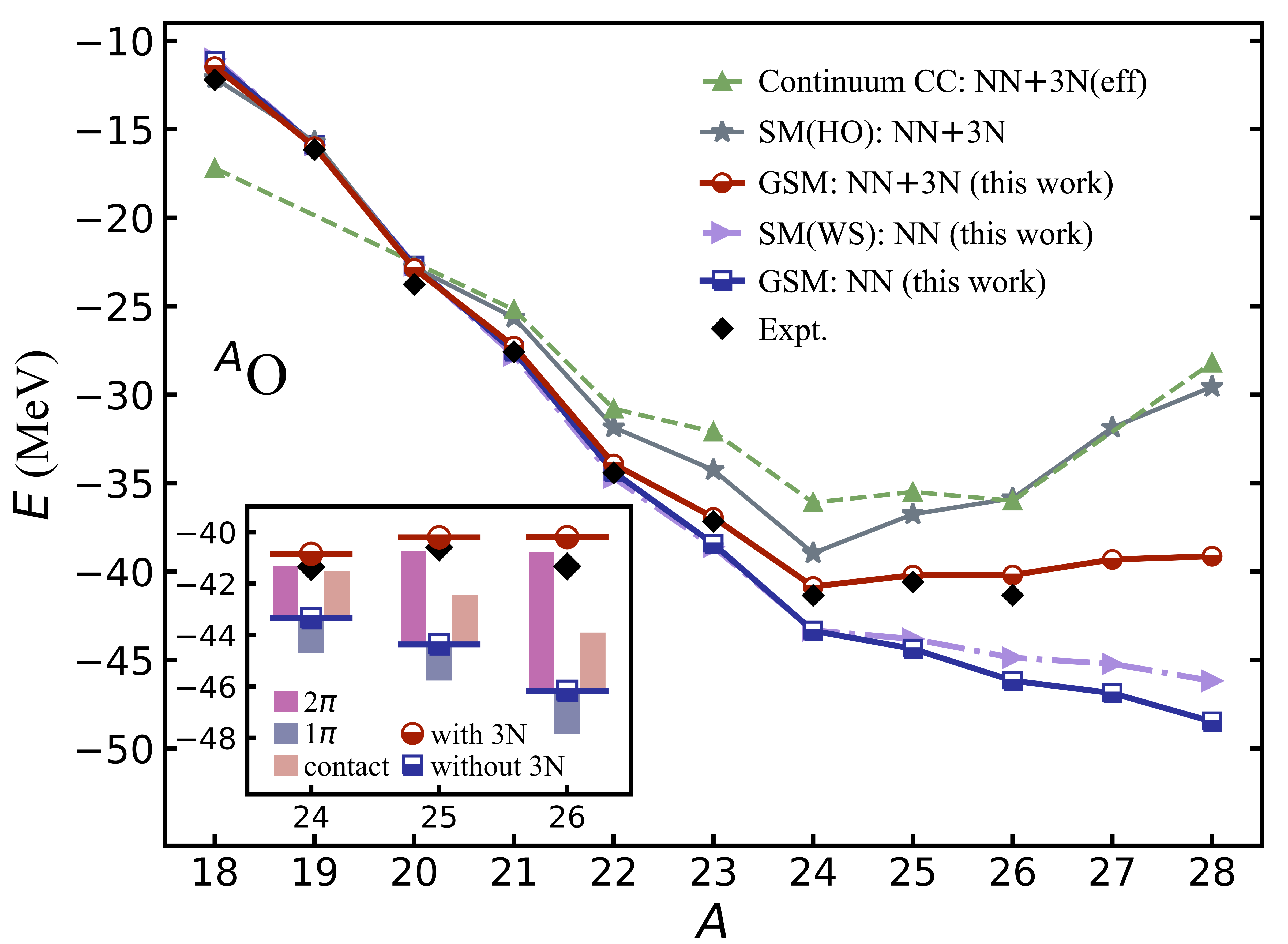}
    \caption{\label{fig:Oxygen_BE}Calculated $^{18-28}$O ground-state energies with respect to the $^{16}$O core \cite{Ma20a}, compared with experimental data and other calculations: 
    conventional SM (HO) with 3NF but without continuum \cite{Otsuka10} and continuum CC with a density-dependent effective 3NF \cite{Hagen2012}. 
    The inset shows the 3NF contributions from 2PE ($2\pi$), 1PE ($1\pi$) and contact terms.}
    \end{center}
\end{figure}

We have calculated the binding energies of oxygen isotopes, shown in Fig. \ref{fig:Oxygen_BE}. The comparison of different calculations given in the figure shows that both 3NF and continuum play important roles in reproducing the experimental binding energies, especially in the vicinity of the dripline. The 3NF gives repulsive contributions to g.s. energies and the effect increases with the increase of the number of valence neutrons. In the inset of Fig. \ref{fig:Oxygen_BE}, the 3NF effect has been dissected in $^{24\text{-}26}$O. We find that the attractive $v^{(1\pi)}_{\text{3N}}$ and repulsive $v^{(\text{ct})}_{\text{3N}}$ terms have similar values but opposite signs. Consequently, their effects are almost canceled out, implying that $v^{(2 \pi)}_{\text{3N}}$ term is responsible for the observed 3NF repulsive effect in this mass region.

\begin{figure}[!ht] 
    \begin{center}
    \includegraphics[width=0.48\textwidth]{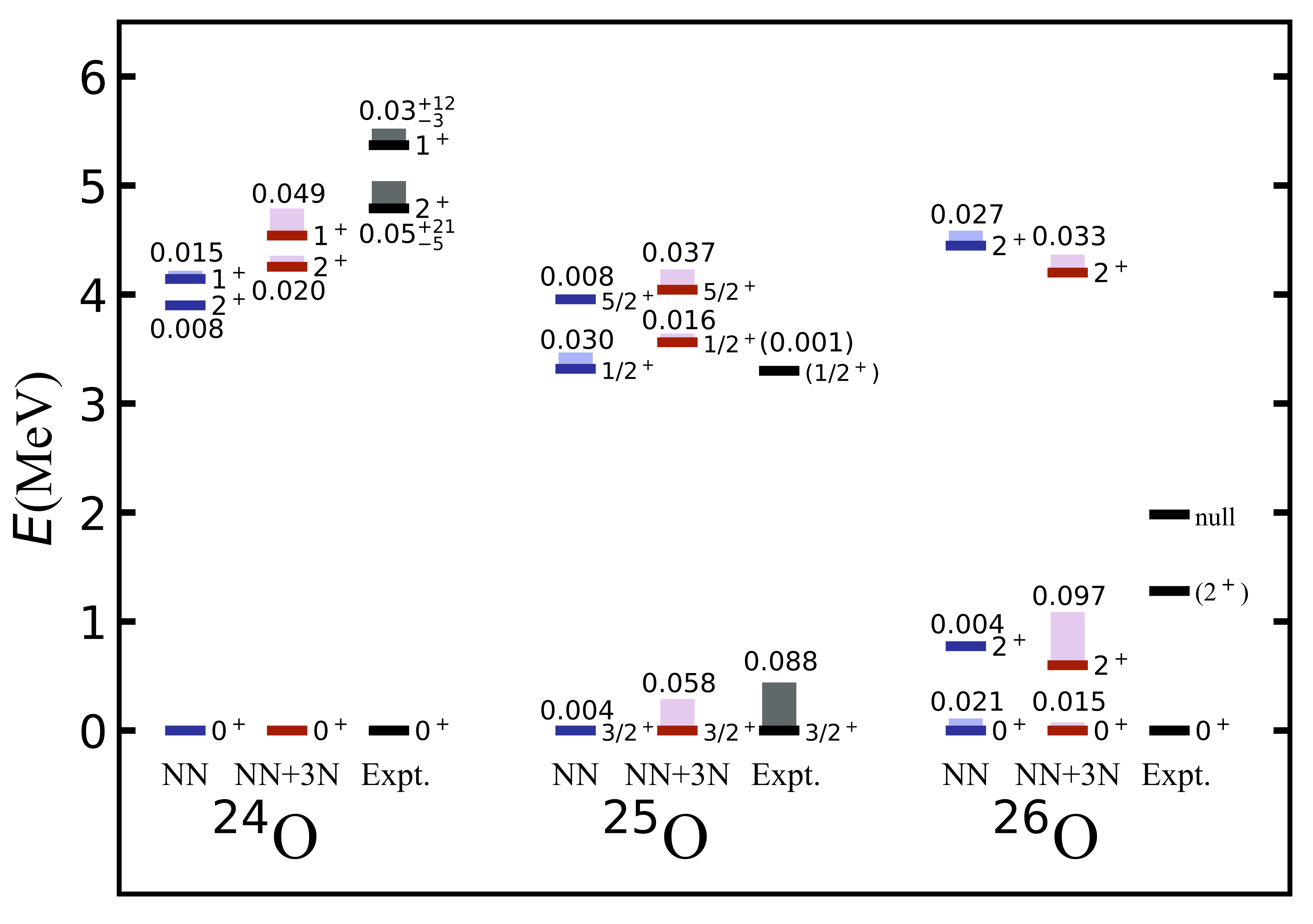}
    \caption{\label{fig:O242526} GSM calculations of excitation spectra for $^{24-26}$O \cite{Ma20a}. The experimental data are taken from \cite{Hoffman09, Jones2017, Lunderberg12, Kondo16}.}
    \end{center}
\end{figure}
Fig. \ref{fig:O242526} displays the calculated excitation spectra of $^{24-26}$O. We see that 3NF also improves spectrum calculations compared with experimental data, especially for unbound resonance states, such as the excited $2^+_1$ and $1^+_1$ in $^{24}$O and the ground state of $^{25}$O. The calculations with 3NF included give better agreements with data in resonance widths. 

\subsubsection{\it 3NF and continuum in proton-rich Borromean $^{17}$Ne}
\label{sec:O_17Ne}
Due to the Coulomb repulsion, the proton dripline is not so far from the stability line compared with the neutron dripline, which has been reached experimentally up to $Z \approx 90$ more than a decade ago~\cite{Thoennessen04}. However, when focusing on the light proton-rich region where the Coulomb barrier is not so high and continuum effect is prominent, interesting phenomena (such as halo and Borromean structures) may emerge. Among these nuclei, the Borromean $^{17}$Ne is of particular interest in both experiment~\cite{Tanaka10b,Sharov17,Charity18} and theory~\cite{Garrido04, Casal16, Parfenova18}. It may bear a similarity to the two-neutron halo nucleus $^6$He which can be described as a three-body $^4$He + 2n system. To see the weakly-bound characteristic in $^{17}$Ne, we performed a realistic GSM calculation with the chiral N$^3$LO 2NF and N$^2$LO 3NF which were discussed already in Section \ref{sec:O_dripline}. 

In this calculation, we choose the closed-shell $^{14}$O as the core. The GSM valence space contains the neutron bound states $\nu\{0p_{1/2}, 0d_{5/2}, 1s_{1/2}, 0d_{3/2}\}$ and the proton resonances $\pi\{1s_{1/2}, 0d_{5/2}\}$ plus continua $\pi\{s_{1/2}, d_{5/2}\}$. To reproduce the inverse positions of proton 1s$_{1/2}$ and 0d$_{5/2}$ orbitals in $^{15}$F, the WS parameters need to be adjusted~\cite{Ma20b}. In detail, we reduce the spin-orbital coupling strength to 2 MeV, and reduce the depth parameter by 2 MeV for the proton WS potential. 
The valence-space Hamiltonian was obtained using MBPT as described in Sec.~\ref{sec:O_dripline}, and diagonalized by the Jacobi-Davidson method. More details can be found in Ref.~\cite{Ma20b}. 

\begin{figure}[!ht]
    \begin{center}
    \includegraphics[width=0.45\textwidth]{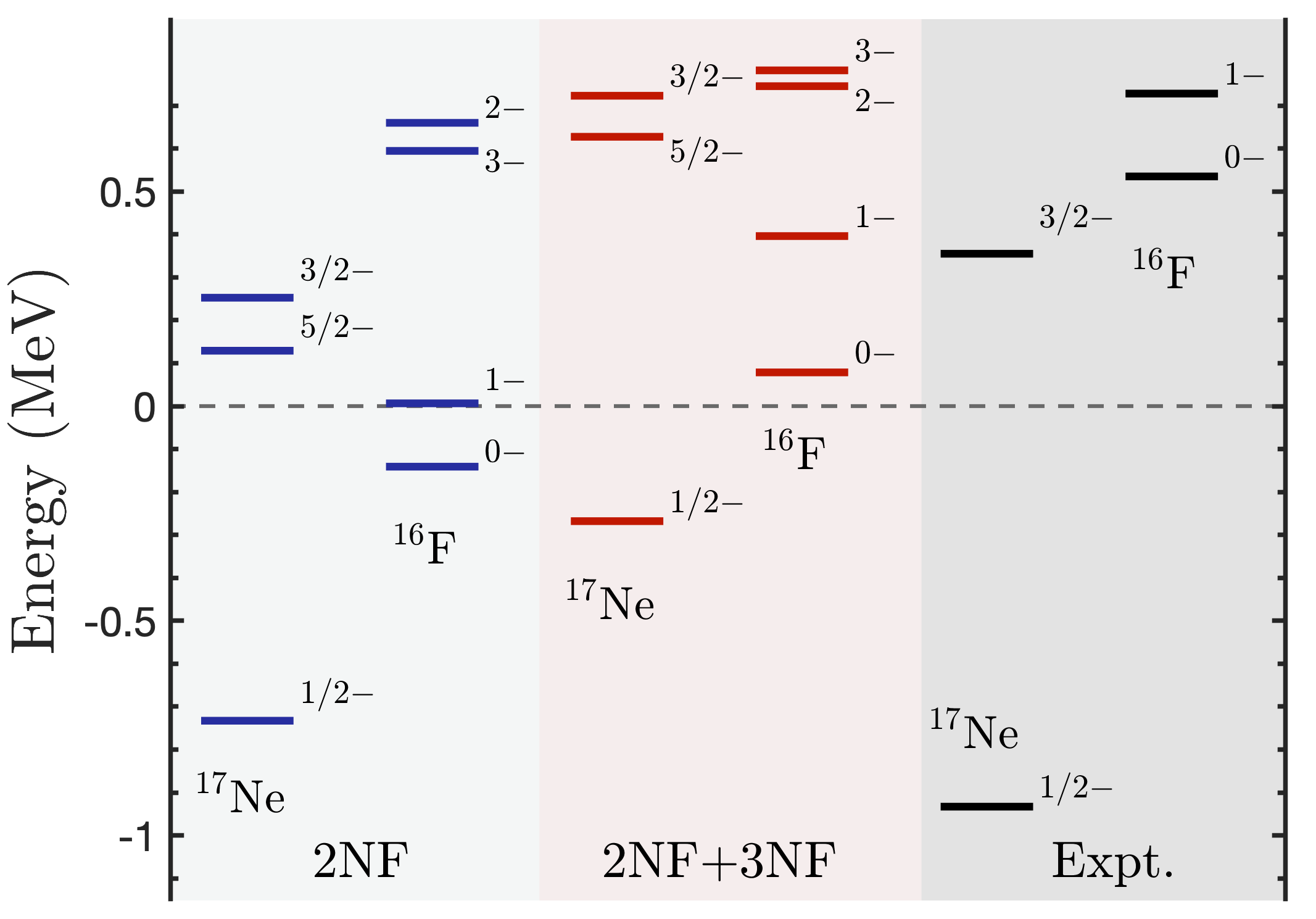}
    \caption{\label{fig:Ne17_spec}Calculated excitation spectra of $^{17}$Ne along with its isotone $^{16}$F, with respect to the $^{15}$O ground-state energy~\cite{Ma20b} . Blue and red lines are the GSM calculations with 2NF only and 2NF+3NF, respectively. Experimental data are taken from Refs.~\cite{Audi12, Sharov17}.}
    \end{center}
\end{figure}

The calculated low-lying levels of $^{17}$Ne and its subsystem $^{16}$F are present in Fig. \ref{fig:Ne17_spec}, with respect to the $^{15}$O ground-state energy. It is seen that the 3NF lifts the whole spectra of $^{17}$Ne and $^{16}$F, which makes $^{16}$F unbound and leads to a Borromean structure of $^{17}$Ne. In the g.s. of $^{17}$Ne, we have found strong configuration mixing with a 54\% s-wave component of $\pi 1s_{1/2}^2\otimes\nu 0p_{1/2}^1$, which is consistent with that in Ref.~\cite{Parfenova18}.

\begin{figure}[!ht]
    \begin{center}
    \includegraphics[width=0.45\textwidth]{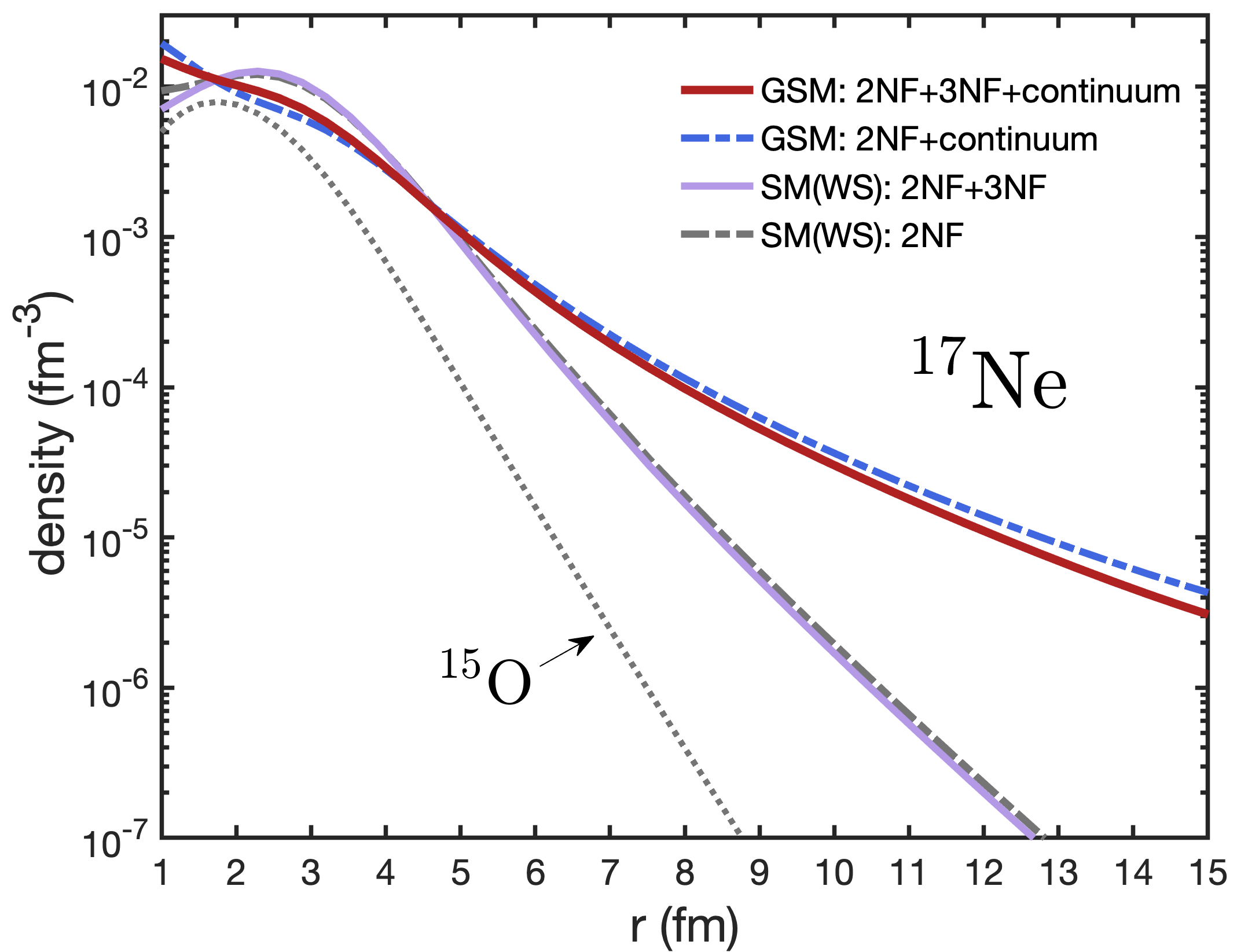}
    \caption{\label{fig:Ne17_density} The $^{17}$Ne density in the valence space, calculated by GSM with 2NF only (blue dot-dashed line) and 2NF+3NF (red line). SM (WS) stands for the conventional SM calculations performed in the non-continuum discrete WS basis, with 2NF+3NF (purple line) and 2NF only (grey dot-dashed line). The $^{15}$O density (black dash line) in the valence space is also displayed for comparison.}
    \end{center}
\end{figure}

The small two-proton separation energy and large weight of $s$-wave component suggest a halo structure in the ground state of $^{17}$Ne. We have calculated the one-body density by GSM, shown in Fig. \ref{fig:Ne17_density}. Compared with $^{15}$O, there exist a long tail in the density of $^{17}$Ne, which is a direct evidence to support the halo nature of $^{17}$Ne. From Fig. \ref{fig:Ne17_density}, we see that 3NF effect on the density is small, while the effect from the continuum coupling is significant. More detailed discussions can be found in Ref.~\cite{Ma20b}.

\subsubsection{\it The calcium isotopes dripline \label{sec-IV.2.2}}
The Ca isotopic chain, from the experimental point of view, presents characteristics similar to the
oxygen one. In analogy to the shell gaps found in correspondence of $N=14$ and 16 for oxygen isotopes, it exhibits two neutron shell closures at $N = 32$ \cite{Huck85,Gade06,Wienholtz13} and 34 \cite{Steppenbeck13} in  addition to the standard ones at $N = 20$ and $28$. However, while oxygen isotopes have been studied experimentally even beyond the neutron dripline \cite{Caesar13,Kondo16}, not even its position is known in Ca isotopic chain. Indeed,
experimental studies have reached only $^{60}$Ca \cite{Tarasov18} with  $N/Z = 2$ and 12 neutrons more than the last stable isotope.

From the theoretical side, calculations for Ca isotopes have been performed by both mean field and microscopic approaches -- as  relativistic Hartree-Bogoliubov  or density functional theories (DFT) in the first case and  SM or {\it ab initio} methods in the second -- but their results provide ambiguous indications about the behavior of the g.s. energies as well as  about the location of the neutron dripline.
As matter of fact, some studies predict the neutron dripline around $^{60}$Ca \cite{Meng02,Hagen12,Holt12,Hergert14} while others pushes it up $^{70}$Ca \cite{Fayans00,Bhattacharya05,Chen15,Neufcourt19,Cao19,Li20,Coraggio20e}. However, the available experimental data, including the recent evidence of a bound  $^{60}$Ca \cite{Tarasov18} and the  mass measurements of heavy odd calcium isotopes \cite{Wienholtz13, Michimasa18}, may be very  helpful to narrow the spread of the theoretical predictions.

\begin{figure}[!h]
    \begin{center}
    \includegraphics[width=0.6\textwidth]{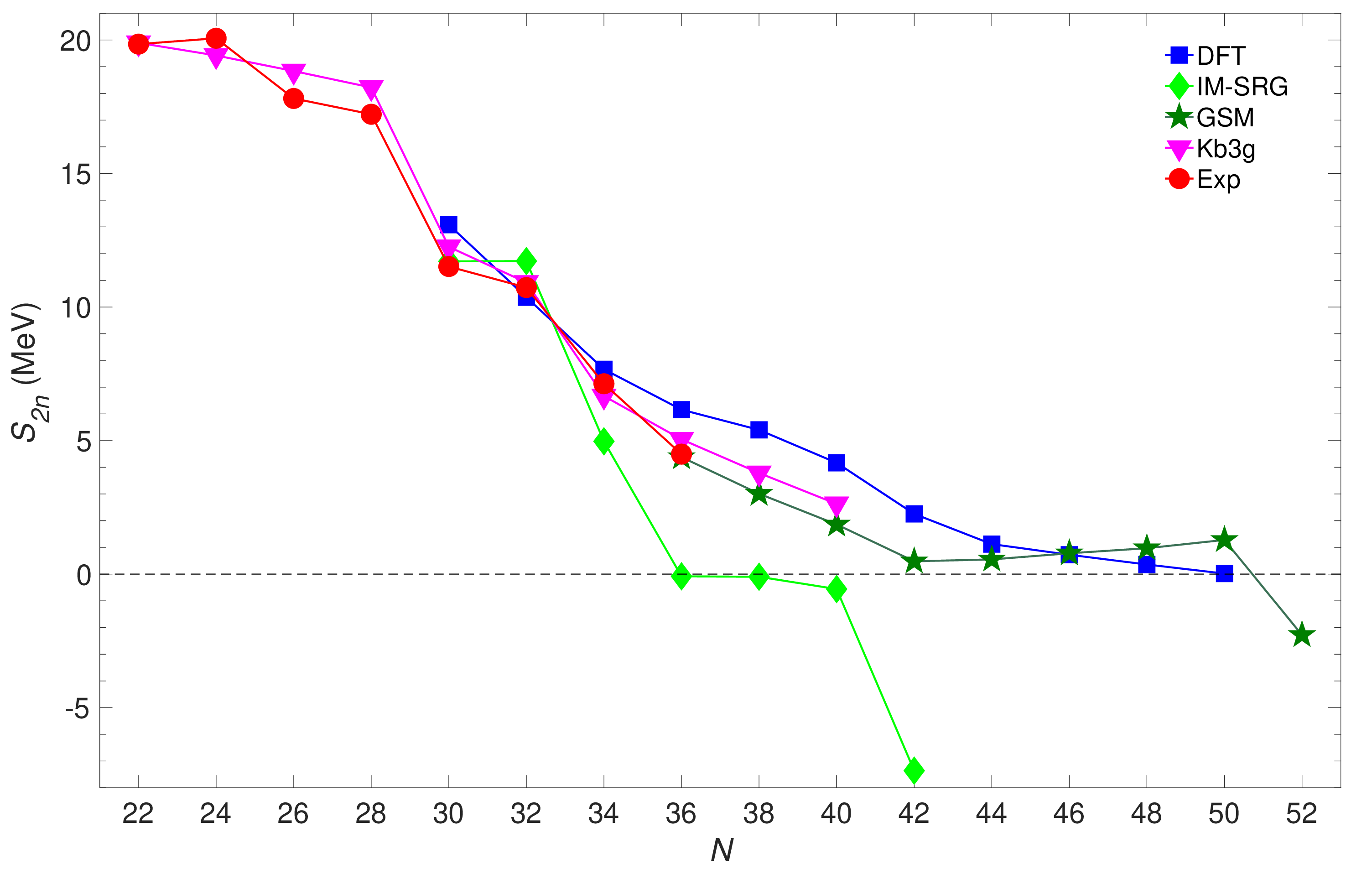}
    \caption{\label{Sec-422_Ca} Experimental \cite{AME21} two-neutron separation energies  as a function of the neutron number $N$ for Ca isotopes compared with the results of a variety on many-body methods. See text for details.}
    \end{center}
\end{figure}

As an illustrative example, we compare in Fig. \ref{Sec-422_Ca} the experimental  two-neutron separation energies  ($S_{2n}$s)   for even Ca isotopes with results from different calculations. We see that DFT using the Skyrme interaction of Ref. \cite{Klupfel09} predicts that $^{60}$Ca is well bound  and the neutron dripline is around  $^{70}$Ca,  as it comes out from other mean field calculations \cite{Fayans00,Bhattacharya05,Chen15, Neufcourt19,Cao19}. It is worth mentioning, however, that the results depend on the symmetry energy, as clearly underlined in Ref. \cite{Chen15}. The last bound Ca isotope was located well beyond $N=40$ also  by GSM   calculations which start from  the CD-Bonn $NN$  potential \cite{Li20}.
On the other hand, the position of the dripline predicted by the IM-SRG approach \cite{Hergert14} is around $^{60}$Ca, as it is confirmed by other microscopic approaches (see Refs. \cite{Hagen12,Holt12}). However, Holt and coworkers \cite{Holt14}, who  updated the SM results of Ref. \cite{Holt12}, have  underlined   the difficulty of a  precise prediction for the dripline  due to the flat evolution of the  g.s. energies beyond $^{60}$Ca. 
Figure \ref{Sec-422_Ca} also reports $S_{2n}$s from SM calculations in the $fp$ valence space based on the KB3G  phenomenological interaction \cite{Poves01}, and we see that the theoretical values nicely overlap the available experimental data and locate between the DFT and GSM curves.

The study of the evolution of the nuclear masses in isotopic chains has pushed forward   our understanding of the nuclear forces, drawing attentions on  new aspects of nuclear forces that develop going from the valley of stability to the limits of existence. In this connection, particular attention is focused on 3NFs, which have shown to be critical in calculations of extreme neutron-rich systems.
The effect of 3NFs on the dripline and on the shell evolution in Ca region has been the subject of investigations only in the last decade or so. The g.s. energies along the Ca isotopic chain have been studied starting from chiral $NN$ and 3N potentials by way of the SM \cite{Holt12,Holt14,Ma19,Coraggio20e},  CC model \cite{Hagen12}, IM-SRG \cite{Hergert14} and  self-consistent Green function theory \cite{Soma20} approaches.  In particular, SM calculations of Ref. \cite{Holt12} - performed within the $fp$ space -   have first evidenced that the inclusion of the three-body component in the derivation of  the effective Hamiltonian leads to a repulsive contribution needed to correct the overbinding obtained when considering the $NN$ force only, and provides a better agreement with the available experimental data, as observed for oxygen nuclei \cite{Otsuka10}.  In Ref. \cite{Holt12}, it was also shown that predictions with $NN$+3N forces are  quite close to those resulting from the phenomenological GXPF1\cite{Honma04} and KB3G \cite{Poves01} $NN$ interactions, which show similar monopole components \cite{Honma04}  although developed  by employing different techniques.   This supports the relevant role of 3N forces in removing deficiencies of the two-body monopole component  arising from $NN$-only theory. A careful analysis of the contribution of chiral 3NFs to the monopole component of the effective SM Hamiltonian is reported in Section \ref{sec-IV.3.2}. 

Here, we shall discuss predictions of Ca dripline based on the realistic SM calculations of Refs. \cite{Ma19,Coraggio20e} performed  in the $0f_{7/2}$,  $0f_{5/2}$, $1p_{3/2}$, $1p_{1/2}$ ($fp$) neutron  space as well as in  the extended  space including the neutron $0g_{9/2}$ orbital ($fpg_{9/2}$). In both cases, a chiral $NN$+3N potential is chosen as starting point to construct the effective SM Hamiltonians, but for the $fp$  space we also report results obtained with the $NN$ force only. The $NN$ and 3N  forces  were derived within the  ChPT framework \cite{Entem02} stopping the perturbative series at  N$^{3}$LO and at N$^{2}$LO, respectively (see Sections \ref{sec-II.2} and \ref{sec-IV.1}). We would like to reiterate  here that  the $NN$ and 3N forces  consistently share the same nonlocal regulator function  and some LECs which are determined by the renormalization procedure described in Ref. \cite{Machleidt11}, while the values of the additional LECs appearing in the 3N force, $c_D$ and $c_E$, are taken from Ref. \cite{Navratil07b}.

\begin{figure}[!h]
    \begin{center}    \includegraphics[width=0.7\textwidth]{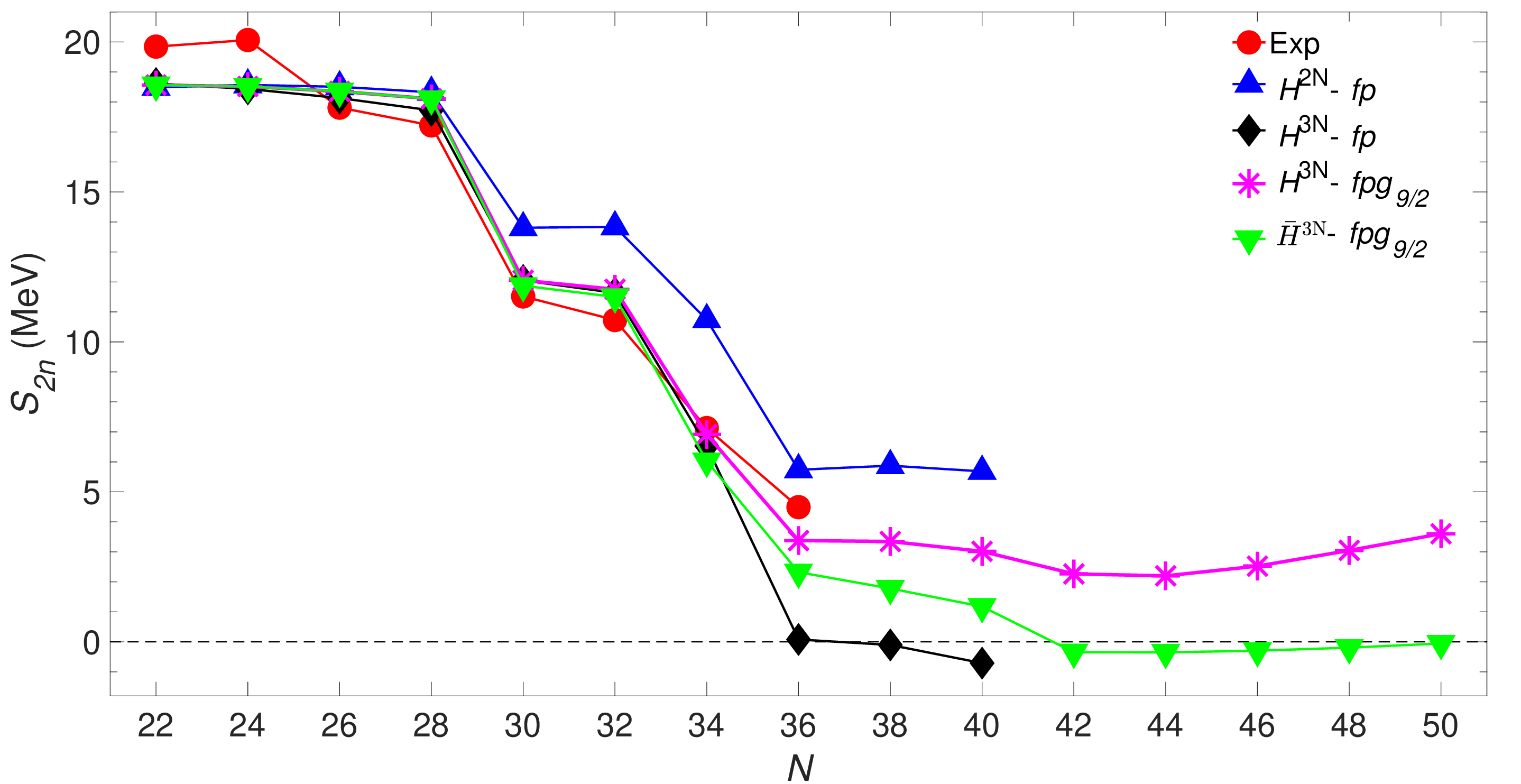}
    \caption{\label{Sec-422_2N_3N} Experimental \cite{AME21} two-neutron separation energies  as a function of the neutron number $N$ for Ca isotopes compared with SM results obtained  within the $fp$  and $fpg_{9/2}$ spaces. Calculations are based on the $NN$ and $NN$+3N forces. See text for details.}
    \end{center}
\end{figure}

The effective Hamiltonians were derived within the framework of the MBPT outlined in Section \ref{perturbativeh} by arresting the $\hat Q$-box expansion of the one- and two-body Goldstone diagrams at third order in the $NN$ potential and at first order in the 3N one, the latter diagrams corresponding to the normal-ordered one- and two-body parts of the 3N force.
Moreover, calculations were carried out, as described in Section \ref{perturbativeh}, by employing 
density-dependent $H_{\rm eff}$s, whose TBMEs change according to the number of valence
nucleons and take into account the interactions via two-body force  of clusters of three-valence  nucleons 
with configurations outside the model space. This means that,  in addition to genuine 3N forces, we consider also induced 3N  contributions,  that come into play for systems  with more than two-valence nucleons.  It is worth mentioning that similar calculations, in both the $fp$ and $fpg_{9/2}$ valence spaces, were performed in \cite{Holt12,Holt14}, where, however, the $NN$ chiral potential was renormalized through the $V_{\rm low-k}$ technique \cite{Bogner02} and the effects of induced three-body contributions in the derivation of $H_{\rm eff}$  were neglected.

 The experimental  $S_{2n}$s  are compared with calculated values in Fig. \ref{Sec-422_2N_3N}. Results within the $fp$ space obtained by using the $NN$ and $NN$+3N force are dubbed, respectively, as   $H^{\rm 2N}-fp$ and  $H^{\rm 3N}-fp$, while $H^{\rm 3N}-fpg_{9/2}$  indicates results  in  the $fp g_{9/2}$ space with the $NN$+3N force. 
By comparing the $H^{\rm 2N}-fp$ and  $H^{\rm 3N}-fp$ results,   we see that both calculations reproduce the rather flat  experimental $S_{2n}$ behavior up to $N=28$. Then,  the measured values are overestimated when  using the $NN$-only force, while, in line with the results of Refs. \cite{Holt12,Holt14}, the repulsion due to the 3NF leads to less bound g.s. energies and improves the agreement with experiment.  However, at $N=36$  a too sudden drop is found by $H^{\rm 3N}-fp$, at variance with the experimental finding, which may be ascribed to  the  missing contributions arising from  the $0g_{9/2}$ orbital.  We find, indeed, that a larger $S_{2n}$, quite close to the experimental value,  is predicted by $H^{\rm 3N}-fpg_{9/2}$ calculations at $N=36$. 
As a matter of fact, when including  both the neutron $0g_{9/2}$ orbital and the 3N force  we are able to well describe the available experimental data, and in particular to predict $^{60}$Ca as a bound system, consistently with the
recent experiment of Ref. \cite{Tarasov18}. We have also found that calcium isotopic chain 
is bound at least up  to $^{70}$Ca in line with the results of   the DFT \cite{Klupfel09} and GSM \cite{Li20} calculations mentioned above, and the recent Bayesian analysis of different DFT calculations \cite{Neufcourt19}.

In Fig. \ref{Sec-422_2N_3N}, labelled by $\bar{H}^{\rm 3N} - fpg_{9/2}$, we have also reported the $S_{2n}$s obtained by considering only a genuine chiral 3N force without accounting for induced  3N contributions.  In this case, the same Hamiltonian derived for the two-body system is adopted for all Ca isotopes.

We see that results of the two Hamiltonians, $H^{\rm 3N} - fpg_{9/2}$ and $\bar{H}^{\rm 3N} - fpg_{9/2}$, almost overlap up to $N=32$, but then  differences between them start to grow and become larger and larger with increasing number of valence neutrons.
Actually, from $N=34$ on the inclusion of induced 3NFs brings an upshift of the two-neutron separation energies evidencing their attractive contribution which counterbalances in part the one arising from a genuine 3NF. 

We may therefore conclude that the effects of these two 3NFs with very different origin are both equally important in determining the dripline of Ca isotopes. It is worth mentioning that similar results are obtained for Ti isotopes as shown in Ref.  \cite{Coraggio21}.

\subsection{{\it Shell evolution and the role of three-body forces} \label{sec-IV.3}}
\subsubsection{{\it Overview: the $fp$ shell region} \label{sec-IV.3.1}}
The access to experimental information for nuclear systems with a large unbalanced  number of neutrons and protons, the so-called exotic nuclei, has  opened  up new possibilities to  advance our understanding 
of nuclear physics. 
One of the key questions that modern research has allowed to be address concerns the robustness of the standard magic numbers, namely the evolution of the shell structure as a function of $N/Z$, in particular when moving far from the stability line and approaching the driplines. 

Significant theoretical and experimental efforts have been devoted in the last four decades to this issue by  investigating nuclear structure properties  along isotopic and isotonic chains.  Experiments  have been performed  with radioactive beams to  identify the disappearance of conventional magic numbers  or  the appearance of new ones,   and at the same time  a number of theoretical papers have  been published aimed at understanding the underlying mechanisms determining such behaviour and the specific role of the various components of the nuclear interaction (see for instance \cite{Sorlin08,Otsuka20}).
 \begin{figure} 
\begin{center}
\includegraphics[width=15.5cm]{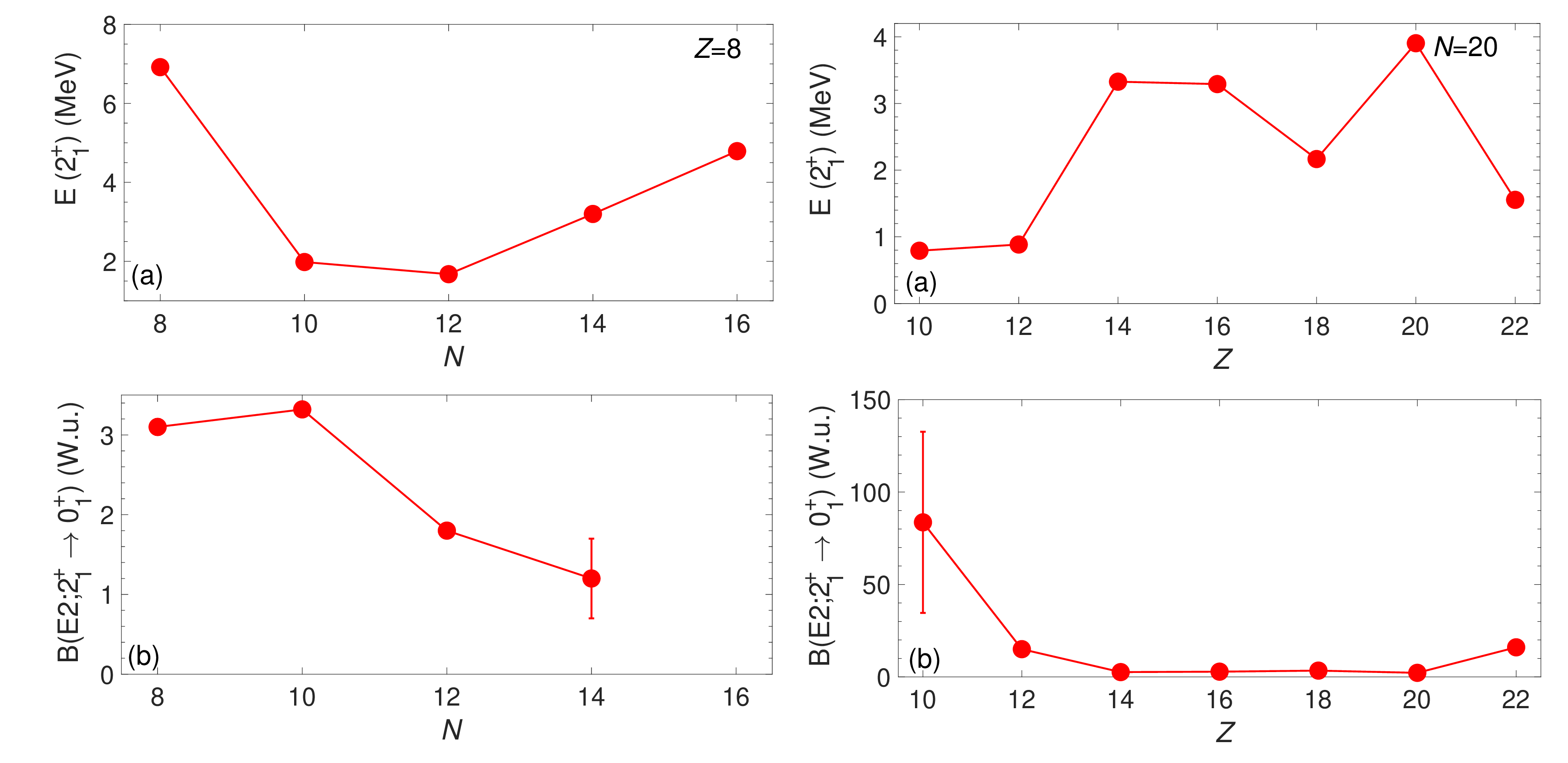}  
\caption{ Experimental (a) $2^{+}_{1}$ excitation energies  and (b) $B(E2, 2^{+}_{1} \rightarrow 0^{+}_{1})$  transition rates  in the  $Z = 8$ isotopes (left) and  $N = 20$ isotones (right).}
\label{sec-IV3_Fig1}
\end{center}
\end{figure}
Noteworthy examples of  disappearance or weakening of canonical magic numbers has been observed in light nuclei at $N = 8$, $N = 20,$ and $N = 28$ for $^{12}$Be, $^{32}$Mg and $^{42}$Si, respectively \cite{Motobayashi95,Navin00,Iwasaki00,Shimoura03,Bastin07,Takeuchi12},  while the onset of new shell closures at $N = 14$, 16 and 
$N = 32$, 34 has been evidenced, respectively,  in neutron-rich  oxygen \cite{Hoffman09,Kanungo09} and calcium isotopes \cite{Huck85,Janssens02,Dinca05,Burger05,Gade06,Wienholtz13,Steppenbeck13,Liu19}. As an illustration, we report in Fig. \ref{sec-IV3_Fig1}
the behaviour of the experimental  excitation energy of the first $2^+$ state and  its $E2$ transition rate to the g.s. in the even  $Z=8$ isotopes and $N=20$ isotones  as a function of increasing  $N$  and $Z$, respectively \cite{ensdf}.

We see that the $2^+$ energy  of  oxygen isotopes in the left-hand side of Fig. \ref {sec-IV3_Fig1} drops by a factor of about three  from  $N = 8$  to 10 before rising up  at $N = 14$, with a more dramatic increase at $N=16$. The behavior of the $2^+$  energy may be seen as a manifestation of the doubly-magic nature of $^{24}$O$_{16}$, corresponding to the complete filling of the  neutron $0d_{5/2}$ and  $1s_{1/2}$ orbitals, and to the appearance of a significant  $0d_{3/2} - 1s_{1/2} $ subshell gap. The increase in energy from $N=12$  to $N=14$ together with a decrease in the  $B(E2, 2^{+}_{1} \rightarrow 0^{+}_{1})$ value testifies the existence of a consistent gap also between the  neutron $0d_{5/2}$ and  $1s_{1/2}$ orbitals.

The evolution of the $N=20$ shell gap is  illustrated on  the right-hand side of Fig. \ref{sec-IV3_Fig1}.  The persistence of the shell closure at  $N = 20$ and the existence of  significant   $\pi 1s_{1/2} - \pi 0d_{5/2} $  and $\pi 0d_{3/2} - \pi 1s_{1/2}$ gaps account for the behavior  of the $2^{+}_1$ energy  and  $B(E2, 2^{+}_{1} \rightarrow 0^{+}_{1})$  transition rate in the $N=20$ isotones  from $^{34}$Si to $^{40}$Ca. However, no SM calculations limited to the $sd$ space can account for the sudden  and strong change from $^{32}$Mg  to $^{34}$Si, where the $2^+$ increases from  $\sim 0.9$ to 3.3 MeV   and the  $B(E2)$ value decreases  by a factor of about 4. This indicated a  collective nature for  $^{32}$Mg, which  can be explained only by taking in  account the correlation energy due to  $2p-2h$ neutron excitations \cite{Caurier98}. These excitations are favoured over  the normal $sd$ configurations by the lowering of the gap between the $sd$ and $fp$ shells, that is produced as soon as protons are removed from the $0d_{5/2}$ orbital. 

 These changes in the shell structure, called shell evolution,   and the  interplay between spherical configurations and deformation  are strictly connected to the monopole component of the interaction (Eq. (\ref{MonI})). In fact, as mentioned in  Section \ref{sec-III.1}, this component  governs the behaviour of the ESPEs  (see Eq. (\ref{ESPE})) by accounting for the variations in the SP energies arising from the residual interaction between the valence  nucleons.
 In this connection,  a particularly appealing  subject is the role of the different components of the nuclear force  in determining the monopole interaction, that  was first raised in 2001  by T. Otsuka and collaborators \cite{Otsuka01b}.  More specifically,  attention has been focused in literature on the evaluation of the contributions originating from the central, vector, tensor components as obtained from the spin-tensor decomposition of the SM
 Hamiltonian \cite{Osnes92}. In Refs. \cite{Otsuka05,Smirnova12}, it has been shown that the  splitting of the spin-orbit partners is essentially due to the tensor component, although any part of the Hamiltonian can give a relevant contribution to the shell evolution.

Another  important question that has been   recently addressed  is the relevance of 3NFs in the shell formation, as well as in the location of the neutron dripline as discussed in Section \ref{sec-IV.2}. In particular,  its  role in the shell formation has been  investigated within the SM and IM-SRG approach for light- and medium-mass nuclei \cite{Otsuka10,Holt12,Holt14,Holt13,Hebeler15,Stroberg16,Simonis16, Stroberg17,Simonis17,Fukui18,Ma19,Coraggio20e,Coraggio21}, focusing mainly  on the oxygen  \cite{Otsuka10, Holt13,Hebeler15,Simonis16,Stroberg17} and calcium \cite{Holt12,Holt14,Simonis16,Ma19,Coraggio20e} isotopic chains. In all these calculations, the effective Hamiltonians are derived from  $NN$ and 3N 
potentials built up within the chiral perturbative theory, while in  \cite{Otsuka10,Tsunoda17}  the Fujita-Miyazawa three-body force is employed to study the low-lying states of neutron-rich nuclei with $Z=8$, $Z=10-14$ and $N\sim 20$.

The main results of  all these works is that 3N forces give rise to a repulsive interaction between the valence particles that improves the agreement with experimental data.
For instance, in Ref. \cite{Otsuka10},   it was  shown that  the 3NF is able  to  correct the strong attractiveness of the monopole interaction between the $0d_{3/2}$ and $0d_{5/2}$ neutron  orbitals, which explains the   change in the location of the neutron dripline from $^{28}$O to the experimentally observed $^{24}$O, as discussed in Section \ref{sec-IV.2.1}.

\begin{figure} 
\begin{center}
\includegraphics[width=12cm]{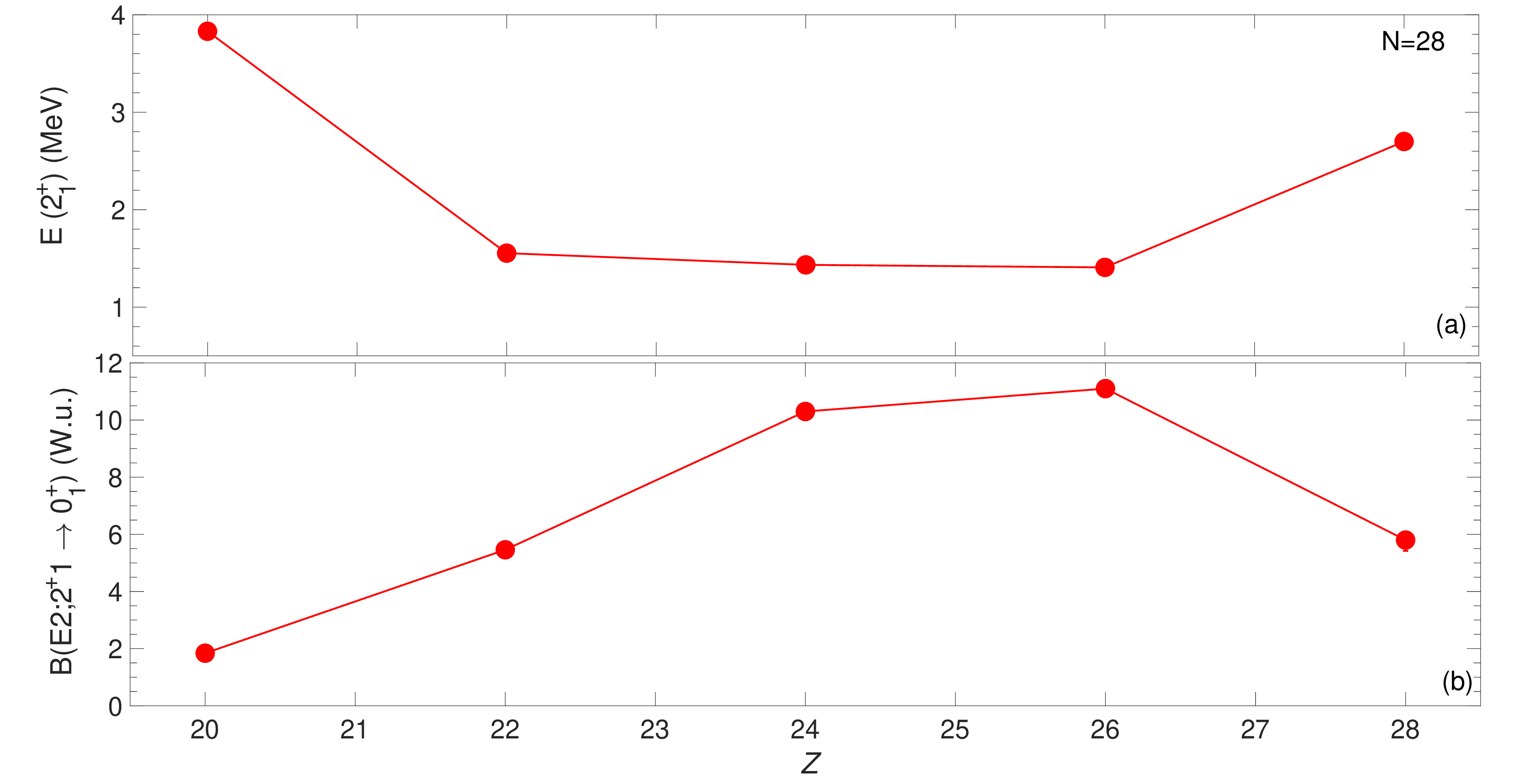}  
\caption{ Experimental (a) $2^{+}_{1}$ excitation energies  and (b) $B(E2, 2^{+}_{1} \rightarrow 0^{+}_{1})$  transition rates  in the  $N = 28$ isotones.}
\label{sec-IV3_Fig2}
\end{center}
\end{figure}

Here, to illustrate   the role of the 3NF in  providing a reliable monopole component of the effective SM Hamiltonian we choose as physics case the nuclei of the $fp$ shell, namely   Ca, Ti, Cr, Fe, Ni isotopes from $N=20$ to 32. 
Within this mass region, two  doubly magic  nuclei, $^{48}$Ca and $^{56}$Ni, are known, while  an enhancement in the collectivity  is observed for nuclei with $22 \leq  Z \leq 26$ and $N=28$, as can be seen in  Fig. \ref{sec-IV3_Fig2} reporting   the experimental $2^{+}_{1}$ excitation energies  and  the   $B(E2, 2^{+}_{1} \rightarrow 0^{+}_{1})$ values for the $N=28$ isotones \cite{ensdf}.
 
This case may be, therefore, a  good testing ground to investigate the  relevance of  3NFs in generating  gaps, and  it may be of particular interest  to study how 3NFs affect the $N=28$ shell closure when the proton   $0f_{7/2 }$ orbital is getting filled.  

 In the next section, on the basis of the results obtained in Ref. \cite{Ma19}, we  discuss   physical quantities as the excitation energies of the yrast  $2^+$ states, the  $B(E2, 2^{+}_{1} \rightarrow 0^{+}_{1})$ values, and the two neutron separation energies focusing on the changes produced by the   the 3N force  on the monopole component of the effective interaction and on the  ESPEs. In Section \ref{sec-IV.3.3},  the 3NF monopole component is analyzed in terms of the   central, vector, and tensor contributions. 

\subsubsection{\it Monopole interaction and effective single-particle energies \label{sec-IV.3.2}}

Calculations have been performed  within the SM framework  by considering   valence protons and neutrons interacting in the valence space composed by the  $0f_{7/2}$, $0f_{5/2}$, $1p_{3/2}$, and $1p_{1/2}$ orbitals outside doubly magic $^{40}$Ca.  The adopted effective Hamiltonians  are derived within the MBPT approach starting from the chiral $NN$-only and  $NN$+3N forces, as described in Section \ref{sec-IV.2.2}. As for the proton-proton channel the Coulomb force is added. We limit to consider nuclei from $N=20$ to 32  to avoid that  our predictions are affected by the choice  of the adopted model space  (see Section \ref{sec-IV.2.2}). 

\begin{figure}[h]
\begin{center}
\includegraphics[width=12cm]{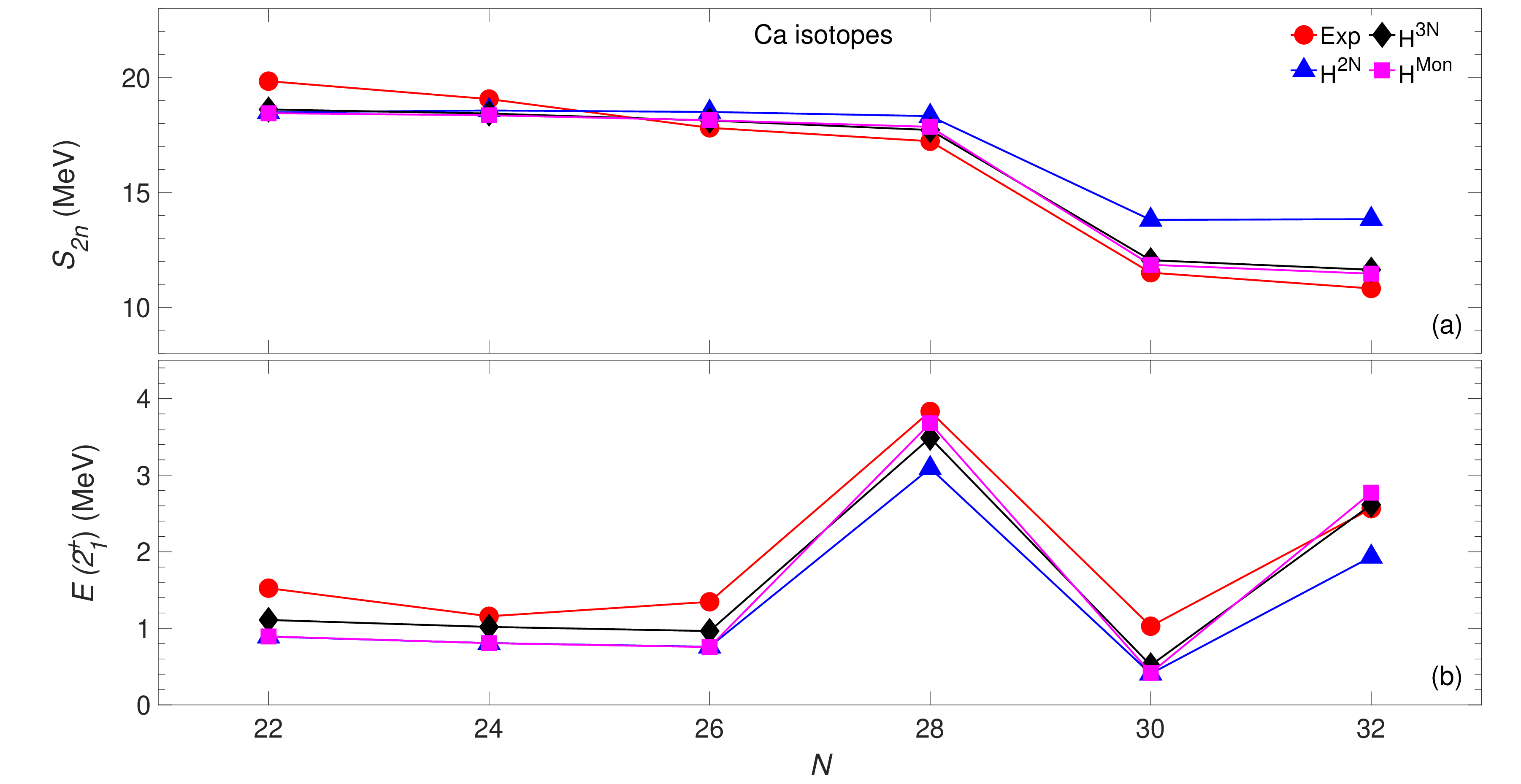}  
\caption{Experimental and calculated  (a) two-neutron separation energies  and (b) $2^{+}_{1}$ excitation energies for calcium isotopes from $N = 22$ to 32. See text for details.}
\label{Sec-432_Ca}
\end{center}
\end{figure}

\begin{figure}
\begin{center}
\includegraphics[width=12cm]{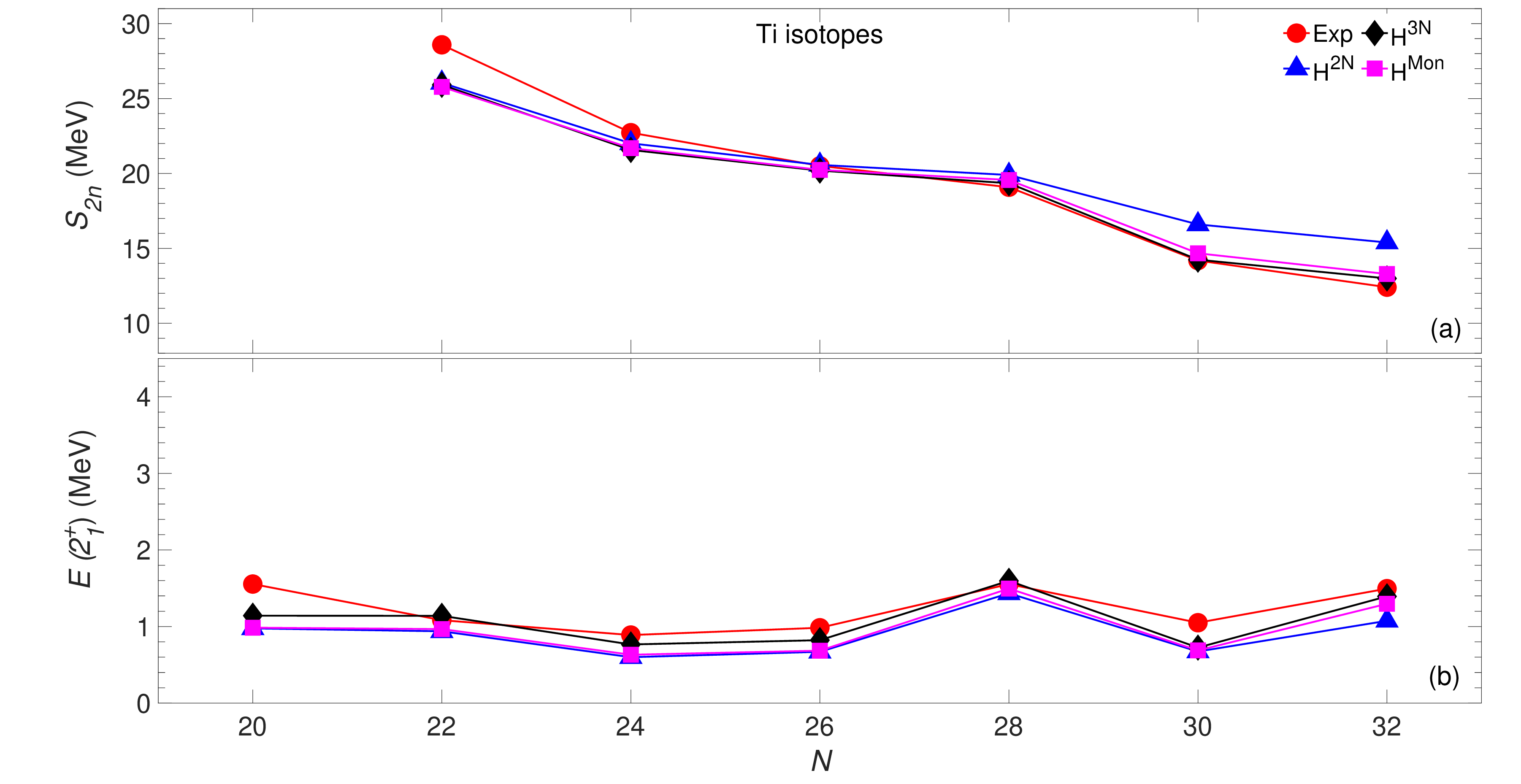}  
\caption{ Experimental and calculated  (a) two-neutron separation energies  and (b) $2^{+}_{1}$ excitation energies for titanium isotopes from $N = 20$ to 32. See text for details.}
\label{Sec-432_Ti}
\end{center}
\end{figure}

\begin{figure}
\begin{center}
\includegraphics[width=12cm]{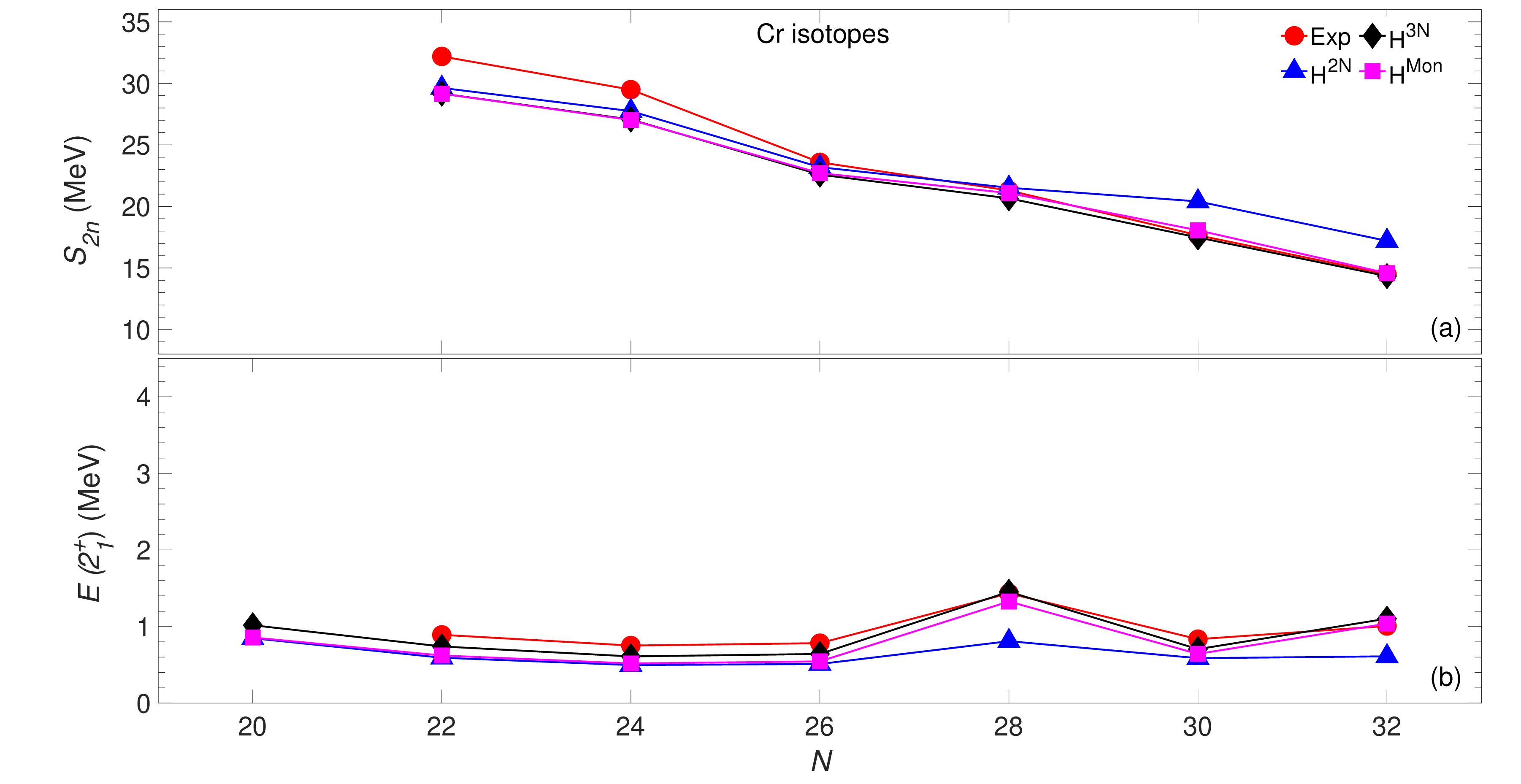}  
\caption{ Experimental and calculated  (a) two-neutron separation energies  and (b) $2^{+}_{1}$ excitation energies for chromium isotopes from $N = 20$ to 32. See text for details.}
\label{Sec-432_Cr}
\end{center}
\end{figure}
In the following, we focus on the $S_{2n}$s  and  excitation energies of the yrast $2^+$ states by comparing   the experimental data  with the results of SM calculations obtained  by employing  effective Hamiltonians derived from the  $NN$ force only ($H^{ \rm 2N}$) and the complete $NN$ +3N force ($H^{\rm 3N}$). Furthermore, to better pinpoint the role of the monopole component and how it is affected by the 3NF we have also introduced  a third Hamiltonian, $H^{\rm mon}$, which is made up by summing the monopole component of $H^{\rm 3N}$  to the multipole component of $H^{\rm 2N}$. 

However, as a preliminary step, we should comment on the SP energies of the adopted effective Hamiltonians. As mentioned in Section \ref{perturbativeh},  the  derivation of $H_{\rm eff}$   for one-valence
nucleon systems  provides the theoretical SP energies,  which are then subtracted from the diagonal matrix elements of  $H_{\rm eff}$  derived
for the two-valence nucleon systems to obtain the  TBMEs of the residual interaction. The
SP energies arising from  $NN$-only theory do not supply a reasonable proton and neutron gap between the $0f_{7/2}$ orbital and the remaing three orbitals \cite{Ma19}, which might prevent the description of the observed shell closure at $Z$, $N=28$.
Therefore, in order to remove the effects due to the SP energies and concentrate on those produced by the two-body part of the effective Hamiltonians,  specifically  on their monopole components,  calculations in all three cases  ($H^{\rm 2N}$,   $H^{\rm 3N}$, and $H^{\rm mon}$) are carried out starting from the same set of SP energies, namely those  derived from the $NN$+3N force.  The values of the neutron and proton SP energies can be found in Ref. \cite{Ma19}.

\begin{figure}
\begin{center}
\includegraphics[width=12cm]{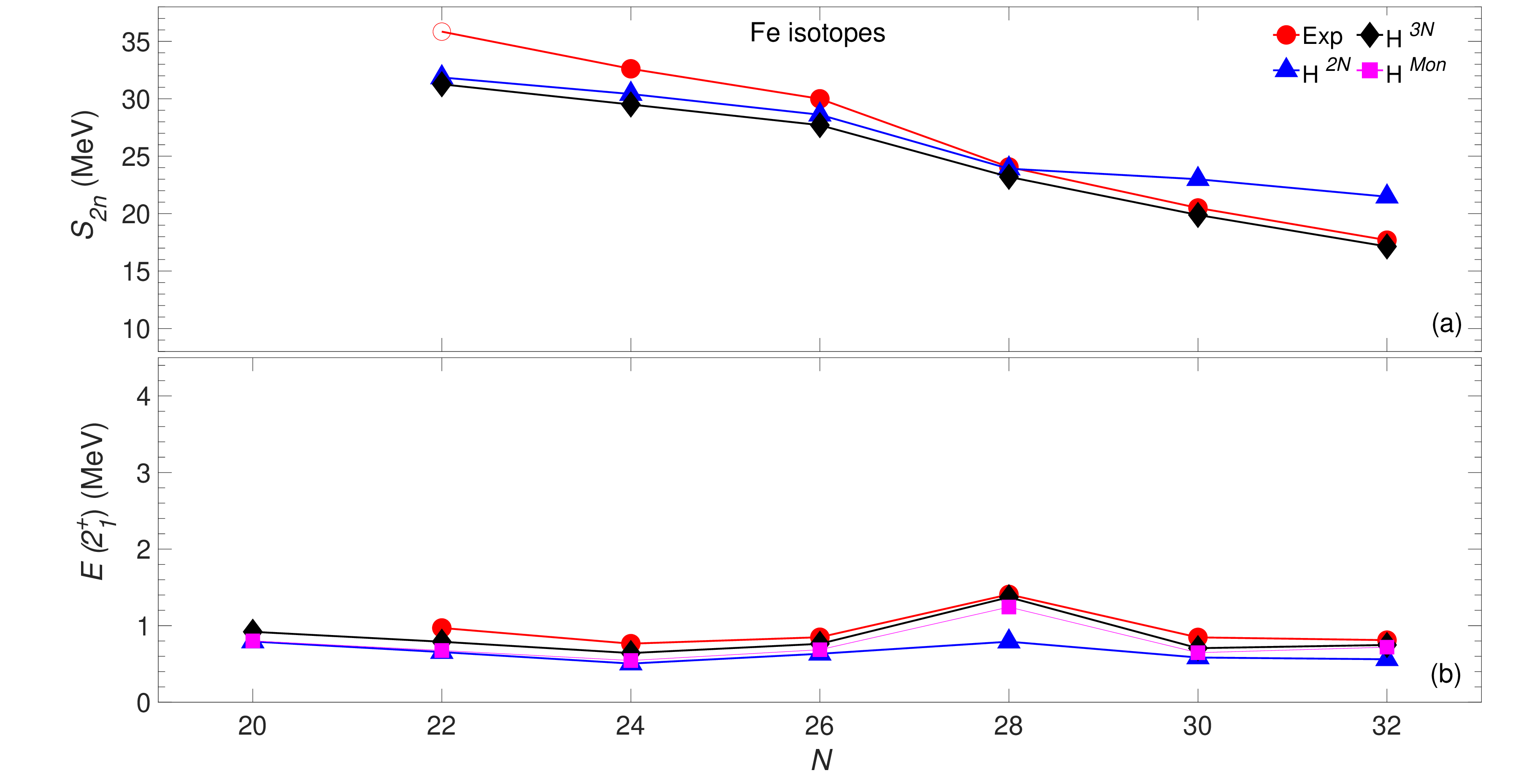}  
\caption{ Experimental and calculated  (a) two-neutron separation energies  and (b) $2^{+}_{1}$ excitation energies for iron isotopes from $N = 20$ to 32. See text for details.}
\label{Sec-432_Fe}
\end{center}
\end{figure}

\begin{figure}
\begin{center}
\includegraphics[width=12cm]{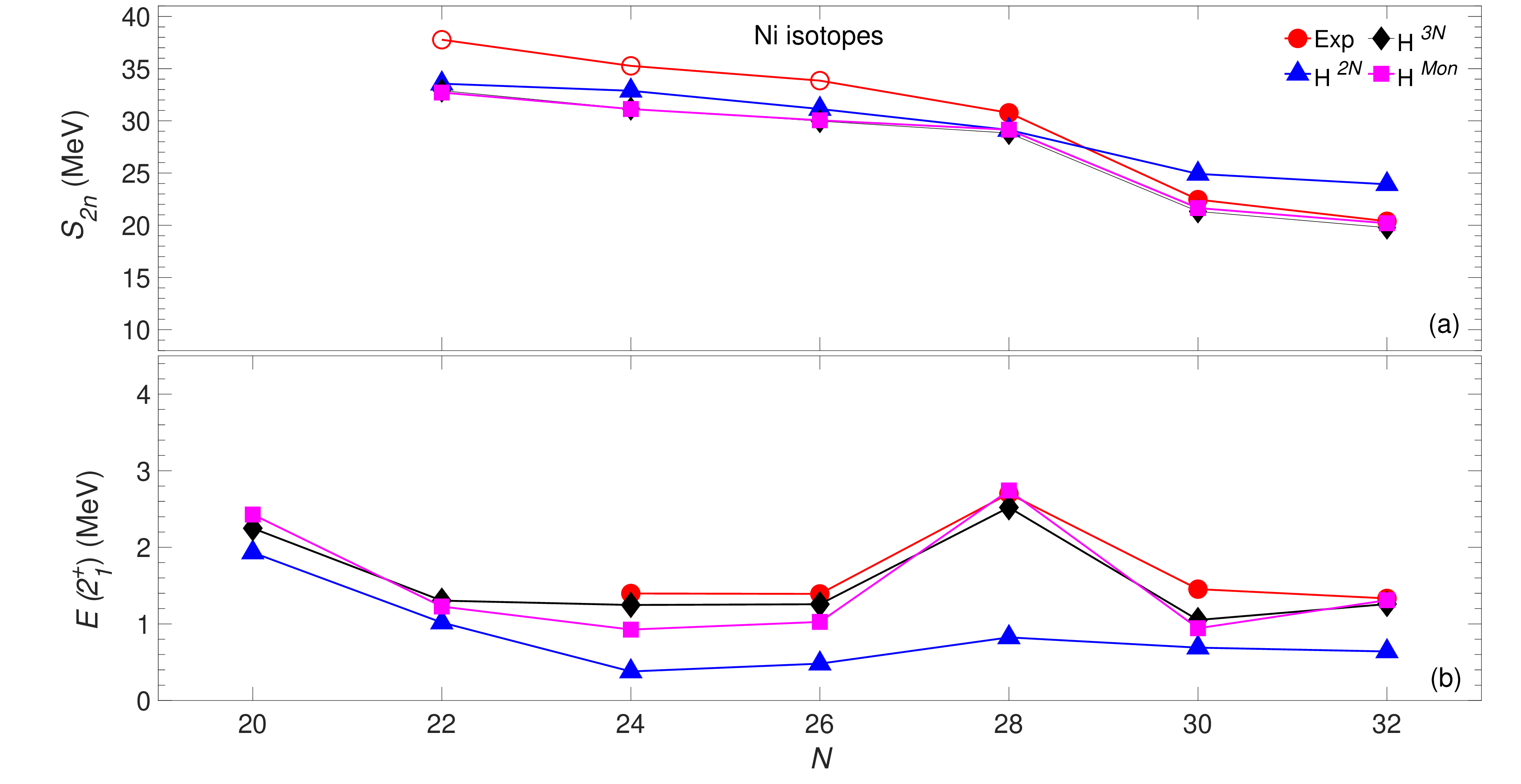}  
\caption{ Experimental and calculated  (a) two-neutron separation energies  and (b) $2^{+}_{1}$ excitation energies for nickel isotopes from $N = 20$ to 32. See text for details.}
\label{Sec-432_Ni}
\end{center}
\end{figure}

In Figs. \ref{Sec-432_Ca} -- \ref{Sec-432_Ni}, the experimental data for the $S_{2n}$ \cite{AME21} and the excitation energies of the yrast $2^+$ states \cite{ensdf} for Ca, Ti, Cr, Fe, and Ni isotopes from $N=20$ to 32  are compared with the theoretical values  obtained with $H^{\rm 2N}$, $H^{\rm 3N}$, and $H^{\rm mon}$. Note that  empty red circles for the experimental  $S_{2n}$ values refer to the estimated values reported in Ref. \cite{AME21}. 

By comparing the theoretical $S_{2n}$s  from $H^{\rm 2N}$ and $H^{\rm 3N}$  we observe, as already pointed out  in Section \ref{sec-IV.2.2}, that  the repulsive contributions of the 3NF is essential to quench the overbinding induced by the $NN$ force only, thus producing  a downshift of the $S_{2n}$ curve and  improving the  agreement with experiment. When these contributions are taken into account, the calculated results follow closely the experimental behaviour for all the considered isotopes, while their omission leads to a bad reproduction of the observed energy drop between $N=28$ and 30. This drop can be seen as a manifestation of the shell closure at $N=28$, corresponding to the filling of the $0f_{7/2}$ neutron orbital, and the deficiency of  $H^{\rm 2N}$  ascribed to the inadequate gap between  the $0f_{7/2}$ and $1p_{3/2}$  neutron ESPE provided by this Hamiltonian.

 On the other hand,  $H^{\rm mon}$ and $H^{\rm 3N}$ give very similar results, thus confirming  the inaccuracy of the  monopole components  obtained by considering the  $NN$ force only, as well as the central role of these components in determining the $S_{2n}$ evolution and the $N = 28$ shell closure.  At the end of this section, we shall analyze  the changes introduced  by the  3NF in  the neutron and proton ESPEs, which only depend on the monopole components of the Hamiltonian.

Similar considerations follow also  from  the excitation energies of the yrast  $2^+$ states.
 As shown in panel (b) of Fig.  \ref{Sec-432_Ca}, the shell closure at  $N = 28$ in Ca isotopes  is very well reproduced by $H^{\rm 3N}$ and $H^{\rm mon}$,
while the $2^{+}_{1}$ state predicted by $H^{\rm 2N}$  lies about 0.7 MeV below the experimental one. 
For Ti isotopes, we see in Fig. \ref{Sec-432_Ti} that  the experimental behavior is, overall, well reproduced by all three SM Hamiltonians, while  for Cr and Fe isotopes  (Figs. \ref{Sec-432_Cr} and \ref{Sec-432_Fe})  the energy gap  at $N=28$  predicted by $H^{\rm 2N}$ is underestimated  by $\sim 0.6$ MeV,  and the difference  increases up $\sim 1.8$ MeV in Ni isotopes (Fig. \ref{Sec-432_Ni}).
Another subshell closure at $N=32$  is observed in Ca,Ti, Cr isotopes, although not so strong  as that at $N=28$, corresponding to the filling of the $1p_{3/2}$ neutron orbital.  Also in this case  results  from  $H^{\rm 2N}$ provide, in general, a less satisfactory agreement with experimental data.

For all  considered isotopic  chains, calculations  with $H^{\rm 2N}$  underestimate the experimental excitation energy of the $2^{+}_{1}$ states
at  both  $N = 28$ and $N=32$   providing too much collectivity. When moving from Ca isotopes with only identical valence nucleons to systems with  $Z >20$,   we see    a change in the closure properties that may  arise from  the collectivity induced from  the proton-neutron channel of the residual interaction. 
 This change reflects on the evolution of the  $N = 28$ shell closure  as a function of $Z$, which  shows  a  lowering of the yrast $2^+$ state and an increase of the $B(E2; 2^{+}_{1} \rightarrow  0^{+}_{1})$ value for nuclei  with $22 \leq  Z \leq 26$.   as discussed at the end  of  Section \ref{sec-IV.3} from the experimental point of view. 

\begin{figure}
\begin{center}
\includegraphics[width=12cm]{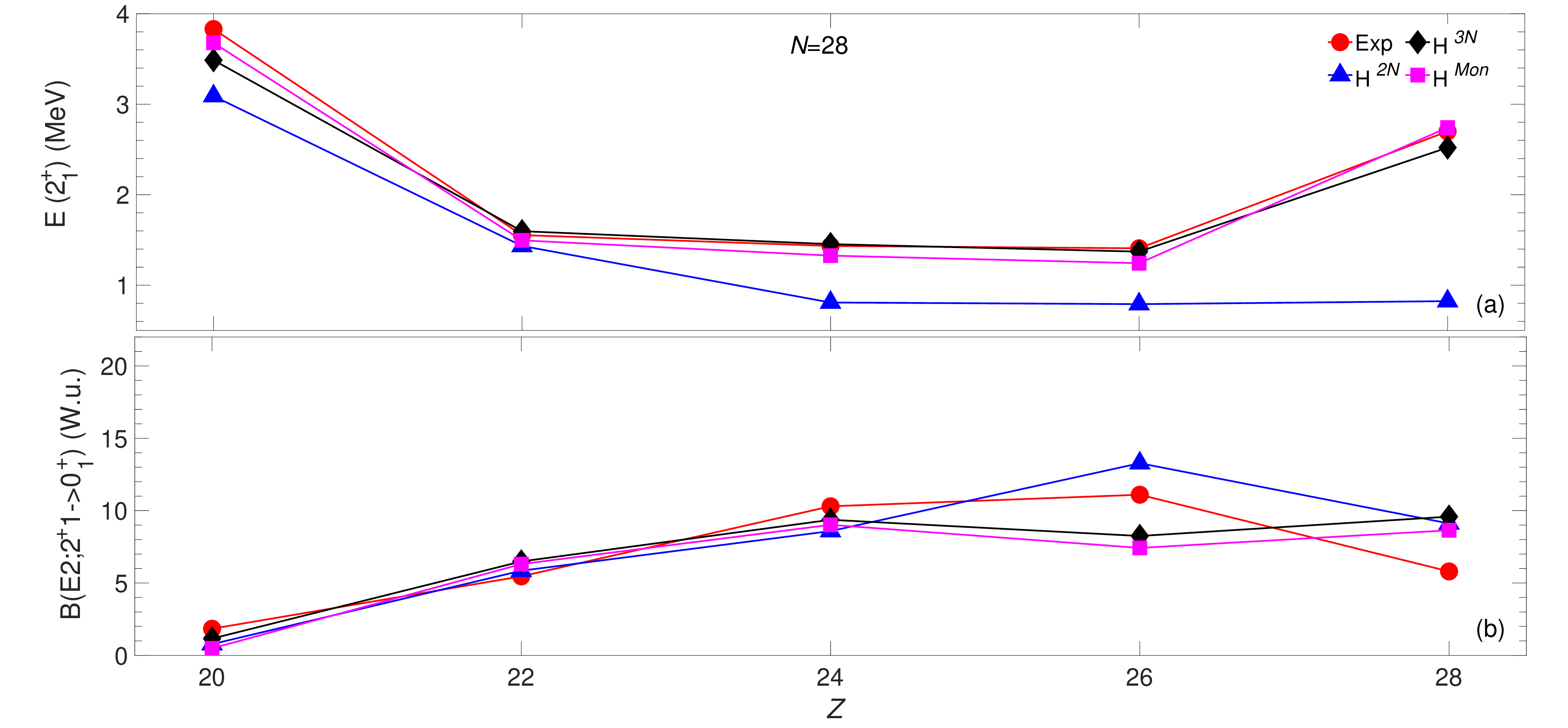}  
\caption{ Experimental and calculated   (a) $2^{+}_{1}$ excitation energies  and  (b) $B(E2, 2^{+}_{1} \rightarrow 0^{+}_{1})$  transition rates  for $N=28$ isotones  from  $Z=20$ to 28. See text for details.}
\label{Sec-432_N28}
\end{center}
\end{figure}
 \begin{figure}
\begin{center}
\includegraphics[width=12cm]{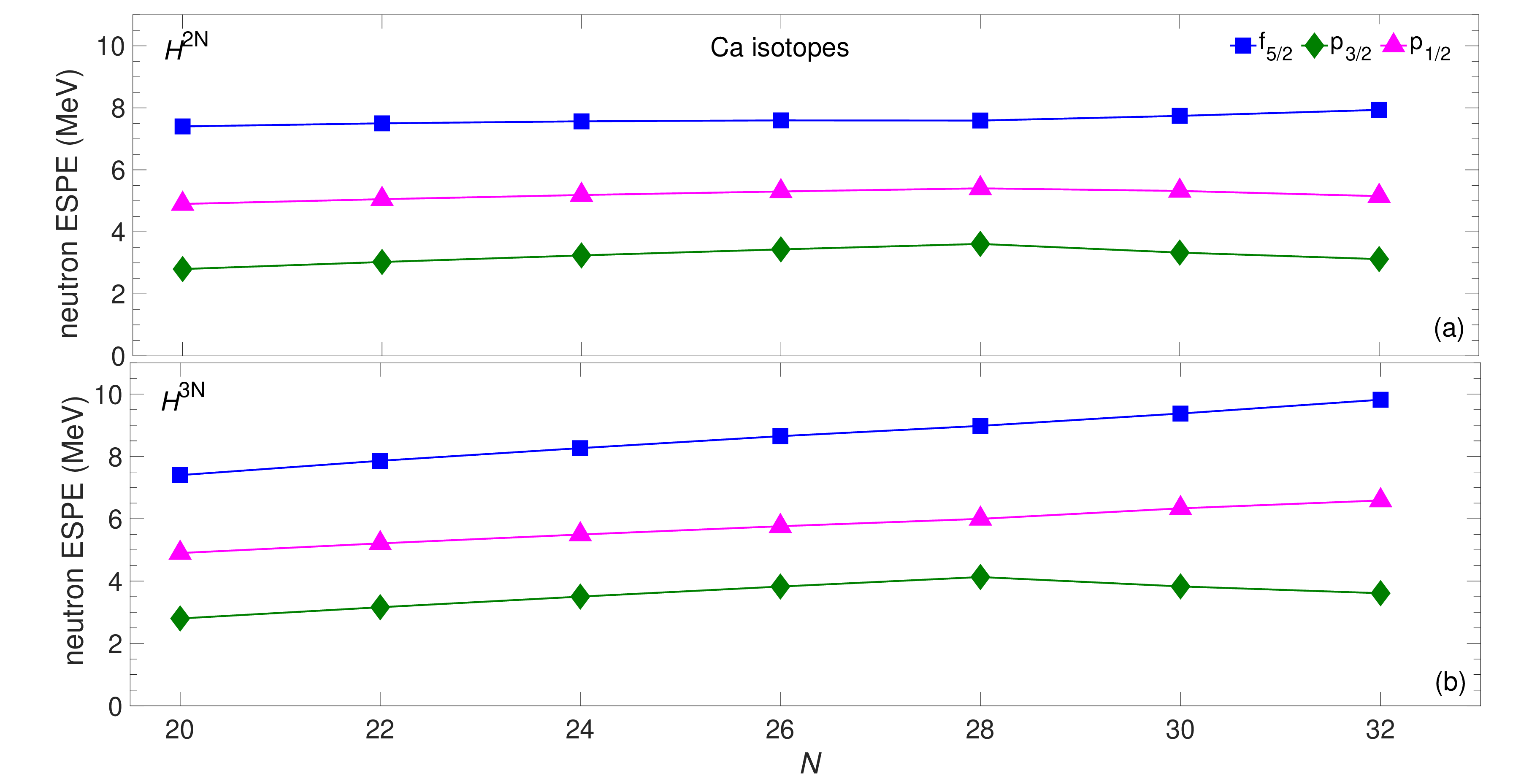}  
\caption{Neutron ESPEs from (a) $H^{\rm 2N}$  and (b) $H^{\rm 3N}$  for calcium 
 isotopes as a function of the neutron number. See text for details.}
\label{Sec-432_ESPECa}
\end{center}
\end{figure}

 In Fig. \ref{Sec-432_N28},  the experimental excitation energies of the $2^{+}_{1}$ states and the  $B(E2; 2^{+}_{1} \rightarrow  0^{+}_{1})$ values are   compared with the  $H^{\rm 2N}$, $H^{\rm 3N}$, and $H^{\rm mon}$ results.  The proton and neutron effective charges  to calculate the $B(E2)$s have been consistently obtained  with the same perturbation approach of the Hamiltonian,  without any empirical adjustment,  as described in Section \ref{perturbativeh}.  

The  collectivity evolution between $^{48}$Ca  and $^{56}$Ni is   well reproduced   by $H^{\rm 3N}$  and $H^{\rm mon}$, but not by $H^{\rm 2N}$. In particular, the latter Hamiltonian is not able to describe the doubly magic nature of $^{56}$Ni.
The monopole component of $H^{\rm 2N}$, responsible for the evolution of the neutron and proton ESPEs,  cannot balance   indeed the collectivity induced by higher multipole components in the proton-neutron channel.

\begin{figure}
\begin{center}
\includegraphics[width=12cm]{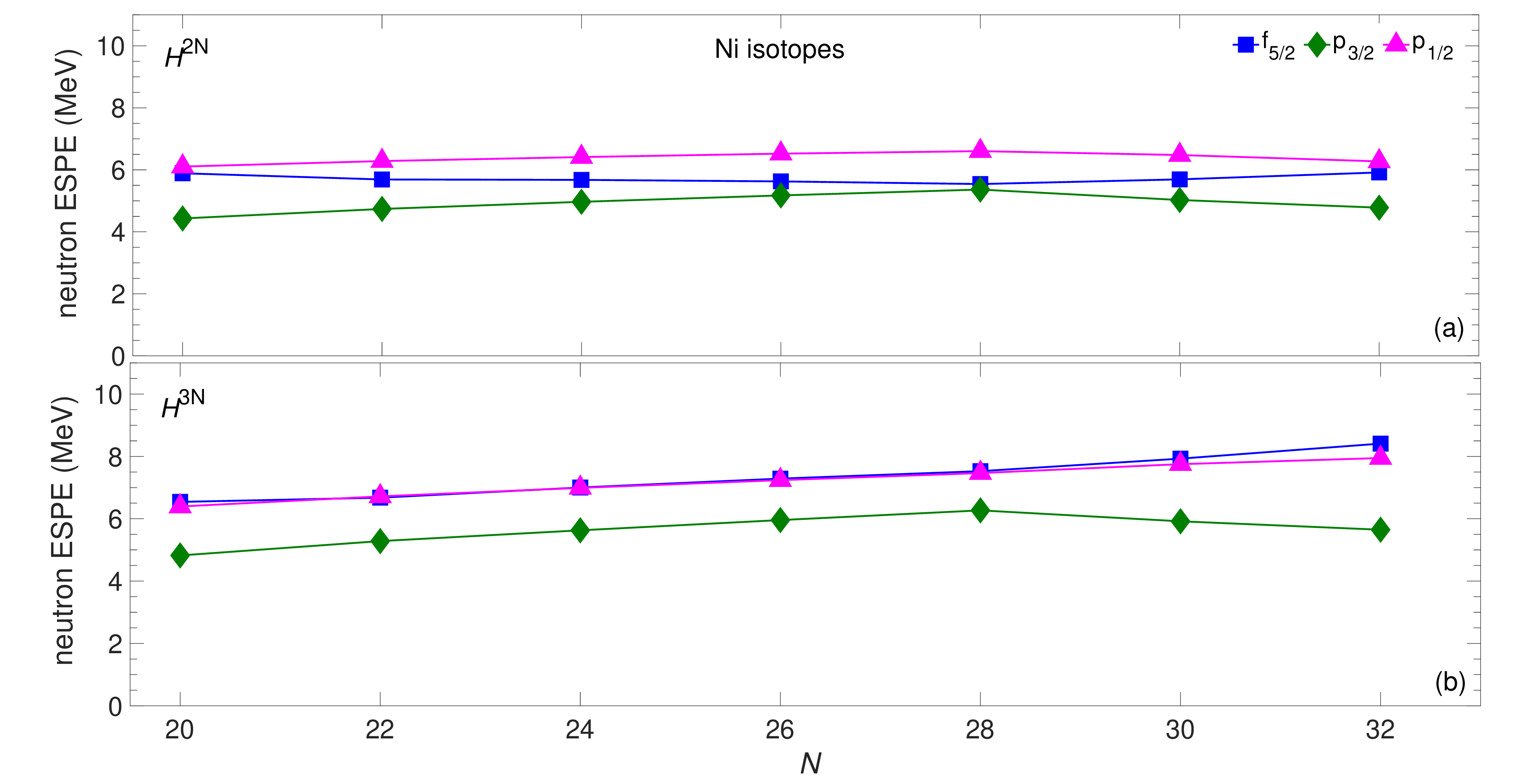}  
\caption{Neutron ESPEs from (a) $H^{\rm 2N}$  and (b) $H^{\rm 3N}$  for nickel 
 isotopes as a function of the neutron number.  See text for details.}
\label{Sec-432_ESPENin}
\end{center}
\end{figure}

\begin{figure}
\begin{center}
\includegraphics[width=12cm]{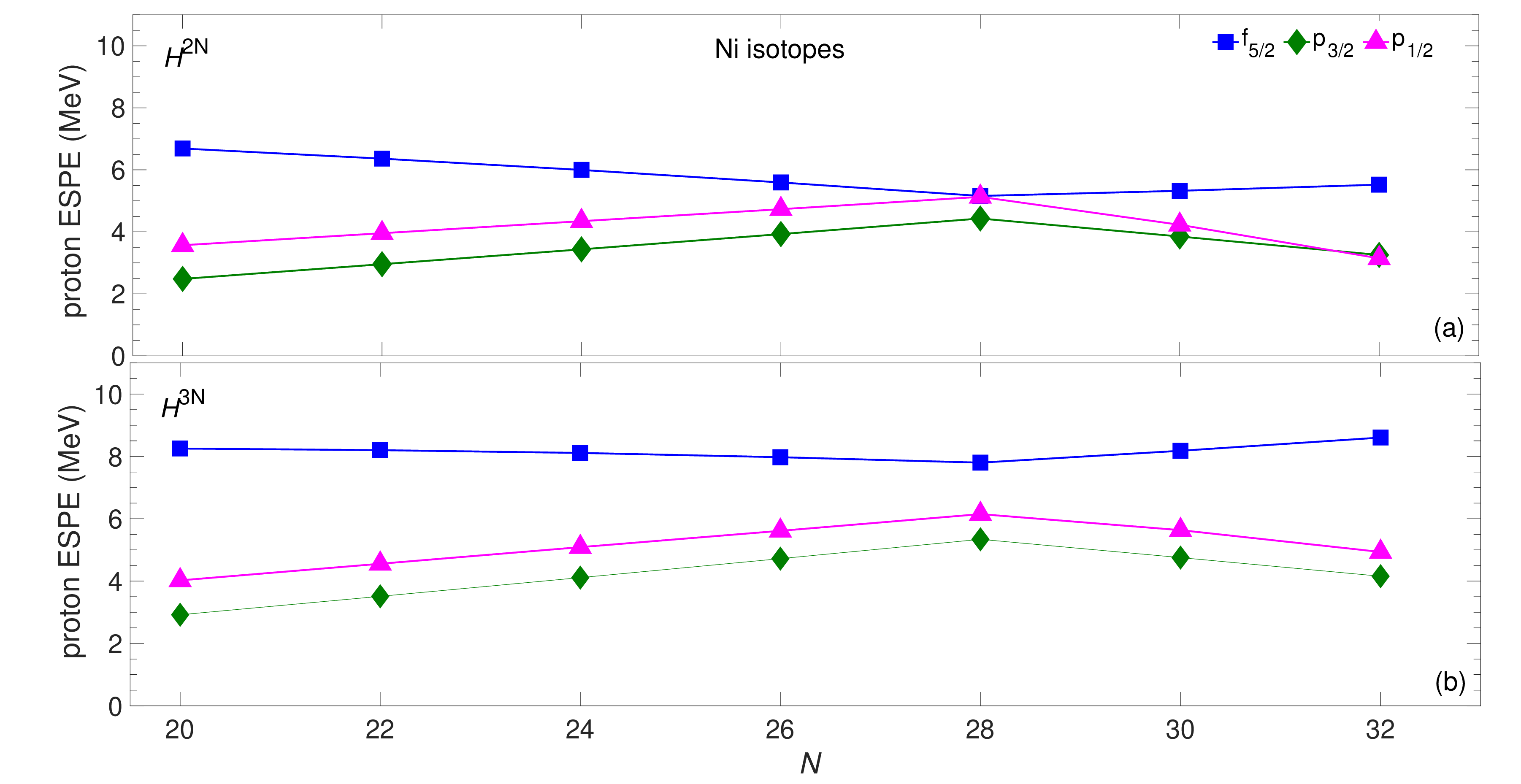}  
\caption{Proton ESPEs from (a) $H^{\rm 2N}$  and (b) $H^{\rm 3N}$  for nickel
 isotopes as a function of the neutron number. See text for details.}
\label{Sec-432_ESPENip}
\end{center}
\end{figure}

To better elucidate this point we examine, in the following, the proton and neutron ESPEs as a function of the number of valence neutrons.  In particular, we compare the neutron ESPEs for Ca isotopes and both neutron and proton ESPEs for Ni isotopes  obtained by employing the monopole component of $H^{\rm 2N}$ and $H^{\rm 3N}$. The ESPEs are defined in Eq. (\ref{ESPE})  with the g.s. occupation numbers, $n^{\tau}_{b}$, fixed by employing the normal filling scheme, namely by putting  the valence nucleons  into the possible lowest orbit one by one. The results are reported in Figs. \ref{Sec-432_ESPECa}, \ref{Sec-432_ESPENin}, \ref{Sec-432_ESPENip} where the ESPEs are referred to the lowest lying $0f_{7/2}$ orbital. It is worth recalling that in all cases the starting SP energies are those derived by adopting the  $NN$+3N force.

From the inspection of Fig. \ref{Sec-432_ESPECa}, for calcium isotopes, we can observe that the inclusion of the 3NF does not affect the general behavior of the neutron ESPEs, but provides  specific features that  give rise to the difference in the results of   the $H^{\rm 2N}$ and  $H^{\rm 3N}$ discussed above. We see, in fact, that the neutron monopole component $H^{\rm 3N}$ produces an increase in the $1p_{3/2} -0f_{7/2}$ energy gap  at $N=28$  inducing a stronger shell closure,  and also a larger
$1p_{1/2} -1p_{3/2}$ splitting in correspondence of the $N=32$ subshell  closure. It is also interesting to note that  a larger energy splitting is found for both pairs  of the $1p$ and $0f$ spin-orbit partners when 3NFs are taken into account. This effect grows with increasing neutron number.

Similar comments can be made for  the neutron and proton ESPEs in Ni isotopes, which are shown in Figs. \ref{Sec-432_ESPENin} and \ref{Sec-432_ESPENip}. It can be seen that  the inclusion of the 3NF provides an increase in the $0f_{7/2} -1p_{3/2}$   and $1p_{1/2} -1p_{3/2}$ splittings  at $N=28$ and  $N=32$, respectively, for  both the neutron and proton ESPEs. In general, the contribution of the 3NF leads  to a substantial  expansion of the orbital separations with respect to the $NN$ force only. In particular,  the $0f_{7/2}-0f_{5/2}$ spin-orbit splitting at $N=28$ increases by about  2 and 3 MeV for neutrons and protons, respectively.  Furthermore, the strong narrowing we observe at $N=28$ for the $0f_{5/2}$ and $1p_{3/2}$ orbitals with the $NN$ force is significantly attenuated by including the 3NF.

To summarize, we have shown  that the monopole component of the 3NF is crucial to correct the behavior of the ESPEs and  smooth the too much collectivity resulting from $H^{\rm 2N}$  thus  leading to results  able to reproduce the experimental  data and the doubly magic nature of $^{56}$Ni. A careful analysis of the difference in the monopole components of $H^{\rm 2N}$ and $H^{3N}$ will be presented in the next section.

\subsubsection {\it Spin-tensor decomposition of the shell-model interaction \label{sec-IV.3.3}}
The difference in the  behavior of the ESPEs resulting when employing the $NN$ force only and the complete $NN$ + 3N force  lies in the  effects produced by the 3N component on the  monopole matrix elements of the  effective SM Hamiltonian.  In order to better substantiate this  statement, we  consider the    $^{56}$Ni case,  whose  neutron and proton shell gaps are strongly affected by the inclusion of the 3NF, as discussed in the previous section.  We therefore compare the monopole matrix elements, namely the centroids of the two density-dependent Hamiltonians, $H^{\rm 2N}$ and $H^{\rm 3N}$, we have used for the calculations of this nucleus. In particular, for the sake of simplicity we focus on the matrix elements  $ {\bar V}_{a b}^ {\tau \tau'} $ (see Eq. (\ref{Mon})) with at least one of  the two indices,   $a$ or $b$,  representing the $0f_{7/2}$ orbital. In fact,  since in the calculation of the ESPEs we adopt the normal filling scheme only these matrix elements  come  into play  for $^{56}$Ni.
\begin{figure}
\begin{center}
\includegraphics[width=16cm]{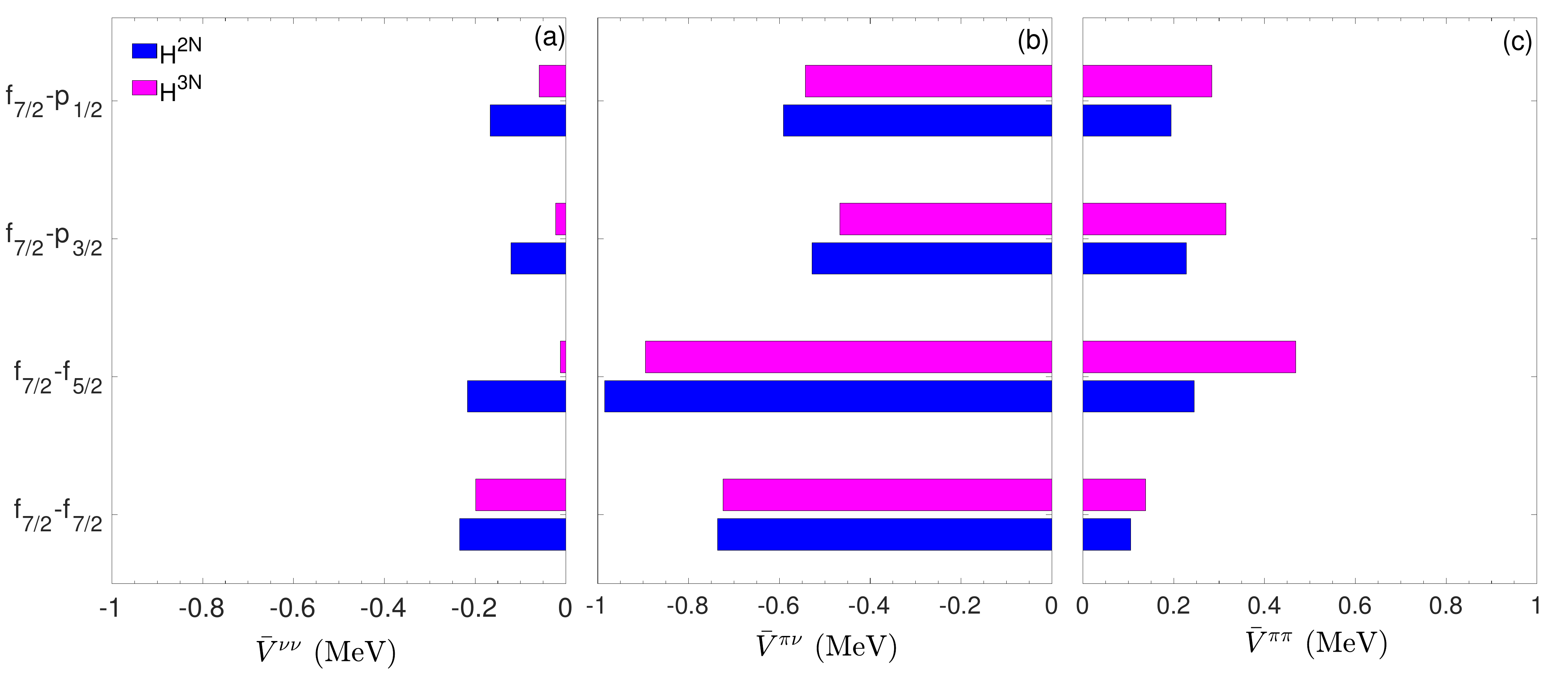}  
\caption{(a) Neutron-neutron, (b) proton-neutron (b), and (c) proton-proton  monopole matrix elements of the effective interactions with and without 3N force. See text for
details}
\label{Sec-433_Vmonf}
\end{center}
\end{figure}
In Fig. \ref {Sec-433_Vmonf}, we report the centroids ${\bar V}_{0f_{7/2} b}^ {\tau \tau'}$ of $H^{\rm 2N}$ and $H^{\rm 3N}$ for the neutron-neutron, proton-neutron,  and proton-proton channels. It is worth mentioning that  neutron-neutron and proton-neutron interactions determine the neutron ESPEs, while the proton ones depend on  proton-proton and proton-neutron interactions. We see that the 3NF provides a repulsive  contribution to all matrix elements, which makes  the neutron-neutron and proton-neutron  matrix elements less attractive and the proton-proton ones more repulsive. However, the size of the contributions  depends on the involved orbitals, ranging  from few tens of keV to about 200 keV, which  produces a  substantial change in the  spacings between the ESPE's  and consequently in the variation  of the shell structure in correspondence of a sizable occupation of  a specific orbital, as it is the case of the $0f_{7/2}$ orbital in $^{56}$Ni.    In all three channels, the changes  produced by $H^{3N}$ are larger for the  ${\bar V}_{0f_{7/2} b}^ {\tau \tau'}$  matrix elements with $b \ne 0f_{7/2}$ than for the diagonal ones. This  leads  to a larger gap between the  $0f_{7/2}$ orbital  and  the remaining orbitals in both the proton and neutron space when the complete $NN$ + 3N force is adopted, which results in   a stronger shell closure at $N=28$ for Ni isotopes as discussed in Section \ref{sec-IV.3.2}.

\begin{figure}
\begin{center}
\includegraphics[width=16cm]{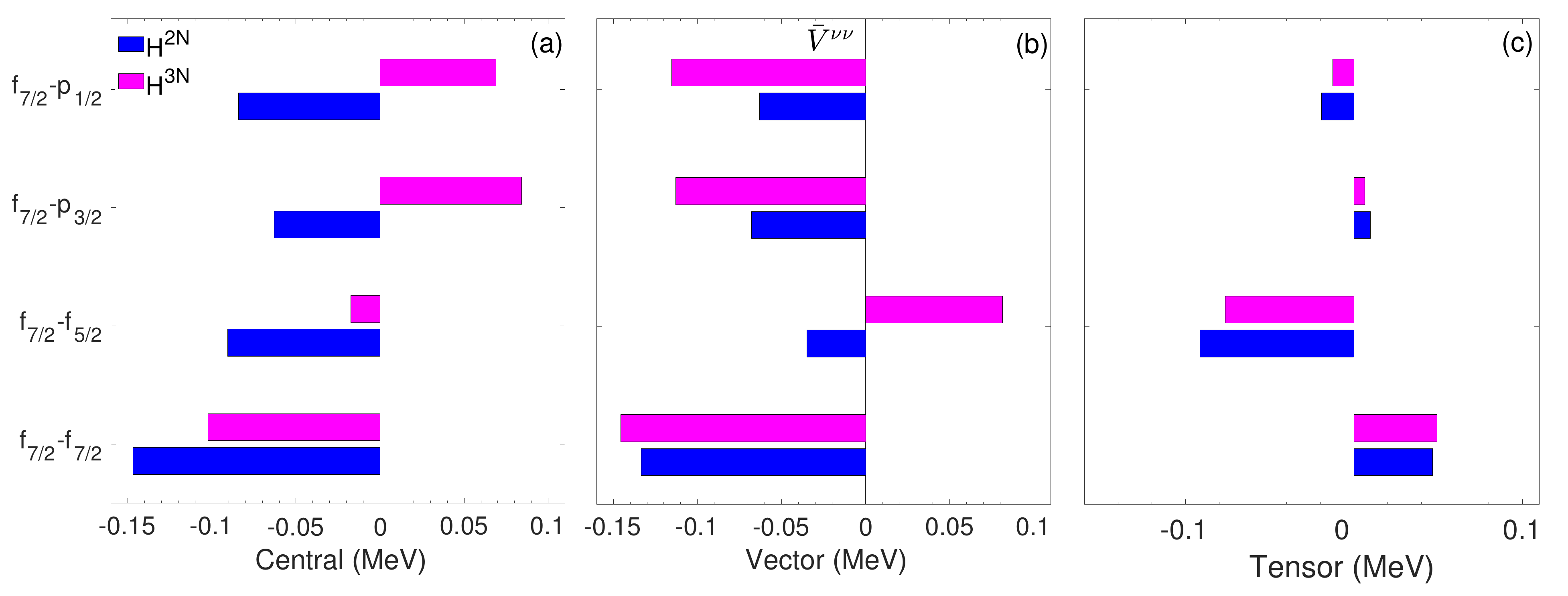}  
\caption{ (a) Central, (b) vector, and  (c) tensor  contributions to ${\bar V}_{0f_{7/2} b}^ {\nu \nu}$  with and without 3N force.  See text for
details.}
\label{Sec-433_Vmondnn}
\end{center}
\end{figure}

It can be also observed that the increased  spacing  between the $0f_{7/2}-0f_{5/2}$ spin-orbit partners and the  $0f_{5/2} - 1p_{3/2}$ orbitals, obtained when including the 3NF, is directly related to the stronger  effects  that this force has on the     ${\bar V}_{0f_{7/2} 0f_{5/2}}^ {\tau \tau'}$ matrix elements as compared to the other ones.

In closing this section, we have found it  instructive to analyze the monopole matrix elements of   $H^{\rm 2N}$ and $H^{\rm 3N}$ in terms of their tensorial structure.   As mentioned in Section \ref{sec-IV.3.1},  several studies, aimed  to   understand the mechanism behind the ESPE variations,   have been devoted to investigate  the role of the central, vector, and tensor components of the effective interactions in governing the shell evolution (see, for instance, Refs. \cite{Otsuka20,Otsuka05,Otsuka01,Smirnova10,Smirnova12}).
 As a main result,  it has been found  that the  behavior of each   ESPE is essentially controlled by the the central component,  while it is the interplay of all the three components to determine the  evolution of the spacings between the ESPEs, with the  tensor one  significantly contributing to the changes in the spin-orbit splittings. 
However, these studies have concerned essentially   phenomenological  effective interactions and microscopic effective interactions derived from the  $NN$ force only. 
In particular,    it has been shown that  empirical adjustments  of the latter interactions introduce,  in general, more  significant changes in the central and vector components than in the tensor  one, especially  for the proton-neutron matrix elements \cite{Smirnova12}.

 Here, we are interested to explicitly investigate the  effects of the 3N force on the effective interaction  to  see if its contribution affects in particular  a specific  component   of the effective interaction,  and  verify the conclusions drawn  from the investigations of empirical adjusted interactions.
To this end,  we shall employ the spin-tensor decomposition to extract the central, vector, tensor contributions from the monopole matrix elements   of our effective interaction,  by following the procedure   presented  in   Ref. \cite{Osnes92} and outlined  below  for the sake of completeness.

Any scalar two-body  interaction $V$  for spin  $1/2$ fermions can be written in terms of spherical tensors  by coupling the spin tensor operators  ($S^{k}$) with the corresponding rank tensors in the configuration space  ($Q^{k}$) as
\begin{figure}
\begin{center}
\includegraphics[width=16cm]{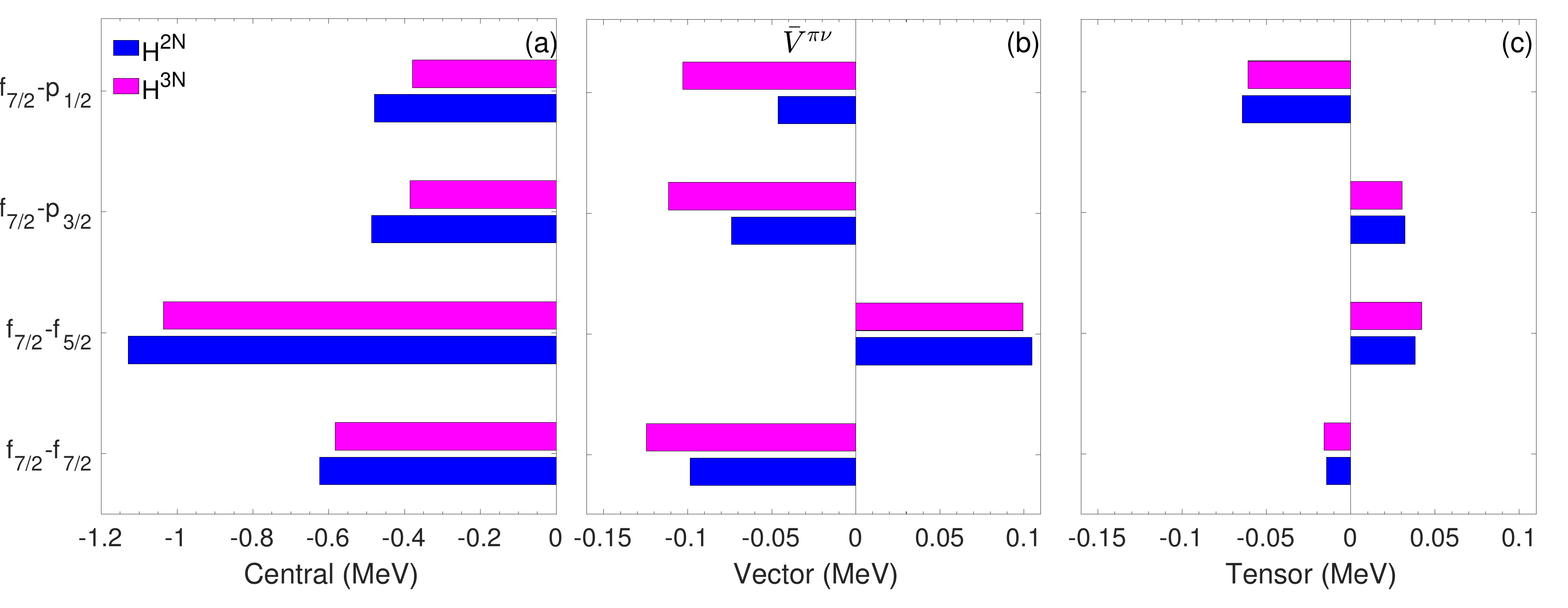}  
\caption {(a) Central, (b) vector, and  (c) tensor contributions to ${\bar V}_{0f_{7/2} b}^ {\pi \nu}$  with and without 3N force. See text for
details. Note that a different scale is used for the vector and tensor components with respect to the central one.}
\label{Sec-433_Vmondpn}
\end{center}
\end{figure}

\begin{equation}
V(1,2)= \sum_{k=0,1,2} (S^{k}  \cdot Q^{k}) = \sum_{k=0,1,2} V^{k},
\label{Dec}
\end{equation}

\noindent
where  $V^{0}$, $V^{1}$, and $V^{2}$ are, respectively,  the central, vector, and tensor components of the interaction $V$. Their matrix elements take the expression

\begin{multline}
\langle a \tau b \tau'; J | V^{k}| c  \tau d\tau';  J \rangle  = \sum_{LL'SS'}   U  
\begin{pmatrix}
l_{a} & 1/2  & j_{a} \\
l_{b} &  1/2  & j_{b} \\
L & S  & J
\end{pmatrix}
U
\begin{pmatrix}
l_{c} & 1/2  & j_{c} \\
l_{d} &  1/2  & j_{d} \\
L' & S ' & J
\end{pmatrix} \\
 \times \hat{k}^2 
\begin{Bmatrix}
L & S & J \\
S'  & L ' & k
\end{Bmatrix} 
\sum_{J'} (-1)^{J'} \hat{J'} \begin{Bmatrix}
L & S & J' \\
S'  & L ' & k
\end{Bmatrix} 
\langle  n_{a}l_{a} \tau n_{b} l_{b} \tau'; L S J' | V | n_{c} \tau l_{c}  n_{d}l_{d}  \tau'; L' S'  J' \rangle, 
\label{Dec2}
\end{multline}

\noindent 
with the coefficients $U$ representing the  generalized $9-j$ symbols 

\begin{equation}
U  
\begin{pmatrix}
l_{a} & 1/2  & j_{a} \\
l_{b} &  1/2  & j_{b} \\
L & S  & J
\end{pmatrix}=
 \hat{j}_a   \hat{j}_b  \hat {L} \hat{S} 
\begin{Bmatrix}
l_{a} & 1/2  & j_{b} \\
l_{b}&  1/2  & j_{b} \\
L & S  & J
\end{Bmatrix}.
\label{Ucoeff}
\end{equation}

\noindent
The $LS$-coupling  matrix elements of $V$ in Eq. \ref{Dec2} are obtained from the $jj$-coupling scheme in the standard way

\begin{align}
&\langle  n_{a}l_{a} \tau n_{b}l_{b} \tau';  L S J | V | n_{c} \tau l_{c}  n_{d}l_{d}  \tau'; L' S'  J \rangle  = 
 \sum_{j_{a} j_{b} j_{c} j_{d}}  U  
\begin{pmatrix}
l_{a} & 1/2  & j_{a} \\
l_{b} &  1/2  & j_{b} \\
L & S  &J
\end{pmatrix} 
U
\begin{pmatrix}
l_{c} & 1/2  & j_{c} \\
l_{d} &  1/2  & j_{d} \\
L' & S ' & J
\end{pmatrix}  \nonumber \\  
  & \quad ~~~~~~~~~~~~~~~~~~~~~~~~~~~~~~\times \langle  a \tau  b \tau';  J | V | c \tau  d  \tau';  J \rangle.
\label{Dec3}
\end{align}

\begin{figure} 
\begin{center}
\includegraphics[width=16cm]{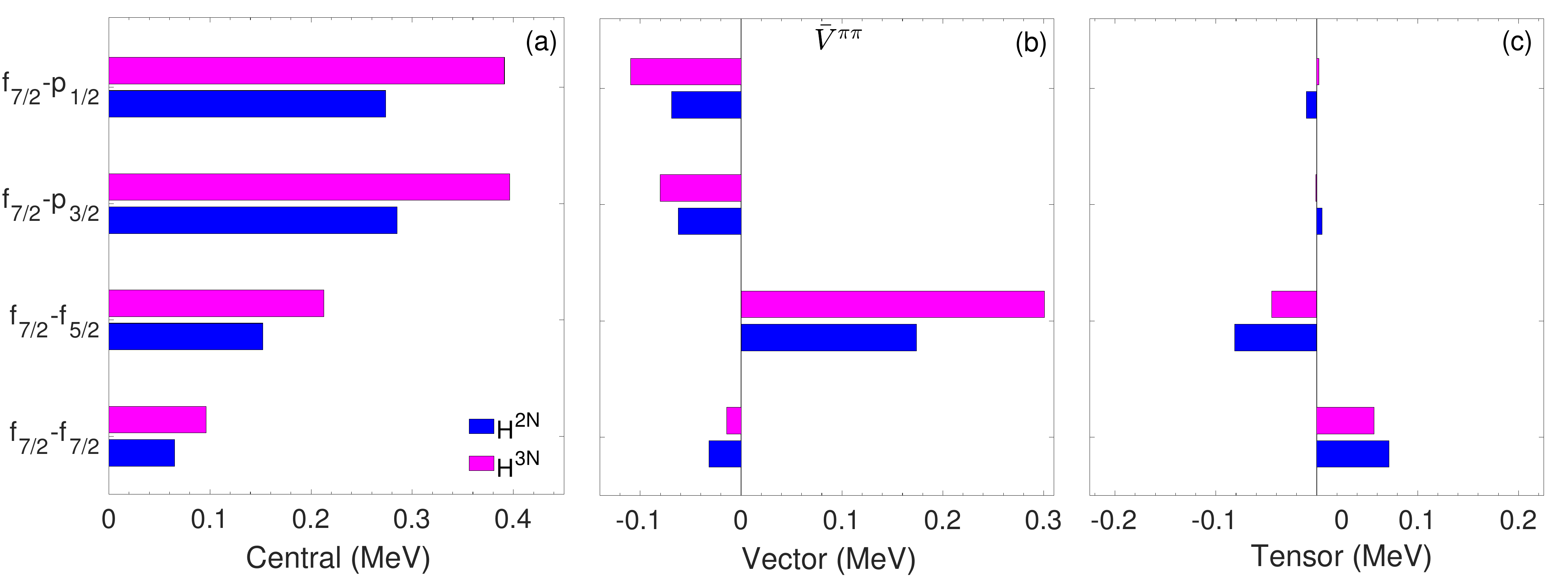}  
\caption{(a)  Central, (b) vector and (c) tensor contributions to ${\bar V}_{0f_{7/2} b}^ {\pi \pi}$  with and without 3N force. See text for
details.}\label{Sec-433_Vmondpp}
\end{center}
\end{figure}

By employing  Eq. \ref{Dec}, the monopole matrix elements $ {\bar V}_{a b}^ {\tau \tau'}$, presented in Fig. \ref{Sec-433_Vmonf}, are decomposed  in their central, vector, tensor contents, which are reported  in Figs. \ref{Sec-433_Vmondnn}, \ref{Sec-433_Vmondpn}, and  \ref{Sec-433_Vmondpp}  for the  the neutron-neutron, proton-neutron, and proton-proton  channels, respectively. 

We see  that the tensor content of all matrix elements  is  rather small with respect to the central and vector ones for  $H^{\rm 2NF}$  as well as for
 $H^{\rm 3NF}$.  As matter of fact,  the tensor components in both cases are of  the order of tens of keV,   and changes   due to the 3NF are limited  to few keV in the vast majority of cases. 
The effects of the 3NF are instead more relevant for the  the central and vector components. However, while the 3NF provides  always  a  repulsive contribution to the central  components,  this is not the case for the vector ones.   The   nature  of these matrix elements is, in fact, enhanced by the inclusion of the 3NF with the exception of ${\bar V}_{0f_{7/2} 0f_{5/2}}^ {\pi \nu}$ which changes from negative to positive.

\begin{table} [h]
\label{Sec-433_Tab}
\begin{center}
\caption{Spin-tensor contents of centroid  differences $\Delta_{b}^{\tau \tau'}$ (in MeV). See text for details.} 
\begin{tabular}{ |l|c|c|c|c|c|c|c|c|}
\hline
\multicolumn{9}{|c|}{$\bar{V}^{\nu \nu}$} \\
\hline
&  \multicolumn{4}{c}{$H^{\rm 2N}$}  & \multicolumn{4}{|c|}{$H^{\rm 3N}$ } \\
\hline
$\beta$ & C &V &T& Tot & C& V & T& Tot \\
\hline
$0f_{5/2}$  & 0.056 & 0.090 & -0.138 & 0.008& 0.085 & 0.227 & -0.126& 0.186 \\
$1p_{3/2}$  & 0.084 & 0.066 & -0.037 &0.113 & 0.187 & 0.033 & -0.043&0.177\\
$1p_{1/2}$  & 0.063 & 0.070 & -0.066&0.067 & 0.171& 0.030 & -0.061&0.140 \\
\hline \hline
\multicolumn{9}{|c|}{$\bar{V}^{\pi \nu}$} \\
\hline
&  \multicolumn{4}{c}{$H^{\rm 2N}$}  & \multicolumn{4}{|c|}{$H^{\rm 3N}$ } \\
\hline
$\beta$ & C &V &T&Tot& C& V & T &Tot\\
\hline
$0f_{5/2}$  & -0.505 &	0.203 &	0.053&-0.249 &-0.453 &	0.224 &	0.058&-0.171 \\
$1p_{3/2}$  & 0.137 &	0.025 &	0.047 &0.209 & 0.198 &	0.013&	0.046&0.257\\
$1p_{1/2}$  & 0.144	& 0.052 &	-0.050 &0.146& 0.204	 & 0.022	& -0.045&0.181 \\
\hline \hline
\multicolumn{9}{|c|}{$\bar{V}^{\pi \pi}$} \\
\hline
&  \multicolumn{4}{c}{$H^{\rm 2N}$}  & \multicolumn{4}{|c|}{$H^{\rm 3N}$ } \\
\hline
$\beta$ & C &V &T& Tot& C& V & T &Tot\\
\hline
$0f_{5/2}$  & 0.087 & 0.206 & -0.153 &0.140&0.117 & 0.315 & -0.101&0.331 \\
$1p_{3/2}$  & 0.220 &	-0.031&	-0.066 &0.123 & 0.300 &	-0.066 &	-0.058&0.176 \\
$1p_{1/2}$  & 0.209 &	-0.037 &	-0.082 & 0.090&0.295	 &-0.095&	-0.055&0.145 \\
\hline
\end{tabular}
\end{center}
\end{table}

As also  evidenced in prior studies, the behavior of the   ESPEs is  largely determined for both $H^{\rm 2N}$  and $H^{\rm 3N}$ by the central and vector components, and in  particular by  the central  monopole proton-neutron interaction. It is this component that is  mainly responsible for pushing down  all the single-particle  orbitals,  and the attenuation of its attractiveness induced by the 3NF leads to a reduction of this phenomenon.   However, in  studying the shell-structure evolution we are  more interested  in the spacings between ESPEs  than in their  absolute  values,  and therefore attention should be focused on the differences between the centroids. These differences, calculated with respect to  ${\bar V}_{0f_{7/2} 0f_{7/2}}^ {\tau \tau'}$, are denoted by 
$\Delta_{b}^{\tau \tau'}$ and reported in Table \ref{Sec-433_Tab}. It  can be seen that  the tensor  content of the  $\Delta_{\beta}^{\tau \tau'}$  is  more relevant as compared to that   of the single centroids and  the central component  loses in part its  dominant role,    all three components of the monopole interaction becoming   important in determining  the energy gaps.

In particular, as observed in Section \ref{sec-IV.3.2}, the inclusion of the 3NF brings an increase of about 2 MeV in the neutron  $0f_{7/2}- 0f_{5/2}$  spin-orbit splitting at $N=28$. As a matter of fact,  the 2NF force only leads to a decrease of the  same amount with respect to the original SP value (7.4 MeV)  that is determined by  the  almost complete balancing of the  central, vector, tensor components  of  $\Delta_{0f_{5/2}}^{\nu \nu}$   and  by the consequent  dominance  of   central monopole proton-neutron  matrix element,  whose attraction is only partially mitigated by the vector and tensor parts. The inclusion of the 3NF provides an increase of the  repulsive neutron-neutron component, arising  essentially  from  the increase of the  vector term,   that  cancels the proton-neutron contribution. A similar mechanism explains the increase of $\sim 3$ MeV  in the proton  $0f_{7/2}- 0f_{5/2}$  spin-orbit splitting resulting from the 3NF.
 
As concerns the  neutron and proton   $0f_{7/2}- 1p_{3/2}$ spacings,  a significant   growth at $N=28$  is produced already by the 2NF only  (see Figs. \ref{Sec-432_ESPENin} and \ref{Sec-432_ESPENip}).   The $\Delta_{b}^{\tau \tau'}$  values of  Table \ref{Sec-433_Tab} evidence  that it is related  to the overall positive interference of all components of the monopole terms,  while the further increase we find with  $H^{\rm 3N}$ comes  out from  the changes  that the 3NF induces in the central terms.

\section{Summary and conclusions}\label{summary}
In this review paper, we have discussed  the state of current developments to account for 3NFs within the SM framework and their impact on our understanding of nuclear structure properties. We have focused on realistic SM calculations with effective interactions derived from the QCD level and shown how crucial these forces are for describing binding energies and formation of shell structure. Particular attention has been devoted to the effects of 3NFs on the monopole component of the SM effective interaction by highlighting how relevant they are in correcting the monopole interaction derived from realistic the 2NF only. 

As widely discussed in the text, starting from 1990s it was realized that the problems arising when using effective SM Hamiltonians derived from $NN$ potentials by means of MBPT should be associated to deficiencies in its monopole component and adjustments were introduced so to obtain results of a quality comparable with the ones provided by phenomenological interactions \cite{Caurier05}. However, the connection with the lack of the 3NF was only suggested few years later \cite{Zuker03}, and subsequently the first SM study including explicitly the effects of 3NFs was carried out for the $sd$ shell nuclei \cite{Otsuka10}.

We have focused on standard  and Gamow shell models employing effective Hamiltonians derived within the MBPT approach from chiral $NN$ and 3N forces, and reported calculations for nuclei ranging from light- to intermediate-mass nuclei. In addition to the effects of genuine chiral 3NFs, we have also discussed the role of induced 3NFs due to the interaction of clusters of three-valence nucleons with configurations outside the model space via the 2NF. Results are presented in Section \ref{applications} and compared with experiment, as well as with  results from other approaches whenever possible and/or useful.  In this section,  we have also discussed in detail the interplay between 3NFs and the coupling with continuum for the description of weakly-bound states in the $sd$ region and the structure of  $^{17}$Ne by means of the realistic GSM. 

In the following,  we shall summarize the main features coming from the results presented in Section \ref{applications}.

1.  We have evidenced the validity of the MBPT in deriving the effective SM Hamiltonian by using as testing ground $p$-shell nuclei.  The perturbative behavior of $H_{\rm eff}$s derived from chiral $NN$ potentials has been assessed by showing that the convergence properties  of the $\hat Q$-box vertex function with respect to the dimension of the intermediate state space and the order-by-order convergence can be taken under control. Then, the quality of the results has been checked by benchmark calculations with the comparison of SM and ab initio NCSM results arising from chiral $NN$-only and $NN$+3N forces.

2. The detailed discussion presented  for $fp$ shell nuclei clearly emphasizes the relevance of the 3NF contribution in determining the location of the neutron dripline and the evolution of the shell structure.  We have shown that genuine 3NFs provide a repulsive interaction, which, although partially counterbalanced by induced 3N forces accounting for excitations outside the valence space, is essential to quench the overbinding of the g.s energies produced by the $NN$ force. Our results for the $S_{2n}$s and the $2^{+}_1$ excitation energies evidence that their effect increases with an increasing number of valence nucleons and is non-negligible in the formation of the shell structure, reflecting, in particular, on the closure properties of $^{58}$Ni. Furthermore, our calculations with a modified interaction, obtained by combing the multipole and monopole components of effective Hamiltonians derived, respectively, from the $NN$ and $NN$+3N potential, soundly confirm that 3NFs affect essentially the monopole component.  It has turned out that monopole component of the 3NF is crucial to correct the behavior of the ESPEs and produce the needed increase in the  $0f_{7/2}$ - $0f_{5/2}$ spin-orbit splitting at $N=28$ for both protons and neutrons. 
The main feature of the 3NF is its repulsive nature, as also evidenced in prior studies.  However, the size of the contribution depends on the involved orbitals which produces a substantial change in the spacings between the ESPEs and consequently in the variation of the shell structure in correspondence of a large occupation of a specific orbital. From our analysis based on the spin-tensor decomposition of the monopole  SM interaction, it has resulted that the  central component  - which  acquires in all channels a significant repulsive contribution from the 3NF  - is mainly responsible for the behavior of the ESPEs. On the other hand, when focusing on the spacings between ESPEs rather than on their absolute values, we have seen that also the vector and tensor components come into play, and therefore the shell structure depends on  the interplay between all the three components. 

3. Calculations performed within the GSM framework, in which continuum and resonance are included, have shown that 3NFs are non-negligible in explaining the dripline position in oxygen chain and the unbound properties of isotopes beyond the dripline as well as the Borromean structure of the proton-rich $^{17}$Ne. 
As in standard SM calculations, the 3NF gives repulsive contributions to the g.s. energies, and its effect increases with the increase of the number of valence neutrons. As a matter of fact, it becomes crucial in the neutron dripline region by pushing up the $0d_{3/2}$ and $1s_{1/2}$ orbitals, when they start to be heavily occupied. Furthermore, a dissection of 3NF clearly evidences the main role of the two-pion exchange term with respect to the one-pion and contact ones which, having the same values but opposite signs, almost cancel out. An improved agreement with experiment and theory is found even in the case of spectra and resonance widths when including the 3NF.
We have found that the repulsive contribution of the 3N force is also essential to explain the Borromean structure of $^{17}$Ne by inducing an increase in energy of $^{16}$F over the threshold of the proton emission.

Important demonstrations of the effect of 3NFs in the SM context are now available, which show their origin and importance in nuclear spectroscopy and explain the empirical modifications one should introduce in effective interactions derived from realistic $NN$ potentials.  A major step in this direction has been the derivation of nuclear potentials in terms of the chiral EFT that provides many-body forces consistently with the nature of the $NN$ interaction.  These are the so-called genuine 3N forces that originates from neglecting subnucleonic degrees of freedom.  However, it is worth mentioning that, as long as we choose to use an inert core, there are also induced 3N forces accounting for excitations outside the valence space, which gives rise to the mass dependence of effective SM interactions.

In the future, important advances for a better understanding of the role of 3NFs  within the realistic SM are the extension of calculations to heavier systems. 
Improvements in our 3NF treatment are also desirable, as the inclusion of higher-order contributions with 3N vertices in the perturbative expansion of the effective Hamiltonian and the development of a technique to account for the 3NF among valence particles.
We also would point out that completely consistent calculations require that all effective operators for general observables, not just the Hamiltonian, should be constructed from $NN+$ 3N potentials derived within the framework of chiral EFT, by including two-body meson-exchange corrections originating from subnucleonic degrees of freedom. 

In closing, it may be interesting to mention as possible development the inclusion of subleading 3NFs beyond the N$^2$LO development. 
While the two-pion-exchnage 3NFs at N$^3$LO are expected to produce rather weak effects as evidenced in the calculations for the nucleon-deuteron system of Ishikawa {\it et al.}~\cite{Ishikawa07}, the short-range 3NFs at N$^3$LO~\cite{https://doi.org/10.48550/arxiv.2302.03468} and N$^4$LO~\cite{PhysRevC.105.054004} could play a non-negligible role in few-body systems. Therefore, the subleading-short-range 3NFs should be considered to include more sizeable contributions.

\section*{Acknowledgements}
This work was supported by the Japan Society for the Promotion of Science KAKENHI under Grant No JP21K13919; the National Key R\&D Program of China under Grant No. 2018YFA0404401; the  National Natural Science Foundation of China under Grants No. 11835001, No. 11921006,  No. 12035001, No. 121051062, and No. 121051063;  the China Postdoctoral Science Foundation under Grants No. BX20200136, No. 2020M682747. The authors thank Nunzio Itaco for helpful comments and fruitful discussions.

\appendix
\section{Three-body matrix elements for shell-model calculations}\label{appendix1}
Here we show the formalism of the three-body matrix elements (MEs) of the chiral 3NF at N$^2$LO. 
First, we describe the three-body states in terms of the HO basis functions,
which enables us to easily factor out the center-of-mass (c.m.) motion of the three-body states.
Next, the antisymmetrization of the three-body states is explained. 
By introducing the Jacobi coordinates, the three-body MEs are reduced to simple forms,
which we call the Jacobi-HO MEs.
Finally, the Jacobi-HO MEs of each term of the chiral 3NF are given.
\subsection{Three-body states}\label{Sec3BMEwf}
First, we define the single-particle state $\Ket{nljm_jm_\tau}$ of a nucleon as
\begin{align}
 \ket{nljm_j m_\tau}
 &=
 \left|\left.\Phi_{nlj m_j}\right>\right. \left|\left.\varphi_{\frac{1}{2}m_\tau}^{(\tau)}\right>\right.,
 \label{spwf1}\\
 \left|\left.\Phi_{nlj m_j}\right>\right.
 &=
 \left|\left.\left[\phi_{nl}\otimes\varphi_{\frac{1}{2}}^{(\sigma)}\right]_{jm_j}\right>\right. ,
 \label{spwf2}
\end{align}
\noindent
where $\Phi_{nlj m_j}$ is specified by the principal quantum number $n$, the orbital angular momentum $l$,
and the total spin $j$. The projections to the $z$ axis in association with  $l$ and $j$ are respectively $m_{l}$ and $m_{j}$.
In Eq.~\eqref{spwf2}, the wave function in the $j$ scheme is obtained by coupling $\phi_{nlm_{l}}$ with the nucleon spin wave function $\varphi_{\frac{1}{2}m_{\sigma}}^{(\sigma)}$. 
The isospin component is expressed by $\varphi_{\frac{1}{2}m_{\tau}}^{(\tau)}$.
Here $m_{\sigma}$ and $m_{\tau}$ are the $z$ components of the nucleon spin and isospin, respectively.
The spatial wave function $\phi_{nlm_{l}}$ is expressed in terms of the HO basis functions:
\begin{align}
 \phi_{nlm_{l}}(\vect{r})=\frac{R_{nl}(r)}{r}Y_{lm_{l}}(\hat{\vect{r}}),
 \label{HOex1}
\end{align}
with the spherical harmonics $Y_{lm_{l}}$, and $\vect{r}$ specifying the position of the nucleon.
The HO-wave function $R_{nl}$ is written by
\begin{align}
 R_{nl}(r)
 =
 \left[\frac{2n!}{b_0^3\Gamma\!\left(l+n+\frac{3}{2}\right)}\right]^{\frac{1}{2}}
 r\left(\frac{r}{b_0}\right)^{l}
 \exp\!\left[-\!\left(\frac{r}{\sqrt{2}b_0}\right)^2\right]
 L_{n}^{l+\frac{1}{2}}\!\left(\frac{r^2}{b_0^2}\right),
 \label{HOsolution}
\end{align}
The so-called size parameter is expressed by $b_0=\sqrt{\hbar/(m_N\omega)}$ 
with the HO-angular velocity $\omega$ and the average nucleon mass $m_N$, 
while $\Gamma$ and $L_{n}^{l+\frac{1}{2}}$ are the gamma function and the Laguerre polynomial, respectively.

Next, we consider three interacting nucleons.
In the following, we explicitly put the subscripts of the quantum numbers to distinguish the particles, $a$, $b$, and $c$.
We symbolically express the single-particle quantum numbers as
$a=\{n_a,l_a,j_a\}$.
Thus, in the $jj$-coupling scheme, the product of the three single-particle wave functions reads
\begin{align}
 &\Ket{\left(a,b\right),c;J_{12}JT_{12}T}
 \nonumber\\
 &\quad=
 (-)^{J}\hat{j}_a\hat{j}_b\hat{j}_c\hat{J}_{12}
 \sum_{\substack{L_{12}S_{12}\\LS}}(-)^{L+S}
 \hat{L}_{12}^2\hat{S}_{12}\hat{L}^2\hat{S}^2
 \begin{Bmatrix}
  l_a    & \frac{1}{2} & j_a \\[5pt]
  l_b    & \frac{1}{2} & j_b \\[5pt]
  L_{12} & S_{12}      & J_{12}
 \end{Bmatrix}
 \begin{Bmatrix}
  L_{12} & S_{12}      & J_{12} \\[5pt]
  l_c    & \frac{1}{2} & j_c    \\[5pt]
  L      & S           & J
 \end{Bmatrix}
 \nonumber\\
 &\quad\times
 \sum_{\substack{n_{12}l_{12}\\\mathcal{N}_{12}\mathcal{L}_{12}}}
 (-)^{l_{12}+l_c+L_{12}}
 \left<\!\Braket{\mathcal{N}_{12}\mathcal{L}_{12}n_{12}l_{12},L_{12}|n_a l_a n_b l_b,L_{12}}\!\right>_{d_1}
 \sum_{\mathcal{J}}(-)^{\mathcal{J}} \hat{\mathcal{J}}^2
  \begin{Bmatrix}
  l_{12} & \mathcal{L}_{12} & L_{12} \\
  l_c    & L                & \mathcal{J}
  \end{Bmatrix}
 \nonumber\\
 &\quad\times
 \sum_{\substack{nl\\\mathcal{N}\mathcal{L}}}
 \left<\!\Braket{\mathcal{N}\mathcal{L}nl,\mathcal{J}|\mathcal{N}_{12}\mathcal{L}_{12} n_cl_c,\mathcal{J}}\!\right>_{d_2}
 \sum_{\substack{\mathcal{K} \mathcal{I}\\I_{12}I}}\hat{\mathcal{K}}^2\hat{\mathcal{I}}\hat{I}_{12}\hat{I}
 \begin{Bmatrix}
  \mathcal{L} & l & \mathcal{J} \\
  l_{12}      & L & \mathcal{K}
 \end{Bmatrix}
 \begin{Bmatrix}
  \mathcal{L} & \mathcal{K} & L \\
  S           & J       & \mathcal{I}
 \end{Bmatrix}
 \begin{Bmatrix}
  l_{12} & l           & \mathcal{K} \\[5pt]
  S_{12} & \frac{1}{2} & S       \\[5pt]
  I_{12} & I           & \mathcal{I}
 \end{Bmatrix}
 \nonumber\\
 &\quad\times
 \sum_{\mathcal{M}_{\mathcal{L}}\mathcal{M}_\mathcal{I}}\left(\mathcal{L} \mathcal{M}_{\mathcal{L}} \mathcal{I} \mathcal{M}_\mathcal{I} | J M_J \right)
 \Ket{\phi_{\mathcal{N}\mathcal{L}\mathcal{M}_\mathcal{L}}}
 \Ket{(n_{12}l_{12}S_{12})I_{12}T_{12}\left(nl\right)I;\mathcal{I}T},
 \label{wfjj3B}
\end{align}
where $\Ket{(n_{12}l_{12}S_{12})I_{12}T_{12}\left(nl\right)I;\mathcal{I}T}$ is the Jacobi-HO state defined by
\begin{align}
 &\Ket{(n_{12}l_{12}S_{12})I_{12}T_{12}\left(nl\right)I;\mathcal{I}T}
 \nonumber\\
 &\quad=
 \Ket{
 \left[\left[\phi_{n_{12}l_{12}}\otimes 
 \left[\varphi_{\frac{1}{2}}^{(\sigma)}\otimes\varphi_{\frac{1}{2}}^{(\sigma)}\right]_{S_{12}}\right]_{I_{12}}
 \otimes
 \left[
 \phi_{nl}
 \otimes
 \varphi^{(\sigma)}_{\frac{1}{2}}
 \right]_{I}
 \right]_{\mathcal{I}\mathcal{M}_\mathcal{I}}}
 \nonumber\\
 &\quad\times
 \Ket{\left[\left[\varphi_{\frac{1}{2}}^{(\tau)}\otimes\varphi_\frac{1}{2}^{(\tau)}\right]_{T_{12}}\otimes\varphi_\frac{1}{2}^{(\tau)}\right]_{TM_{T}}}.
 \label{wfJacobiHO}
\end{align}
The total angular momentum (total isospin) and its projection to the $z$ axis
are represented by $J$ and $M_J$ ($T$ and $M_T$), respectively.
We also introduce $J_{12}$ ($T_{12}$), 
which are obtained by coupling $j_a$ and $j_b$ (the isospins of $a$ and $b$). 
The harmonic-oscillator bracket, $\left<\!\Braket{\cdots|\cdots}\!\right>_{d_n}$,
originates from the Talmi transformation~\cite{Talmi52,Brody60,Moshinsky60},
and its explicit expression is given, for example, in Refs.~\cite{TrlifajPhysRevC.5.1534,BUCK1996387,KAMUNTAVICIUS2001191}.
The subscript $d_n$ specifies the mass ratio relevant to the Talmi transformation of the $n$-body system, namely, $d_1=1$ and $d_2=2$.
One can see that the c.m. motion described by $\Ket{\phi_{\mathcal{N}\mathcal{L}\mathcal{M}_\mathcal{L}}}$ is factored out.

\subsection{Antisymmetrization}\label{Sec3BMEAntisym}
The three-body antisymmetrizer is given by
\begin{align}
 \hat{\mathcal{A}}_{3}
 =
 \frac{1}{3!}
 \left[
 \mathbb{1}-\hat{\mathcal{P}}_{ab}-\hat{\mathcal{P}}_{bc}-\hat{\mathcal{P}}_{ca}
 +\hat{\mathcal{P}}_{ab}\hat{\mathcal{P}}_{bc}+\hat{\mathcal{P}}_{ab}\hat{\mathcal{P}}_{ca}
 \right],
 \label{OpeAntisym3}
\end{align}
where $\mathbb{1}$ is the unity operator and $\hat{\mathcal{P}}_{ab}$ is the permutation operator with respect to the particles $a$ and $b$.
Using Eq.~\eqref{wfjj3B}, the antisymmetrized $jj$-coupled state reads
\begin{align}
 &\Ket{\left(a,b\right),c;J_{12}JT_{12}T}_A
 \nonumber\\
 &\quad=
 \sqrt{6}\hat{\mathcal{A}}_3\Ket{\left(a,b\right),c;J_{12}JT_{12}T}
 \nonumber\\
 &\quad=
 (-)^{J}\hat{j}_a\hat{j}_b\hat{j}_c\hat{J}_{12}
 \sum_{\substack{L_{12}S_{12}\\LS}}(-)^{L+S}
 \hat{L}_{12}^2\hat{S}_{12}\hat{L}^2\hat{S}^2
 \begin{Bmatrix}
  l_a    & \frac{1}{2} & j_a \\[5pt]
  l_b    & \frac{1}{2} & j_b \\[5pt]
  L_{12} & S_{12}      & J_{12}
 \end{Bmatrix}
 \begin{Bmatrix}
  L_{12} & S_{12}      & J_{12} \\[5pt]
  l_c    & \frac{1}{2} & j_c    \\[5pt]
  L      & S           & J
 \end{Bmatrix}
 \nonumber\\
 &\quad\times
 \sum_{\substack{n_{12}l_{12}\\\mathcal{N}_{12}\mathcal{L}_{12}}}
 (-)^{l_{12}+l_c+L_{12}}
 \left<\!\Braket{\mathcal{N}_{12}\mathcal{L}_{12}n_{12}l_{12},L_{12}|n_a l_a n_b l_b,L_{12}}\!\right>_{d_1}
 \sum_{\mathcal{J}}(-)^{\mathcal{J}} \hat{\mathcal{J}}^2
  \begin{Bmatrix}
  l_{12} & \mathcal{L}_{12} & L_{12} \\
  l_c    & L                & \mathcal{J}
  \end{Bmatrix}
 \nonumber\\
 &\quad\times
 \sum_{\substack{nl\\\mathcal{N}\mathcal{L}}}
 \left<\!\Braket{\mathcal{N}\mathcal{L}nl,\mathcal{J}|\mathcal{N}_{12}\mathcal{L}_{12} n_cl_c,\mathcal{J}}\!\right>_{d_2}
 \sum_{\substack{\mathcal{K} \mathcal{I}\\I_{12}I}}\hat{\mathcal{K}}^2\hat{\mathcal{I}}\hat{I}_{12}\hat{I}
 \begin{Bmatrix}
  \mathcal{L} & l & \mathcal{J} \\
  l_{12}      & L & \mathcal{K}
 \end{Bmatrix}
 \begin{Bmatrix}
  \mathcal{L} & \mathcal{K} & L \\
  S           & J       & \mathcal{I}
 \end{Bmatrix}
 \begin{Bmatrix}
  l_{12} & l           & \mathcal{K} \\[5pt]
  S_{12} & \frac{1}{2} & S       \\[5pt]
  I_{12} & I           & \mathcal{I}
 \end{Bmatrix}
 \nonumber\\
 &\quad\times
 \sum_{\mathcal{M}_{\mathcal{L}}\mathcal{M}_\mathcal{I}}\left(\mathcal{L} \mathcal{M}_{\mathcal{L}} \mathcal{I} \mathcal{M}_\mathcal{I} | J M_J \right)
 \Ket{\phi_{\mathcal{N}\mathcal{L}\mathcal{M}_\mathcal{L}}}
 \Ket{i_1;\mathcal{I}T}_A,
 \label{antisymjjWF1}
\end{align}
with
\begin{align}
 \Ket{i;\mathcal{I}T}_A
 =
 \sqrt{6}\hat{\mathcal{A}}_3
 \Ket{i;\mathcal{I}T},
 \label{antisymJacHO1}
\end{align}
where the Jacobi-HO state $\Ket{i;\mathcal{I}T}$ is defined by Eq.~\eqref{wfJacobiHO},
and we simplify the set of quantum numbers:
\begin{align}
 i = \left\{n_{12},l_{12},S_{12},I_{12},T_{12},n,l,I\right\}.
 \label{simplei0}
\end{align}

Now our task is to antisymmetrize the Jacobi-HO states.
To this end, we expand the antisymmetrizer $\hat{\mathcal{A}}_3$ using the spectral decomposition as
\begin{align}
 \hat{\mathcal{A}}_3
 =
 \sum_\eta \epsilon_{\eta}
 \Ket{\eta}\Bra{\eta}
 \label{SpExp1},
\end{align}
where $\Ket{\eta}$ are the eigenfunctions of $\hat{\mathcal{A}}_3$
characterized by a quantum number $\eta$.
The eigenvalue $\epsilon_\eta$ should be 0 or 1 because the antisymmetrizer $\hat{\mathcal{A}}_3$ is idempotent.
The eigenstates corresponding to $\epsilon_\eta=1$ form physical antisymmetrized states,
while the other eigenstates with $\epsilon_\eta=0$ give spurious states~\cite{Navratil99,Navratil00c}.
By selecting the physical states only, Eq.~\eqref{antisymJacHO1} can be written as
\begin{align}
 \Ket{i;\mathcal{I}T}_A
 &=
\sqrt{6}
 \sum_j
 D_{ij}^{(\mathcal{I}T)}
 \Ket{j;\mathcal{I}T},
 \label{antisymJacHO3}\\
 D_{ij}^{(\mathcal{I}T)}
 &=
 \sum_{\eta}^{N_{\rm P}}
 C_{\eta}^{i(\mathcal{I}T)}
 C_{\eta}^{j(\mathcal{I}T)*},
 \label{Dcoeff1}\\
 C_{\eta}^{i(\mathcal{I}T)}
 &=
 \Braket{\eta| i;\mathcal{I}T},
 \label{Ccoeff1}
\end{align}
with
\begin{align}
 j=\left\{n_{12}',l_{12}',S_{12}',I_{12}',T_{12}',n',l',I'\right\}.
 \label{simplej0}
\end{align}
Here, $\Ket{\eta}$ is expanded in terms of the partially antisymmetrized states,
i.e., the antisymmetrization only for the first two particles ($ab$) is considered.
Thus, the condition $(-)^{l_{12}'+S_{12}'+T_{12}'}=-1$ is always satisfied.
How to evaluate the number of the physical states $N_{\rm P}$ is given later.

The coefficient $C_{\eta}^{i(\mathcal{I}T)}$ is computed as follows.
Since $\Ket{\eta}$ satisfies the eigenvalue equation,
\begin{align}
 \left(\hat{\mathcal{A}}_3-\epsilon_\eta\right)\Ket{\eta}
 =0,
 \label{eigenEq3}
\end{align}
we obtain
\begin{align}
 \sum_j
 \Braket{i;\mathcal{I}T
 \left|\left(\hat{\mathcal{A}}_3-\epsilon_\eta\right)\right|
 j;\mathcal{I}T}
 C_{\eta}^{j(\mathcal{I}T)*}
 =0.
 \label{eigenEq4}
\end{align}
Thus, $C_{\eta}^{j(\mathcal{I}T)*}$ is obtained by diagonalizing the 
antisymmetrizer matrix, the matrix element of which is given by
\begin{align}
 \mathcal{A}_{ij}
 &=
 \Braket{i;\mathcal{I}T
 \left|\hat{\mathcal{A}}_3\right|
 j;\mathcal{I}T}
 =
 \frac{1}{3}
 \Braket{i;\mathcal{I}T
 \left|\left(\mathbb{1}-2\hat{\mathcal{P}}_{bc}\right)\right|
 j;\mathcal{I}T},
 \label{Amatij1}
\end{align}
and
\begin{align}
 \Braket{i;\mathcal{I}T \left|\hat{\mathcal{P}}_{bc}\right| j;\mathcal{I}T}
 &=
 (-)^{l_{12}+l_{12}'}
 \hat{I}_{12}\hat{I}_{12}'\hat{I}\hat{I}'
 \hat{S}_{12}\hat{S}_{12}'\hat{T}_{12}\hat{T}_{12}'
 \begin{Bmatrix}
  \frac{1}{2} & \frac{1}{2} & T_{12} \\[5pt]
  \frac{1}{2} & T           & T_{12}'
 \end{Bmatrix}
 \nonumber\\
 &\times
 \sum_{\substack{\lambda\sigma}}
 \hat{\lambda}^2\hat{\sigma}^2
 \begin{Bmatrix}
  \frac{1}{2} & \frac{1}{2} & S_{12} \\[5pt]
  \frac{1}{2} & \sigma      & S_{12}'
 \end{Bmatrix}
 \begin{Bmatrix}
  l_{12}  & S_{12}      & I_{12} \\[5pt]
  l       & \frac{1}{2} & I      \\[5pt]
  \lambda & \sigma      & \mathcal{I}
 \end{Bmatrix}
 \begin{Bmatrix}
  l_{12}'  & S_{12}'     & I_{12}' \\[5pt]
  l'       & \frac{1}{2} & I'      \\[5pt]
  \lambda  & \sigma      & \mathcal{I}
 \end{Bmatrix}
 \nonumber\\
 &\times
 \left<\!\Braket{n_{12}l_{12}nl,\lambda|n_{12}'l_{12}'n'l',\lambda}\!\right>_{d_3}.
 \label{MEPbc3}
\end{align}
Note that the Jacobi-HO states are orthonormal:
$\Braket{i;\mathcal{I}T | j;\mathcal{I}T}=\delta_{ij}$.
The mass ratio in the harmonic-oscillator bracket is now $d_3=1/3$.
The number of the physical states $N_{\rm P}$ is given as a sum of the eigenvalues, 
i.e., the trace of the antisymmetrizer matrix after the diagonalization.

\subsection{Structures of three-body matrix elements}\label{Sec3BMESt}
\subsubsection{$JT$-coupled three-body matrix elements}\label{Sec3BMEStJT}
From Eq.~\eqref{antisymjjWF1}, the antisymmetrized $JT$-coupled matrix elements for the three-body interaction $V_{3N}$ is given by
\begin{align}
 &{}_{\substack{\\A}\!}\Braket{\left(d,e\right),f;J_{12}'JT_{12}'T
 \left|V_{3N}\right|\left(a,b\right),c;J_{12}JT_{12}T}_{A}
 \nonumber\\
 &\qquad=
 \sum_{\substack{n_{12}l_{12}S_{12}I_{12}\\nlI}}
 \sum_{\substack{n_{12}'l_{12}'S_{12}'I_{12}'\\n'l'I'}}
 \sum_{\mathcal{I}}\,
 {}_{\substack{\\A}\!}\Braket{\kappa';\mathcal{I}T \left|V_{3N}\right|\kappa;\mathcal{I}T}_A
 \nonumber\\
 &\qquad\times
 \sum_{\mathcal{N}\mathcal{L}}
 T_{abcJ_{12}J\mathcal{I}\mathcal{N}\mathcal{L}}^{n_{12}l_{12}S_{12}I_{12}nlI}
 ~
 T_{defJ_{12}'J\mathcal{I}\mathcal{N}\mathcal{L}}^{n_{12}'l_{12}'S_{12}'I_{12}'n'l'I'},
 \label{A3bME1}
\end{align}
where the coefficient $T_{abcJ_{12}J\mathcal{I}\mathcal{N}\mathcal{L}}^{n_{12}l_{12}S_{12}I_{12}nlI}$,
called the $T$ coefficient~\cite{PhysRevC.73.064002,PhysRevC.90.024325}, is defined by
\begin{align}
 T_{abcJ_{12}J\mathcal{I}\mathcal{N}\mathcal{L}}^{n_{12}l_{12}S_{12}I_{12}nlI}
 &=
 (-)^{l_c+l_{12}+J}
 \hat{j}_a\hat{j}_b\hat{j}_c\hat{J}_{12}\hat{S}_{12}\hat{I}_{12}\hat{I}\hat{\mathcal{I}}
 \sum_{L_{12}LS\mathcal{J}}(-)^{L_{12}+L+S+\mathcal{J}}
 \hat{L}_{12}^2\hat{L}^2\hat{S}^2\hat{\mathcal{J}}^2
 \nonumber\\
 &\times
 \begin{Bmatrix}
  l_{12} & \mathcal{L}_{12} & L_{12}   \\
  l_c    & L                & \mathcal{J}
 \end{Bmatrix}
 \begin{Bmatrix}
  l_a    & \frac{1}{2} & j_a \\[5pt]
  l_b    & \frac{1}{2} & j_b \\[5pt]
  L_{12} & S_{12}      & J_{12}
 \end{Bmatrix}
 \begin{Bmatrix}
  L_{12} & S_{12}      & J_{12} \\[5pt]
  l_c    & \frac{1}{2} & j_c    \\[5pt]
  L      & S           & J
 \end{Bmatrix}
 \nonumber\\
 &\times
 \sum_{\mathcal{N}_{12}\mathcal{L}_{12}}
 \left<\!\Braket{\mathcal{N}_{12}\mathcal{L}_{12}n_{12}l_{12},L_{12}|n_a l_a n_b l_b,L_{12}}\!\right>_{d_1}
 \left<\!\Braket{\mathcal{N}\mathcal{L}nl,\mathcal{J}|\mathcal{N}_{12}\mathcal{L}_{12} n_cl_c,\mathcal{J}}\!\right>_{d_2}
 \nonumber\\
 &\times
 \sum_{\mathcal{K}}
 \hat{\mathcal{K}}^2
 \begin{Bmatrix}
  \mathcal{L} & l & \mathcal{J} \\
  l_{12}      & L & \mathcal{K}
 \end{Bmatrix}
 \begin{Bmatrix}
  \mathcal{L} & \mathcal{K} & L \\
  S           & J       & \mathcal{I}
 \end{Bmatrix}
 \begin{Bmatrix}
  l_{12} & l           & \mathcal{K} \\[5pt]
  S_{12} & \frac{1}{2} & S       \\[5pt]
  I_{12} & I           & \mathcal{I}
 \end{Bmatrix}.
 \label{Tcoef}
\end{align}
Note that Eq.~\ref{Tcoef} can be simplified by using the 12-$j$ symbol of the first kind~\cite{PhysRevC.105.014302}.
Through diagonalizing the antisymmetrizer, the antisymmetrized Jaobi-HO matrix element
is expressed by
\begin{align}
 {}_{\substack{\\A}\!}\Braket{\kappa';\mathcal{I}T \left|V_{3N}\right|\kappa;\mathcal{I}T}_A
 =
 6\sum_{\bar\kappa\bar\kappa'}D_{\kappa\bar\kappa}^{(\mathcal{I}T)}D_{\kappa'\bar\kappa'}^{(\mathcal{I}T)}
 \Braket{\bar\kappa';\mathcal{I}T \left|V_{3N}\right|\bar\kappa;\mathcal{I}T}.
 \label{A3bMERed1}
\end{align}
The indices $\kappa$, $\bar\kappa$, $\kappa'$, and $\bar\kappa'$, respectively, stand for
\begin{align}
 \kappa &= \left\{n_{12},l_{12},S_{12},I_{12},T_{12},n,l,I\right\}
 \qquad\quad
 \Bigl((-)^{l_{12}+S_{12}+T_{12}}=-1\Bigr),
 \label{indexp0}\\
 \bar\kappa &= \left\{\bar n_{12},\bar l_{12},\bar S_{12},\bar I_{12},\bar T_{12},\bar n,\bar l,\bar I\right\}
 \qquad\quad
 \Bigl((-)^{\bar l_{12}+\bar S_{12}+\bar T_{12}}=-1\Bigr),
 \label{indexp}\\
 \kappa' &= \left\{n_{12}',l_{12}',S_{12}',I_{12}',T_{12}',n',l',I'\right\}
 \qquad\quad
 \Bigl((-)^{l_{12}'+S_{12}'+T_{12}'}=-1\Bigr),
 \label{indexq0}\\
 \bar\kappa' &= \left\{\bar n_{12}',\bar l_{12}',\bar S_{12}',\bar I_{12}',\bar T_{12}',\bar n',\bar l',\bar I'\right\}
 \qquad\quad
 \Bigl((-)^{\bar l_{12}'+\bar S_{12}'+\bar T_{12}'}=-1\Bigr),
 \label{indexq}
\end{align}
where $\bar\kappa$ and $\bar\kappa'$ are the quantum numbers originating from the expansion by Eq.~\eqref{antisymJacHO3}.

Once the three-body MEs are obtained withing the $JT$-coupled scheme by Eq.~\eqref{A3bME1}, those within the proton-neutron formalism can be constructed via
\begin{align}
{}_{\substack{\\A}\!}\Braket{\left(d,e\right),f;J_{12}'J
 \left|V_{3N}\right|\left(a,b\right),c;J_{12}J}_{A}
&=
 \sum_{TT_{12}T_{12}'}
 \left.\left(\frac{1}{2} m_{\tau_a} \frac{1}{2} m_{\tau_b} \right| T_{12} M_{T_{12}} \right)
 \left.\left(T_{12} M_{T_{12}} \frac{1}{2} m_{\tau_c} \right| T M_{T} \right)
 \nonumber\\
 &\times
 \left.\left(\frac{1}{2} m_{\tau_d} \frac{1}{2} m_{\tau_e} \right| T_{12}' M_{T_{12}}' \right)
 \left.\left(T_{12}' M_{T_{12}}' \frac{1}{2} m_{\tau_f} \right| T M_{T} \right)
 \nonumber\\
 &\times
{}_{\substack{\\A}\!}\Braket{\left(d,e\right),f;J_{12}'JT_{12}'T
 \left|V_{3N}\right|\left(a,b\right),c;J_{12}JT_{12}T}_{A}.
 \label{pn3BMEs}
\end{align}

\subsubsection{Chiral three-body potentials and nonlocal regularization}\label{Sec3BMEStNLReg}
The three-nucleon force appears at N$^2$LO of the chiral EFT, as explained in Sec.~\ref{sec-II.2}.
It consists of the two-pion exchange (2PE) term, the one-pion exchange plus the two-body contact (1PE) term,
and the three-body contact term, as shown in Fig.~\ref{n2lo3f}.
In the momentum space, the potential $v_{3N}$ of the operator $V_{3N}$ is
explicitly given by Eqs.~\eqref{eq_3nf_nnloa} to~\eqref{eq_3nf_nnlod}.

Now we can show that the antisymmetrized three-body MEs can be simplified by
\begin{align}
 {}_{\substack{\\A}\!}\Braket{\kappa';\mathcal{I}T \left|V_{3N}\right|\kappa;\mathcal{I}T}_A
 &=
 3 \,{}_{\substack{\\\\A}\!}\Braket{\kappa';\mathcal{I}T
 \left|W_{3N}\right|\kappa;\mathcal{I}T}_A
\nonumber\\
 &=18\sum_{\bar\kappa\bar\kappa'}D_{\kappa\bar\kappa}^{(\mathcal{I}T)}D_{\kappa'\bar\kappa'}^{(\mathcal{I}T)}
 \Braket{\bar\kappa';\mathcal{I}T \left|W_{3N}\right|\bar\kappa;\mathcal{I}T}.
 \label{3NMEsimple1}
\end{align}
This is owing to the symmetry of the three-nucleon force with respect to the permutation of particles.
In the following sections, the explicit form of $\Braket{\bar\kappa';\mathcal{I}T \left|W_{3N}\right|\bar\kappa;\mathcal{I}T}$ is presented.
The reduced operator $W_{3N}$ is defined by
\begin{align}
 \Braket{\vect{p}_a',\vect{p}_b',\vect{p}_c'\left|W_{3N}\right|\vect{p}_a,\vect{p}_b,\vect{p}_c}
 &=
 w_{3N}\left(\vect{q}_a,\vect{q}_b,\vect{q}_c\right)\delta\!\left(\vect{q}_a+\vect{q}_b+\vect{q}_c\right),
 \label{MEv3N2}\\
 w_{3N}\left(\vect{q}_a,\vect{q}_b,\vect{q}_c\right)
 &=w_{3N}^{(2\pi)}\left(\vect{q}_a,\vect{q}_b,\vect{q}_c\right)
 +w_{3N}^{(1\pi)}\left(\vect{q}_a,\vect{q}_b,\vect{q}_c\right)
 +w_{3N}^{\rm (ct)}\left(\vect{q}_a,\vect{q}_b,\vect{q}_c\right).
 \label{chiral3bSum2}
\end{align}
The two-pion exchange potential $w_{3N}^{(2\pi)}$ is given by
\begin{align}
 w_{3N}^{(2\pi)}\left(\vect{q}_a,\vect{q}_b,\vect{q}_c\right)
 &=
 w_{3N}^{(2\pi;c_1)}\left(\vect{q}_a,\vect{q}_b,\vect{q}_c\right)
 +w_{3N}^{(2\pi;c_3)}\left(\vect{q}_a,\vect{q}_b,\vect{q}_c\right)
 +w_{3N}^{(2\pi;c_4)}\left(\vect{q}_a,\vect{q}_b,\vect{q}_c\right),
 \label{w2pi1}\\
 w_{3N}^{(2\pi;c_1)}\left(\vect{q}_a,\vect{q}_b,\vect{q}_c\right)
 &=
 -\frac{1}{(2\pi)^6}\frac{g_A^2c_1m_\pi^2}{f_\pi^4}
 \frac{\left(\vect{\sigma}_b\cdot\vect{q}_b\right)\left(\vect{\sigma}_c\cdot\vect{q}_c\right)}
 {\left(q_b^2+m_\pi^2\right)\left(q_c^2+m_\pi^2\right)}
 \vect{\tau}_b\cdot\vect{\tau}_{c},
 \label{w2pic1}\\
 w_{3N}^{(2\pi;c_3)}\left(\vect{q}_a,\vect{q}_b,\vect{q}_c\right)
 &=
 \frac{1}{(2\pi)^6}\frac{g_A^2c_3}{2f_\pi^4}
 \frac{\left(\vect{\sigma}_b\cdot\vect{q}_b\right)\left(\vect{\sigma}_c\cdot\vect{q}_c\right)}
 {\left(q_b^2+m_\pi^2\right)\left(q_c^2+m_\pi^2\right)}
 \left(\vect{q}_b\cdot\vect{q}_c\right)
 \left(\vect{\tau}_b\cdot\vect{\tau}_{c}\right),
 \label{w2pic3}\\
 w_{3N}^{(2\pi;c_4)}\left(\vect{q}_a,\vect{q}_b,\vect{q}_c\right)
 &=
 \frac{1}{(2\pi)^6}\frac{g_A^2c_4}{4f_\pi^4}
 \frac{\left(\vect{\sigma}_b\cdot\vect{q}_b\right)\left(\vect{\sigma}_c\cdot\vect{q}_c\right)}
 {\left(q_b^2+m_\pi^2\right)\left(q_c^2+m_\pi^2\right)}
 \left\{\left(\vect{q}_b\times\vect{q}_c\right)\cdot\vect{\sigma}_{a}\right\}
 \left\{\left(\vect{\tau}_b\times\vect{\tau}_c\right)\cdot\vect{\tau}_{a}\right\},
 \label{w2pic4}
\end{align}
while the other potentials $w_{3N}^{(1\pi)}$ and $w_{3N}^{\rm (ct)}$ are written as
\begin{align}
 w_{3N}^{(1\pi)}\left(\vect{q}_a,\vect{q}_b,\vect{q}_c\right)
 &=
 -\frac{1}{(2\pi)^6}\frac{g_Ac_D}{4f_\pi^4\Lambda_\chi}
 \frac{\left(\vect{\sigma}_c\cdot\vect{q}_c\right)\left(\vect{\sigma}_b\cdot\vect{q}_c\right)}{q_c^2+m_\pi^2}\vect{\tau}_b\cdot\vect{\tau}_c,
 \label{w1pi1}\\
 w_{3N}^{\rm (ct)}\left(\vect{q}_a,\vect{q}_b,\vect{q}_c\right)
 &=
 \frac{1}{(2\pi)^6}\frac{c_E}{f_\pi^4\Lambda_\chi}
 \vect{\tau}_a\cdot\vect{\tau}_b.
 \label{wct1}
 \end{align}
Note that $w_{3N}$ is one component of Eqs.~\eqref{eq_3nf_nnloa} to~\eqref{eq_3nf_nnlod}.
Also pay attention that our potential $w_{3N}$ contains a prefactor $1/(2\pi)^6$,
which originates from our convention of the normalization:${\Braket{\vect{p}_a' | \vect{p}_a}=\delta(\vect{q}_a)}$.
See Refs.~\cite{Navratil07b,Coon81} for more details.

It is convenient to define the Jacobi-HO momenta as
\begin{align}
 \vect{k}=\frac{1}{\sqrt{2}}\left(\vect{p}_a-\vect{p}_b\right),\quad
 \vect{K}=\sqrt{\frac{2}{3}}\left[\frac{1}{2}\left(\vect{p}_a+\vect{p}_b\right)-\vect{p}_c\right],\quad
 \vect{K}_0=\frac{1}{\sqrt{3}}\left(\vect{p}_a+\vect{p}_b+\vect{p}_c\right).
 \label{Jacobimom1}
 \end{align}
Then, one finds that the transferred momenta are expressed with the Jacobi-HO momenta:
\begin{align}
 \vect{q}_a
 &=\frac{1}{\sqrt{3}}\left(\vect{K}_0'-\vect{K}_0\right)
 +\frac{1}{\sqrt{2}}\left(\vect{k}'-\vect{k}\right)
 +\frac{1}{\sqrt{6}}\left(\vect{K}'-\vect{K}\right),
 \label{momtraJacobi1-2}\\
 \vect{q}_b
 &=\frac{1}{\sqrt{3}}\left(\vect{K}_0'-\vect{K}_0\right)
 -\frac{1}{\sqrt{2}}\left(\vect{k}'-\vect{k}\right)
 +\frac{1}{\sqrt{6}}\left(\vect{K}'-\vect{K}\right),
 \label{momtraJacobi2-2}\\
 \vect{q}_c
 &=\frac{1}{\sqrt{3}}\left(\vect{K}_0'-\vect{K}_0\right)
 -\sqrt{\frac{2}{3}}\left(\vect{K}'-\vect{K}\right).
 \label{momtraJacobi3-2}
\end{align}
Thus, the operator $W_{3N}$ is given as
\begin{align}
 W_{3N}
 &=
\int\!\!\!\!\int\!\!\!\!\int\!\!\!\!\int\!\!\!\!\int\!\!\!\!\int
 d\vect{k} d\vect{K} d\vect{K}_0 d\vect{k}' d\vect{K}' d\vect{K}_0'
 \Ket{\vect{k}',\vect{K}',\vect{K}_0'}
 \nonumber\\
 &\quad\times
 w_{3N}\left(\vect{q}_a,\vect{q}_b,\vect{q}_c\right)\delta\!\left(\sqrt{3}\left[\vect{K}_0'-\vect{K}_0\right]\right)
 \Bra{\vect{k},\vect{K},\vect{K}_0}
 \nonumber\\
 &=
 \frac{1}{\left(\sqrt{3}\right)^3}
 \int\!\!\!\!\int\!\!\!\!\int\!\!\!\!\int
 d\vect{k} d\vect{K} d\vect{k}' d\vect{K}'
 \Ket{\vect{k}',\vect{K}'} w_{3N}^{(\rm c.m.)}\left(\vect{q}_a,\vect{q}_b,\vect{q}_c\right) \Bra{\vect{k},\vect{K}}.
 \label{OpeW3N1}
\end{align}
Here $w_{3N}^{(\rm c.m.)}\left(\vect{q}_a,\vect{q}_b,\vect{q}_c\right)=
\left.w_{3N}\left(\vect{q}_a,\vect{q}_b,\vect{q}_c\right)\right|_{\vect{K}_0=\vect{K}_0'}$.
Below we adopt always the condition $\vect{K}_0=\vect{K}_0'$, and therefore, we omit the superscript ${(\rm c.m.)}$
from $w_{3N}$ for simplicity.
It should be paid attention that, in Eq.~\eqref{OpeW3N1}, we have the prefactor $1/\left(\sqrt{3}\right)^3$ originating
from the delta function in Eq.~\eqref{MEv3N2}.

Since the chiral EFT is valid only in the low-momentum region, we introduce the regularization to suppress the high-momentum component of the potential.
In our approach, the nonlocal regulator~\cite{Epelbaum2006654} depending on the sum of the Jacobi momenta is employed.
Note that, consistently, we employ the chiral two-nucleon potential nonlocally regularized~\cite{Entem03,Machleidt11}.
Moreover, in the nonlocal regularization, there is an advantage
over the local regularization that
the Fierz rearrangement freedom holds exactly~\cite{Lynn16,Huth96,Piarulli20}.
The nonlocal regulator has the form,
\begin{align}
 u_{\nu_0}\!\left(k,K,\Lambda_0\right)&=
 \exp\!\left[-\!\left(\frac{k^2+K^2}{2\Lambda_0^2}\right)^{\!\!\nu_0}\right].
 \label{nonlocreg}
\end{align}
The cutoff momentum $\Lambda_0$ and the power $\nu_0$ must be fixed consistently with the LECs, $c_D$ and $c_E$.
Thus, the nonlocally regularized potential reads
\begin{align}
 w_{3N}(\vect{q}_a,\vect{q}_b,\vect{q}_c)
 &\to
 u_{\nu_0}\!\left(k',K',\Lambda_0\right)
 w_{3N}(\vect{q}_a,\vect{q}_b,\vect{q}_c)
 u_{\nu_0}\!\left(k,K,\Lambda_0\right).
 \label{2piPotReg}
\end{align}

\subsubsection{Contact term}\label{Sec3BMECt}
Here, we formulate the Jacobi-HO matrix element $\Braket{\bar\kappa';\mathcal{I}T \left|W_{3N}^{\rm (ct)}\right|\bar\kappa;\mathcal{I}T}$ of the contact term.
Using~\eqref{OpeW3N1}, we obtain 
\begin{align}
 &\Braket{\bar\kappa';\mathcal{I}T \left|W_{3N}^{\rm (ct)}\right|\bar\kappa;\mathcal{I}T}
 \nonumber\\
 &\quad=
 \frac{1}{2\sqrt{3}\pi^4}\frac{c_E}{f_\pi^4\Lambda_\chi}
 \delta_{\bar l_{12}0} \delta_{\bar l0} \delta_{\bar l_{12}'0} \delta_{\bar l'0}
 \delta_{\bar S_{12}\bar I_{12}} \delta_{\bar S_{12}' \bar I_{12}'} \delta_{\bar I \frac{1}{2}} \delta_{\bar I'\frac{1}{2}}
 \delta_{\bar S_{12}\bar S_{12}'}\delta_{\bar T_{12}\bar T_{12}'}
 (-)^{\bar T_{12}+1}
 \begin{Bmatrix}
  \frac{1}{2} & \frac{1}{2} & \bar T_{12} \\[5pt]
  \frac{1}{2} & \frac{1}{2} & 1
 \end{Bmatrix}
 \nonumber\\
 &\quad\times
 \int\!\!\!\!\int dk dK k K P_{\bar n_{12}0}\!\left(k\right) P_{\bar n0}\!\left(K\right)
 u_{\nu_0}\!\left(k,K,\Lambda_0\right)
 \nonumber\\
 &\quad\times
 \int\!\!\!\!\int dk'_{1}dK'_{1} k'_{1}K'_{1} P_{\bar n'_{12}0}\!\left(k'_{1}\right) P_{\bar n'0}\!\left(K'_{1}\right)
 u_{\nu_0}\!\left(k',K',\Lambda_0\right),
 \label{redmatContReg1}
\end{align}
where HO-wave function in the momentum space is written as
\begin{align}
 P_{nl}(k)
 &=
 \left(\frac{2}{\pi}\right)^{\!\!\frac{1}{2}}k \int dr r j_{l}(kr)R_{nl}\!\left(r,b_0\right)
 =(-)^{n}R_{nl}\!\left(k,\frac{1}{b_0}\right).
 \label{momHO1}
\end{align}
Here $j_l$ is the spherical Bessel function and we explicitly put the argument $b_0$ in $R_{nl}$ defined by Eq.~\eqref{HOsolution}.
The momentum integration in Eq.~\eqref{redmatContReg1} needs to be carried out numerically.

\subsubsection{One-pion exchange plus contact term}\label{Sec3BME1pi}
The Jacobi-HO matrix element $\Braket{\bar\kappa';\mathcal{I}T \left|W_{3N}^{(1\pi)}\right|\bar\kappa;\mathcal{I}T}$ of the 1PE term
is slightly complicated:
\begin{align}
 &\Braket{\bar\kappa';\mathcal{I}T \left|W_{3N}^{(1\pi)}\right|\bar\kappa;\mathcal{I}T}
 \nonumber\\
 &\quad=
 \frac{\sqrt{3}}{4\pi^4}
 \frac{g_Ac_D}{f_\pi^4\Lambda_\chi}
 \delta_{\bar l_{12}0}\delta_{\bar l_{12}'0}\delta_{\bar S_{12}\bar I_{12}}\delta_{\bar S_{12}'\bar I_{12}'}
 i^{\bar l+\bar l'}(-)^{\bar l +\bar I+\mathcal{I}+T+\frac{1}{2}}
 \hat{\bar S}_{12}\hat{\bar S}'_{12}\hat{\bar I}\hat{\bar I}'\hat{\bar T}_{12}\hat{\bar T}'_{12}
 \nonumber\\
 &\quad\times
 \begin{Bmatrix}
  \bar T_{12} & \bar T_{12}' & 1 \\[5pt]
  \frac{1}{2} & \frac{1}{2}  & \frac{1}{2}
 \end{Bmatrix}
 \begin{Bmatrix}
  \bar T_{12} & \bar T_{12}' & 1 \\[5pt]
  \frac{1}{2} & \frac{1}{2}  & T
 \end{Bmatrix}
 \begin{Bmatrix}
  \bar S_{12} & \bar S_{12}' & 1 \\[5pt]
  \frac{1}{2} & \frac{1}{2}  & \frac{1}{2}
 \end{Bmatrix}
 \begin{Bmatrix}
  \bar S_{12} & \bar S_{12}' & 1 \\[5pt]
  \bar I'     & \bar I       & \mathcal{I}
 \end{Bmatrix}
 \nonumber\\
 &\quad\times
 \sum_{\lambda_0\lambda_1\lambda_2}
 \hat{\lambda}_0 \widehat{\lambda_0-\lambda_1}\binom{2\lambda_0+1}{2\lambda_1}^{\!\frac{1}{2}}
 \left( 1 0 1 0 | \lambda_0 0\right)
 \left( \lambda_0-\lambda_1, 0 \lambda_2 0 | \bar l 0\right)
 \left( \lambda_1 0 \lambda_2 0 | \bar l' 0\right)
 \nonumber\\
 &\quad\times
 \begin{Bmatrix}
  \lambda_0-\lambda_1 & \lambda_1 & \lambda_0 \\
  \bar l'             & \bar l    & \lambda_2
 \end{Bmatrix}
 \begin{Bmatrix}
  \frac{1}{2} & \bar l'   & \bar I' \\[5pt]
  \frac{1}{2} & \bar l    & \bar I  \\[5pt]
  1           & \lambda_0 & 1
 \end{Bmatrix}
 \nonumber\\
 &\quad\times
 \int \!\!\!\!\int\!\!\!\!\int\!\!\!\!\int dkdk' dK dK' k k'
 K^{\lambda_0-\lambda_1+1}K'^{\lambda_1+1} f_{\lambda_2}^{(\lambda_0)}(K,K')
 \nonumber\\
 &\quad\times
 P_{\bar n_{12}0}\!\left(k\right) P_{\bar n_{12}'0}\!\left(k'\right)
 P_{\bar n\bar l}\!\left(K\right) P_{\bar n'\bar l'}\!\left(K'\right)
 u_{\nu_0}\!\left(k,K,\Lambda_0\right)
 u_{\nu_0}\!\left(k',K',\Lambda_0\right),
 \label{redmat1piReg1}
\end{align}
with the binomial coefficient $\binom{n_1}{n_2} = n_1!/\left[\left(n_1-n_2\right)!n_2!\right]$.
The function $f_{\lambda_2}^{(\lambda_0)}$ originating from the multipole expansion of the propagator, 
together with the factor $(2/3)^{\lambda_0/2}q_c^{2-\lambda_0}$ coming from $\left(\vect{\sigma}_c\cdot\vect{q}_c\right)\left(\vect{\sigma}_b\cdot\vect{q}_c\right)$ is given by
\begin{align}
 f_{\lambda_2}^{(\lambda_0)}(K,K')
 &=
 \frac{\hat{\lambda}_2^2}{2}\int_{-1}^{1}dw P_{\lambda_2}(w)
 \frac{\left(\frac{2}{3}\right)^{\frac{\lambda_0}{2}}q_c^{2-\lambda_0}}{q_c^2+m_\pi^2},
 \label{MPEfun1pi}
\end{align}
where $P_{\lambda_2}$ is the Legendre polynomial as a function of $w=\cos \theta$ with the angle $\theta$ between $\vect{K}$ and $\vect{K}'$.

\subsubsection{Two-pion exchange term}\label{Sec3BME2pi}
The 2PE potentials given by Eqs.~\eqref{w2pic1} to~\eqref{w2pic4} depend on two momenta, $\vect{q}_b$ and $\vect{q}_c$,
which make the matrix elements cumbersome.
After some manipulations, one can derive the Jacobi-HO matrix elements as
\begin{align}
 &\Braket{\bar\kappa';\mathcal{I}T \left|W_{3N}^{(2\pi;c_1)}\right|\bar\kappa;\mathcal{I}T}
 \nonumber\\
 &\quad=
 3c_1m_\pi^2
 S_{\bar\kappa\bar\kappa'}^{\mathcal{I}T}
 \begin{Bmatrix}
  \bar S_{12} & \bar S_{12}' & 1 \\[5pt]
  \frac{1}{2} & \frac{1}{2}  & \frac{1}{2}
 \end{Bmatrix}
 \begin{Bmatrix}
  \bar T_{12} & \bar T_{12}' & 1 \\[5pt]
  \frac{1}{2} & \frac{1}{2}  & \frac{1}{2}
 \end{Bmatrix}
 \nonumber\\
 &\quad\times
 \sum_{\substack{\lambda_b \lambda_c \\\lambda_b'\lambda_b''}}
 (-)^{\lambda_b+1}
 \sum_{\substack{\lambda_1\lambda_2\lambda_3\\\lambda_3'\lambda_3''}}
 I_{\bar n_{12} \bar l_{12} \bar n \bar l \bar n_{12}' \bar l_{12}' \bar n' \bar l'L_b=2,L_c=2, L_b'=1,L_c'=1}
       ^{\nu_0\lambda_b\lambda_c\lambda_b'\lambda_b''\lambda_1\lambda_2\lambda_3\lambda_3'\lambda_3''}\!\left(\Lambda_0\right)
 \nonumber\\
 &\quad\times
 \sum_{l_1}(-)^{l_1}\hat{l}_1^2
 X_{\bar\kappa\bar\kappa' \mathcal{I}, L_0=1,L_b'=1,L_c'=1,l_0=\lambda_b,l_1}
 ^{\lambda_b\lambda_c\lambda_b'\lambda_b''\lambda_1\lambda_2\lambda_3\lambda_3'\lambda_3''},
 \label{ME2pic1-2Reg}\\
 \nonumber\\
 &\Braket{\bar\kappa';\mathcal{I}T \left|W_{3N}^{(2\pi;c_3)}\right|\bar\kappa;\mathcal{I}T}
 \nonumber\\
 &\quad=
 \frac{\sqrt{3}}{2}c_3
 S_{\bar\kappa\bar\kappa'}^{\mathcal{I}T}
 \begin{Bmatrix}
  \bar S_{12} & \bar S_{12}' & 1 \\[5pt]
  \frac{1}{2} & \frac{1}{2}  & \frac{1}{2}
 \end{Bmatrix}
 \begin{Bmatrix}
  \bar T_{12} & \bar T_{12}' & 1 \\[5pt]
  \frac{1}{2} & \frac{1}{2}  & \frac{1}{2}
 \end{Bmatrix}
 \nonumber\\
 &\quad\times
 \sum_{L_bL_c} \hat{L}_b\hat{L}_c
 \left(1 0 1 0 | L_b 0\right) \left(1 0 1 0 | L_c 0\right)
 \sum_{\substack{\lambda_b \lambda_c \\\lambda_b'\lambda_b''}}
 \sum_{\substack{\lambda_1\lambda_2\lambda_3\\\lambda_3'\lambda_3''}}
 I_{\bar n_{12} \bar l_{12} \bar n \bar l \bar n_{12}' \bar l_{12}' \bar n' \bar l' L_bL_c,L_b'=L_b,L_c'=L_c}
       ^{\nu_0\lambda_b\lambda_c\lambda_b'\lambda_b''\lambda_1\lambda_2\lambda_3\lambda_3'\lambda_3''}\!\left(\Lambda_0\right)
 \nonumber\\
 &\quad\times
 \sum_{l_0l_1}
 \hat{l}_0^2\hat{l}_1^2
 \begin{Bmatrix}
  L_b-\lambda_b & \lambda_b & L_b \\
  1             & 1         & l_0
 \end{Bmatrix}
 \begin{Bmatrix}
  l_0 & l_1 & 1         \\
  L_c & 1   & \lambda_b
 \end{Bmatrix}
 X_{\bar\kappa\bar\kappa' \mathcal{I}, L_0=1,L_b'=L_b,L_c'=L_c,l_0l_1}
 ^{\lambda_b\lambda_c\lambda_b'\lambda_b''\lambda_1\lambda_2\lambda_3\lambda_3'\lambda_3''},
 \label{ME2pic3-1Reg}\\
 \nonumber\\
 &\Braket{\bar\kappa';\mathcal{I}T \left|W_{3N}^{(2\pi;c_4)}\right|\bar\kappa;\mathcal{I}T}
 \nonumber\\
 &\quad=
 9\sqrt{3}c_4
 (-)^{\bar l_{12}'+1}S_{\bar\kappa\bar\kappa'}^{\mathcal{I}T}
 \begin{Bmatrix}
  \frac{1}{2} & \frac{1}{2} & \bar T_{12}' \\[5pt]
  \frac{1}{2} & \frac{1}{2} & \bar T_{12}  \\[5pt]
  1           & 1           & 1
 \end{Bmatrix}
 \nonumber\\
 &\quad\times
 \sum_{L_0L_bL_c} \hat{L}_0^2\hat{L}_b\hat{L}_c
 \left(1 0 1 0 | L_b 0\right) \left(1 0 1 0 | L_c 0\right)
 \begin{Bmatrix}
  L_0 & L_b & 1 \\
  1   & 1   & 1
 \end{Bmatrix}
 \begin{Bmatrix}
  \frac{1}{2} & \frac{1}{2} & \bar S_{12}' \\[5pt]
  \frac{1}{2} & \frac{1}{2} & \bar S_{12}  \\[5pt]
  1           & 1           & L_0
 \end{Bmatrix}
 \nonumber\\
 &\quad\times
 \sum_{\substack{\lambda_b \lambda_c \\\lambda_b'\lambda_b''}}
 \sum_{\substack{\lambda_1\lambda_2\lambda_3\\\lambda_3'\lambda_3''}}
 I_{\bar n_{12} \bar l_{12} \bar n \bar l \bar n_{12}' \bar l_{12}' \bar n' \bar l' L_bL_c,L_b'=L_b,L_c'=L_c}
       ^{\nu_0\lambda_b\lambda_c\lambda_b'\lambda_b''\lambda_1\lambda_2\lambda_3\lambda_3'\lambda_3''}\!\left(\Lambda_0\right)
 \nonumber\\
 &\quad\times
 \sum_{l_0l_1}
 \hat{l}_0^2\hat{l}_1^2
 \begin{Bmatrix}
  L_b-\lambda_b & \lambda_b & L_b \\
  1             & L_0       & l_0
 \end{Bmatrix}
 \begin{Bmatrix}
  l_0 & l_1 & 1 \\
  L_c & 1   & \lambda_b
 \end{Bmatrix}
 X_{\bar\kappa\bar\kappa'\mathcal{I}, L_0L_b'=L_b,L_c'=L_c,l_0l_1}
 ^{\lambda_b\lambda_c\lambda_b'\lambda_b''\lambda_1\lambda_2\lambda_3\lambda_3'\lambda_3''}.
 \label{ME2pic4-1Reg}
\end{align}
The coefficients involved in Eqs.~\eqref{ME2pic1-2Reg} to~\eqref{ME2pic4-1Reg} are defined by
\begin{align}
 S_{\bar\kappa\bar\kappa'}^{\mathcal{I}T}
 =
 \left[\frac{g_A}{\left(\pi f_\pi\right)^2}\right]^2
 i^{\bar l_{12}+\bar l_{12}'+\bar l+\bar l'}
 (-)^{\bar S_{12}+\bar I_{12}'-\bar I+\mathcal{I}+\bar T_{12}+\bar T_{12}'+T+\frac{1}{2}}
 \hat{\bar S}_{12}\hat{\bar S}_{12}'\hat{\bar I}_{12}\hat{\bar I}_{12}'\hat{\bar I}\hat{\bar I}'\hat{\bar T}_{12}\hat{\bar T}_{12}'
 \begin{Bmatrix}
  \bar T_{12} & \bar T_{12}' & 1 \\[5pt]
  \frac{1}{2} & \frac{1}{2}  & T
 \end{Bmatrix},
 \label{Scommon1}
\end{align}
\begin{align}
 &I_{\bar n_{12} \bar l_{12} \bar n \bar l \bar n_{12}' \bar l_{12}' \bar n' \bar l' L_bL_cL_b'L_c'}
    ^{\nu_0\lambda_b\lambda_c\lambda_b'\lambda_b''\lambda_1\lambda_2\lambda_3\lambda_3'\lambda_3''}\!\left(\Lambda_0\right)
 \nonumber\\
 &\quad=
 3^{-\frac{\lambda_b}{2}}(-)^{\lambda_b+\lambda_c+\lambda_b'+\lambda_b''+\lambda_1+\lambda_2+\lambda_3+\lambda_3'+\lambda_3''}
 \Widehat{L_b'-\lambda_b}\Widehat{L_c'-\lambda_c}\Widehat{L_b'-\lambda_b-\lambda_b'}\Widehat{\lambda_b-\lambda_b''}
 \Widehat{\lambda_3-\lambda_3'}\Widehat{\lambda_3-\lambda_3''}
 \nonumber\\
 &\quad\times
 \left[\binom{2L_b'+1}{2\lambda_b}\binom{2L_c'+1}{2\lambda_c}\binom{2(L_b'-\lambda_b)+1}{2\lambda_b'}\binom{2\lambda_b+1}{2\lambda_b''}\binom{2\lambda_3+1}{2\lambda_3'}\binom{2\lambda_3+1}{2\lambda_3''}\right]^{\frac{1}{2}}
 \nonumber\\
 &\quad\times
 \int\!\!\!\!\int\!\!\!\!\int\!\!\!\!\int dk dK dk' dK'
 k^{L_b'-\lambda_b-\lambda_b'+\lambda_3-\lambda_3'+1}K^{L_c'-\lambda_c+\lambda_b-\lambda_b''+\lambda_3-\lambda_3''+1}
 k'^{\lambda_b'+\lambda_3'+1}K'^{\lambda_c+\lambda_b''+\lambda_3''+1}
 \nonumber\\
 &\quad\times
 f_{\lambda_1\lambda_2\lambda_3}^{(L_bL_c)}(k,k',K,K')
 P_{\bar n_{12}\bar l_{12}}\!\left(k\right)P_{\bar n\bar l}\!\left(K\right)
 P_{\bar n_{12}'\bar l_{12}'}\!\left(k'\right)P_{\bar n'\bar l'}\!\left(K'\right)
 u_{\nu_0}\!\left(k,K,\Lambda_0\right)
 u_{\nu_0}\!\left(k',K',\Lambda_0\right),
 \label{Icommon2}
\end{align}
\begin{align}
 &X_{\bar\kappa\bar\kappa'\mathcal{I} L_0L_b'L_c'l_0l_1}
   ^{\lambda_b\lambda_c\lambda_b'\lambda_b''\lambda_1\lambda_2\lambda_3\lambda_3'\lambda_3''}
 \nonumber\\
 &\quad=
 \sum_{l_2l_3} \hat{l}_2\hat{l}_3
 \left(L_c'-\lambda_c,0,\lambda_b-\lambda_b'',0|l_2 0\right)
 \left(\lambda_c 0 \lambda_b''0|l_3 0\right)
 \begin{Bmatrix}
  \lambda_b-\lambda_b'' & \lambda_b'' & \lambda_b \\
  L_c'-\lambda_c        & \lambda_c   & L_c'      \\
  l_2                   & l_3         & l_1
 \end{Bmatrix}
 \nonumber\\
 &\quad\times
 \sum_{\substack{\lambda\lambda'\\\Lambda\Lambda'}}
 \hat{\lambda}\hat{\lambda}'\hat{\Lambda}\hat{\Lambda}'
 \left(L_b'-\lambda_b-\lambda_b', 0 \lambda 0 | \bar l_{12} 0\right)
 \left(\lambda_b' 0 \lambda' 0 | \bar l_{12}' 0\right)
 \left(l_2 0 \Lambda 0 | \bar l 0\right)
 \left(l_3 0 \Lambda' 0 | \bar l' 0\right)
 \nonumber\\
 &\quad\times
 \left(\lambda_2 0,\lambda_3-\lambda_3',0 | \lambda 0\right)
 \left(\lambda_2 0 \lambda_3' 0 | \lambda' 0\right)
 \left(\lambda_1 0, \lambda_3-\lambda_3'', 0| \Lambda 0\right)
 \left(\lambda_1 0 \lambda_3'' 0 | \Lambda' 0\right)
 \nonumber\\
 &\quad\times
 \begin{Bmatrix}
  \lambda_3-\lambda_3' & \lambda_3' & \lambda_3 \\
  \lambda'             & \lambda    & \lambda_2
 \end{Bmatrix}
 \begin{Bmatrix}
  \lambda_3-\lambda_3'' & \lambda_3'' & \lambda_3 \\
  \Lambda'              & \Lambda     & \lambda_1
 \end{Bmatrix}
 \nonumber\\
 &\quad\times
 \sum_{L_1L_2L_3}
 (-)^{L_1+L_2+L_3}
 \hat{L}_1^2\hat{L}_2^2\hat{L}_3^2
 \begin{Bmatrix}
  \bar I_{12} & \bar I_{12}' & L_1 \\
  \bar I'     & \bar I       & \mathcal{I}
 \end{Bmatrix}
 \begin{Bmatrix}
  L_0       & L_b'-\lambda_b & l_0 \\
  \lambda_3 & L_1            & L_2
 \end{Bmatrix}
 \begin{Bmatrix}
  1         & l_1 & l_0 \\
  \lambda_3 & L_1 & L_3
 \end{Bmatrix}
 \nonumber\\
 &\quad\times
 \begin{Bmatrix}
  \bar S_{12}' & \bar l_{12}' & \bar I_{12}' \\[5pt]
  \bar S_{12}  & \bar l_{12}  & \bar I_{12}  \\[5pt]
  L_0          & L_2          & L_1
 \end{Bmatrix}
 \begin{Bmatrix}
  \frac{1}{2} & \bar l' & \bar I' \\[5pt]
  \frac{1}{2} & \bar l  & \bar I  \\[5pt]
  1           & L_3     & L_1
 \end{Bmatrix}
 \begin{Bmatrix}
  L_b'-\lambda_b-\lambda_b' & \lambda_b'   & L_b'-\lambda_b \\[5pt]
  \lambda                   & \lambda'     & \lambda_3      \\[5pt]
  \bar l_{12}               & \bar l_{12}' & L_2
 \end{Bmatrix}
 \begin{Bmatrix}
  l_2     & l_3      & l_1       \\[5pt]
  \Lambda & \Lambda' & \lambda_3 \\[5pt]
  \bar l  & \bar l'  & L_3
 \end{Bmatrix},
  \label{Xcommon1}
\end{align}
as well as 
\begin{align}
 f_{\lambda_1\lambda_2\lambda_3}^{(L_bL_c)}(k,k',K,K')
 &=
 \frac{\hat{\lambda}_1^2\hat{\lambda}_2^2\hat{\lambda}_3^2}{8}
 \int_{-1}^{1} \int_{-1}^{1}  \int_{-1}^{1} dw_1 dw_2 dw_3
  P_{\lambda_1}(w_1) P_{\lambda_2}(w_2) P_{\lambda_3}(w_3)
  \nonumber\\
  &\times
  \left(\left|\vect{k}-\vect{k}'\right| \left|\vect{K}-\vect{K}'\right| \right)^{-\lambda_3}
  \frac{2^{-\frac{L_b}{2}}\left(\frac{2}{3}\right)^{\frac{L_c}{2}} q_b^{2-L_b}q_c^{2-L_c}}{\left(q_b^2+m_\pi^2\right)\left(q_c^2+m_\pi^2\right)},
 \label{c3f123}
\end{align}
which originates from the triple-fold multipole expansion with respect to
$w_1=\cos\theta_1$, $w_2=\cos\theta_2$, and $w_3=\cos\theta_3$ respectively defined by the angles
$\theta_1$, $\theta_2$, and $\theta_3$ between $\vect{K}$ and $\vect{K}'$, $\vect{k}$ and $\vect{k}'$,
and $\vect{K}-\vect{K}'$ and $\vect{k}-\vect{k}'$.

\bibliographystyle{elsarticle-num_noURL}
\bibliography{review_1}

\end{document}